\documentclass[a4paper,11pt]{article}
\pdfoutput=1 
\usepackage{jheppub} 
\usepackage[T1]{fontenc} 
\usepackage{amsmath,amssymb,epsfig,txfonts,relsize}

\allowdisplaybreaks

\definecolor{nicered}{rgb}{0.7,0.1,0.1}
\definecolor{nicegreen}{rgb}{0.1,0.5,0.1}
\definecolor{niceblue}{rgb}{0.0,0.1,0.7}
\hypersetup{colorlinks,citecolor=nicegreen,linkcolor=nicered,urlcolor=niceblue}

\def\bm#1{\mbox{\boldmath$#1$\unboldmath}}      

\title{Searching for heavy Higgs bosons \\ in the $\bm{t \bar t Z}$ and $\bm{t b W}$ final states}

\author[1,2]{Ulrich Haisch}

\author[3]{and Giacomo Polesello}

\affiliation[1]{Rudolf Peierls Centre for Theoretical Physics,
   University of Oxford, OX1 3NP Oxford, United Kingdom}
    
\affiliation[2]{CERN, Theoretical Physics Department,  CH-1211 Geneva 23, Switzerland}

\affiliation[3]{INFN, Sezione di Pavia, Via Bassi 6, 27100 Pavia, Italy}

\emailAdd{ulrich.haisch@physics.ox.ac.uk}
\emailAdd{giacomo.polesello@cern.ch}

\abstract{
\phantom{iii} In the context of two-Higgs doublet models, we explore the possibility of searching for heavy Higgs bosons in the $t \bar t Z$ and $t bW$ final states. We develop realistic analysis strategies and in the case of the $t \bar t Z$ channel provide a detailed evaluation of the new-physics reach  at the 14 TeV LHC. We find that already with an integrated luminosity of $300 \, {\rm fb}^{-1}$ searches for the $t \bar t Z$ signature can provide statistically significant constraints at low values of $\tan \beta$ for heavy Higgs masses in the range from around $450 \, {\rm GeV}$ to $1150 \, {\rm GeV}$. Future searches for heavy Higgses in the $tbW$ final state are also expected to be able to probe parts of this parameter space, though the precise constraints turn out to depend sensitively on  the assumed systematics on the shape of the $t \bar t$ background.}

\preprint{CERN-TH-2018-155}

\begin{document} 

\maketitle

\section{Introduction}
\label{sec:introduction}

The most important accomplishment of the LHC~Run-1  physics programme has been the discovery of a new spin-0 resonance $h$ with a mass of  around $125 \, {\rm GeV}$ in 2012~\cite{Aad:2012tfa,Chatrchyan:2012xdj}. In the last five years the LHC~Higgs programme has matured, providing precise measurements of processes   such as $pp \to h \to \gamma \gamma$ and $pp \to h \to ZZ^\ast \to  \ell^+  \ell^- \ell^+  \ell^-$ (see~\cite{Sirunyan:2017exp,Aaboud:2017oem,Aaboud:2018xdt,Sirunyan:2018ouh} for the latest LHC results at $\sqrt{s} = 13 \, {\rm TeV}$) with standard~model~(SM) rates of around $100 \, {\rm fb}$ and $5 \, {\rm fb}$, respectively. 

The finding that the $125 \, {\rm GeV}$ spin-0 resonance has properties   close to the one expected for the SM~Higgs~\cite{Khachatryan:2016vau} implies that if additional Higgs bosons exist in nature such states can only  be slightly  mixed with the $h$. An extended Higgs sector has thus to be approximately aligned, either via decoupling or via alignment without decoupling. While in the former case the extra spin-0 particles might be too heavy to be accessible at the LHC, in the latter case the additional  Higgs bosons can have masses at or not far above the electroweak~(EW) scale without being in conflict with any other observation. In the case of alignment without decoupling, direct searches for extra  Higgs-like particles are hence particularly well-motivated as they can provide complementary information with respect to the LHC programme of precision Higgs measurements. 

The existing ATLAS and CMS searches for heavy neutral CP-even (CP-odd) Higgses $H$ ($A$) cover by now a wide range of final states  (cf.~\cite{Aad:2015pla,CMS-PAS-HIG-16-007} for  LHC~Run-1  summaries), and their results are routinely interpreted in the context of two-Higgs doublet models~(2HDMs) or the minimal supersymmetric SM~(MSSM). Well-studied  channels are $pp \to (b \bar b) H/A \to (b \bar b) \, \tau^+ \tau^-$~\cite{Aaboud:2017sjh,Sirunyan:2018zut} and  $pp \to b \bar b H/A \to b \bar b b \bar b$~\cite{Khachatryan:2015tra,CMS-PAS-HIG-16-025,Sirunyan:2018taj}, which provide the leading direct constraints on the parts of the 2HDM and MSSM parameter space where the $H/A$ couplings to taus and bottom quarks are $\tan \beta$-enhanced. If the Higgs sector is not fully decoupled/aligned, the processes $pp \to H \to WW$~\cite{Khachatryan:2015cwa,Aaboud:2017gsl} and  $pp \to H \to  ZZ$~\cite{Aaboud:2017itg,Aaboud:2017rel,Sirunyan:2018qlb} can provide important bounds as well. Other interesting modes are $pp \to H/A \to A/HZ \to b \bar b \ell^+ \ell^-$~\cite{Khachatryan:2016are,CMS-PAS-HIG-16-010,Aaboud:2018eoy} since these channels have non-zero rates even in the exact decoupling/alignment limit. Depending on the model realisation, useful  information can  also be obtained  from $p p \to A \to hZ \to b \bar b/\tau^+ \tau^- \ell^+ \ell^-$~\cite{Aaboud:2017cxo,Khachatryan:2015tha}, $pp \to H \to hh \to WW\gamma\gamma/b \bar b b \bar b$~\cite{ATLAS-CONF-2016-071,Sirunyan:2018zkk}, $pp \to H/A \to \gamma \gamma$~\cite{Khachatryan:2016yec,Aaboud:2017yyg} and $pp \to H/A \to Z\gamma$~\cite{Aaboud:2017uhw,Sirunyan:2017hsb}.

All the channels mentioned so far have  in common that they only have limited sensitivity to additional Higgses with masses above the top threshold, in particular if  the $H/A \to t \bar t$ branching ratio is sizeable as it  happens to be the case  in the MSSM at low and moderate $\tan \beta$. In order to gain sensitivity to new-physics scenarios of the latter kind the channels $pp \to H/A \to t \bar t$, $pp \to t \bar t H/A \to t \bar t t \bar t$ and $pp \to b \bar b H/A \to b \bar b t \bar t$ have been proposed~(see~\cite{Djouadi:2015jea,Craig:2015jba,Hajer:2015gka,Gori:2016zto,Alvarez:2016nrz} for instance) and first experimental searches for the $t \bar t$~\cite{Aaboud:2017hnm},  $t \bar t t \bar t$~\cite{Sirunyan:2017roi,Aaboud:2018xuw} and $b \bar b t \bar t$~\cite{Aaboud:2018xuw} final state have been  carried out recently. While, at first sight, all three signatures seem to offer good prospects for probing heavy Higgs bosons,  it turns out that in practice they all suffer certain limitations. In the case of $pp \to H/A \to t \bar t$, interference effects between the signal and the SM $t \bar t$ background~\cite{Gaemers:1984sj,Dicus:1994bm,Bernreuther:1997gs,Frederix:2007gi,Hespel:2016qaf} represent a serious obstacle, while for what concerns the searches for $pp \to t \bar t H/A \to t \bar t t \bar t$ and $pp \to b \bar b H/A \to b \bar b t \bar t$ the small signal-over-background ratio is in general an issue. As a result,  a very good experimental understanding of the systematic uncertainties plaguing the overwhelming~$t \bar t$~background is crucial in order for the $t \bar t$, $ t \bar t t \bar t$ and $b \bar b t \bar t$ final states to provide statistically significant constraints at low to moderate values of $\tan \beta$. 

In  this work two novel search strategies for neutral Higgs particles with masses above the top-quark threshold are devised. The first strategy exploits the $t \bar t Z$ final state and is based on the isolation of the irreducible $t \bar tZ$ process from other SM~backgrounds, followed by the discrimination of the signal from SM $t \bar tZ$  production using the distinctive kinematic features of the new-physics signal.  As a compromise  between purity and statistics, we consider final states where  the $Z$ boson decays into charged leptons, and only one of  the two top quarks decays semileptonically. The examined final state thus involves three charged leptons,~i.e.~a~pair of same-flavour  leptons compatible with the $Z \to \ell^+ \ell^-$ decay and one charged lepton from $t \to b W \to b \ell \nu$, missing transverse energy~$(E_{T, \rm miss})$ associated to the neutrinos  from top decays and four jets, two of which are produced via bottom-quark fragmentation. The invariant masses of the $t \bar t Z$ and $t \bar t$ systems~($m_{t \bar t Z}$ and $m_{t \bar t}$) can be experimentally reconstructed and their distributions are peaked at the masses of the heavy Higgs bosons appearing in  $H/A \to A/H Z \to t \bar t Z$. To separate signal from background  a shape fit to the distribution of the variable $\Delta m \equiv m_{t \bar t Z} - m_{t \bar t}$ can be used, since  this spectrum is smoothly falling with~$\Delta m$ in the case of  the SM $t \bar t Z$ background. An observable that provides additional information while being well measurable is the $Z$-boson transverse momentum~($p_{T,Z}$). The shape of the $p_{T,Z}$ spectrum of the $t \bar t Z$ signal is in fact predicted to be Jacobian with an endpoint that is related to the structure of the massive two-body phase describing the $H/A \to A/H Z$ transition. We will show that by using  the experimental informations on $\Delta m$ and $p_{T,Z}$, future searches for the $t \bar tZ$ final state should allow to set unique bounds on the~part of the 2HDM parameter space that features heavy Higgses with masses above the top  threshold and small values of  $\tan \beta$. 

Our second search strategy targets the $tbW$ final state. We point out that there is a number of kinematic handles that can be used to separate  the $tbW$ signal from the $t \bar t$ background. In the case of the two-lepton final state,  one can exploit the invariant masses of the $b \bar b$ and $b \bar b \ell$ systems~($m_{b\bar b}$ and $m_{b \bar b \ell}$) since the corresponding distributions have kinematic endpoints, while in the one-lepton final state the Breit-Wigner peaks in the invariant mass spectra of the $tb$ and $tbW$ systems~($m_{tb}$ and $m_{tbW}$) can be harnessed.  Based on these observations, we sketch  the main ingredients of an actual  two-lepton analysis.  In our exploratory study, the signal-over-background ratio however turns out to be at most a few percent for the considered 2HDM realisations, making it difficult to determine the precise LHC reach of the proposed two-lepton analysis. Similar statements also apply to the one-lepton case. A~full exploration of the potential of the $tbW$ final state is therefore left to the ATLAS and CMS collaborations once they have collected data in excess of $300 \, {\rm fb}^{-1}$. 

The outline of this  article is as follows. Section~\ref{sec:interactions} discusses the structure of the relevant Higgs interactions and the resulting decay modes, while the anatomy of the $t \bar t Z$ and $t b W$ signal is studied in~Section~\ref{sec:anatomyttZ} and~\ref{sec:anatomytbW}, respectively. A~concise description of  our Monte Carlo (MC) generation and detector simulation  is given  in Section~\ref{sec:montecarlo}.  The actual analysis strategies  are detailed in Sections~\ref{sec:strategyttZ}~and~\ref{sec:strategytbW}. In Section~\ref{sec:results} we present our numerical results and examine the new-physics sensitivity of the $t \bar t Z$~signature at upcoming LHC runs. We conclude in Section~\ref{sec:conclusions}. Supplementary material is provided in Appendices~\ref{app:pTZfit} and~\ref{app:discovery}.

\section{Heavy Higgs interactions and decays}
\label{sec:interactions}

The addition of the second Higgs doublet in 2HDMs leads to five physical spin-0 states: two neutral CP-even ones ($h$ and $H$), one neutral CP-odd state ($A$), and the remaining two carry electric charge of $\pm 1$ and are degenerate in mass ($H^\pm$). Following standard practice, we identify the $125 \, {\rm GeV}$ resonance discovered at the LHC with  the $h$ field, denote the angle that mixes the neutral CP-even states  by~$\alpha$, and define $\tan \beta$ to be the ratio of the Higgs vacuum expectation values~(VEVs). 

The tree-level couplings of the Higgses $h,H$ to EW gauge bosons satisfy in  all 2HDMs  with a CP-conserving Higgs potential the relations
\begin{equation} \label{eq:ghHVV}
g_{hVV} \propto s_{\beta - \alpha} \,, \qquad 
g_{HVV} \propto c_{\beta - \alpha} \,,
\end{equation}
where $V = W, Z$ and we have used the shorthand notation $s_{\beta - \alpha} \equiv \sin \left (\beta - \alpha \right )$ and $c_{\beta - \alpha} \equiv \cos \left (\beta - \alpha \right )$.  Notice that in the so-called alignment limit,~i.e.~$\alpha \to \beta - \pi/2$, the interactions of $h$ with EW gauge bosons resembles those in the SM while the couplings between $H$ and $W$-boson or $Z$-boson pairs vanish identically.  The consistency of the LHC Higgs measurements with SM predictions requires that  any 2HDM Higgs sector is close to the alignment limit, meaning that small values of $c_{\beta - \alpha}$ are experimentally favoured. 

The combinations of mixing angles appearing in (\ref{eq:ghHVV}) also govern the interactions between two Higgses and one EW gauge boson. Explicitly one has 
\begin{equation} \label{eq:ghHAZ}
g_{hAZ} \propto c_{\beta - \alpha} \,, \qquad 
g_{HAZ} \propto s_{\beta - \alpha} \,, \qquad 
g_{HH^\pm W^{\mp}} \propto s_{\beta - \alpha}  \,, \qquad 
g_{AH^\pm W^{\mp}} \propto s_{\beta - \alpha}  \,.
\end{equation}
Notice that the first relation leads to a suppression of the $A \to h Z$ decay rate in the alignment limit. In contrast, the decay rate $H \to A Z$ ($A \to H Z$) is unsuppressed for $c_{\beta - \alpha} \to 0$ and can be large if this channel is kinematically allowed, i.e.~$M_H > M_A + M_Z$ ($M_A > M_H + M_Z$). Like $g_{HAZ}$ also  $g_{HH^\pm W^{\mp}}$ and $g_{AH^\pm W^{\mp}}$ are non-vanishing in the alignment limit, and in consequence the decays $H \to H^\pm W^\mp$ and $A \to H^\pm W^\mp$ are phenomenologically relevant  if they are open.

\begin{figure}[!t]
\begin{center}
\includegraphics[width=0.95\textwidth]{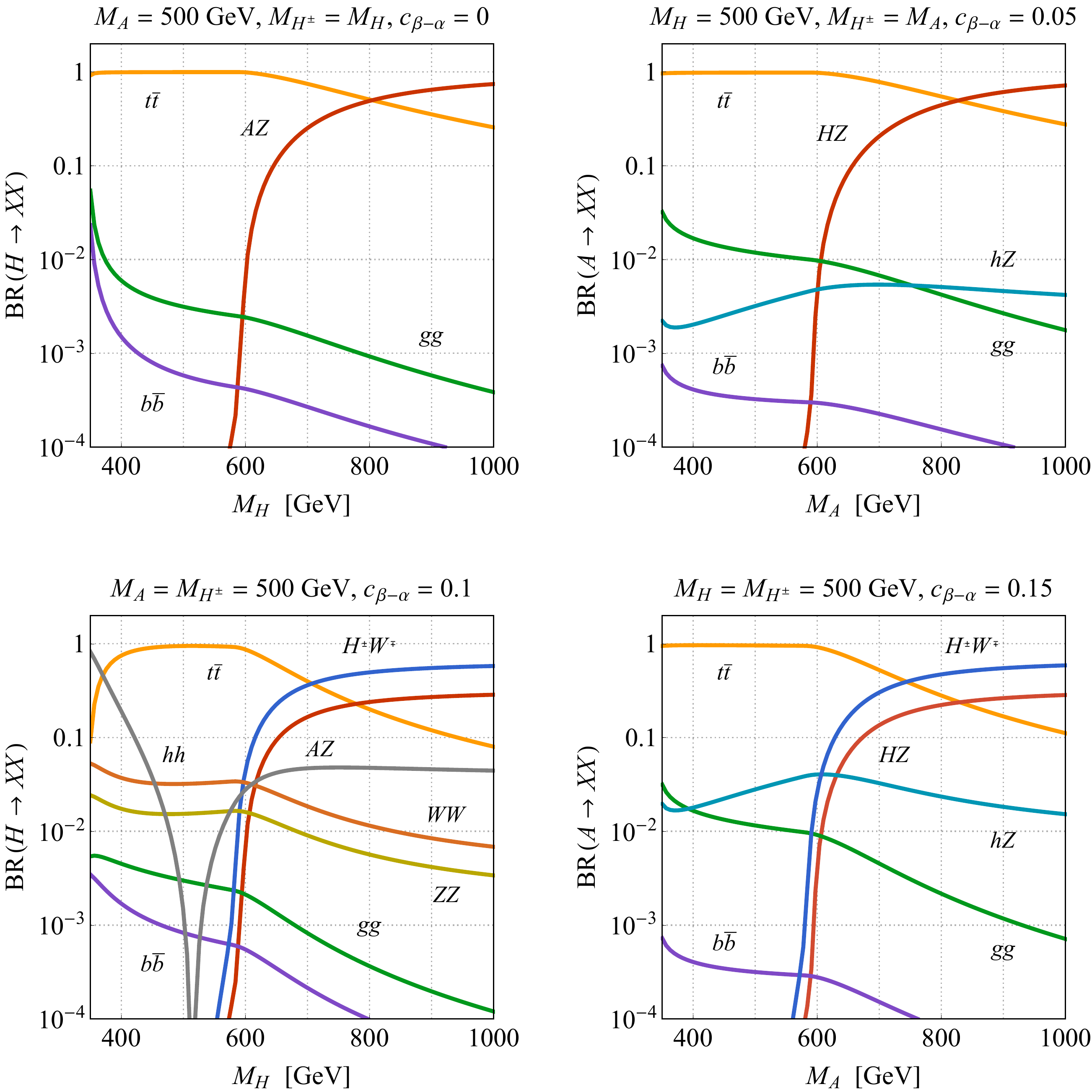}  
\vspace{2mm}
\caption{\label{fig:branchingratios} Branching ratios of the heavy Higgses $H$ and $A$ in the type-II~2HDM.  The shown results all correspond to $t_\beta = 1$, while different parameter choices have been used for $M_H, M_A, M_{H^\pm}$ and $c_{\beta -\alpha}$ as indicated in the headline of the panels. 
}
\end{center}
\end{figure}

In order to tame dangerous tree-level flavour-changing neutral currents the Yukawa interactions in 2HDMs have to satisfy the natural flavour conservation hypothesis~\cite{Glashow:1976nt,Paschos:1976ay}. Depending on which fermions couple to which Higgs doublet, one can divide the resulting 2HDMs into four different types. While the Higgs couplings  to light fermions turn out to be model dependent, the  couplings of $h$, $H$ and $A$  to top quarks take in all four cases the generic form 
\begin{equation}  \label{eq:ghHAtt}
g_{ht \bar t} \propto s_{\beta - \alpha} \,, \qquad 
g_{Ht \bar t} \propto c_{\beta - \alpha} - \frac{s_{\beta -  \alpha}}{t_\beta} \,, \qquad 
g_{At \bar t}\propto \frac{1}{t_\beta} \,,
\end{equation}
where we have introduced the abbreviation $t_\beta \equiv  \tan \beta$. These expressions imply that in the alignment limit the coupling of $h$ to top quarks becomes SM-like, while the top couplings of $H$, $A$ are both $t_\beta$ suppressed. The only  charged Higgs coupling to fermions relevant to our work is the one to  right-handed anti-top and left-handed bottom quarks.  This coupling resembles the form of $g_{A t \bar t}$, and in consequence the charged Higgs decays dominantly via $H^+ \to t \bar b$ if this channel is open. 

The magnitudes of 2HDM couplings that involve more than two Higgses depend on the precise structure of the full scalar potential. For what concerns the coupling $g_{Hhh}$ that describes the self-coupling between a $H$ and two $h$, it turns out that it is homogenous in $c_{\beta - \alpha}$, and therefore vanishes in the alignment limit in pure 2HDMs (see~\cite{Craig:2013hca} for example). In fact, in the limit $c_{\beta - \alpha} \to 0$ and $M_{H^\pm} > M_{H} > v, M_h$ with $v \simeq 246 \, {\rm GeV}$  the Higgs~VEV and assuming that the quartic couplings~$\lambda_i$ that appear in the scalar potential are of order 1, the $g_{Hhh}$ coupling behaves approximately as $g_{Hhh} \propto c_{\beta -\alpha} \, M_{H^\pm}^2/v$.  It~follows that for a sufficiently large mass splitting $M_{H^\pm} - M_H >0$,  the partial decay width $\Gamma \left (H \to hh \right ) \propto g_{Hhh}^2/M_H$ can be numerically relevant in pure 2HDMs. In contrast, in the MSSM the trilinear $Hhh$ coupling  scales  as $g_{Hhh} \propto  M_Z^2/v \hspace{0.25mm} s_{4 \beta}$  in the limit $\alpha \to \beta - \pi/2$. The coupling $g_{Hhh}$ is hence non-zero in the alignment (or decoupling) limit of the~MSSM, but since $\Gamma \left (H \to hh \right ) \propto g_{Hhh}^2/M_H$ while  $\Gamma \left (H \to t \bar t \right ) \propto g_{Ht\bar t}^2 \hspace{0.75mm} M_H$, the branching ratio of $H \to hh$ is always small for Higgs masses $M_H$ sufficiently above the top threshold.  

The above discussion suggests that close to the alignment limit the decay pattern of the heavy Higgses $H$ and $A$ is rather simple in all 2HDMs. To   corroborate this statement we show in Figure~\ref{fig:branchingratios} the branching ratios of $H$ and $A$ for  four  type-II 2HDM benchmark models. The different benchmarks thereby cover values of $c_{\beta - \alpha}$ that range from the pure alignment limit $c_{\beta -\alpha} = 0$ to the case of maximally allowed misalignment, which amounts to around $c_{\beta -\alpha} = 0.15$ in the type-II 2HDM after~LHC Run-1 (see for instance \cite{CMS-PAS-HIG-16-007}). Our calculation of the branching ratios is based on the formulas and results given in~\cite{Djouadi:1995gv,Djouadi:2005gi,Djouadi:2005gj,Alves:2017snd,Haisch:2017gql,Haisch:2018kqx}. From the upper left panel one observes that for $M_H \lesssim 600 \, {\rm GeV}$ the decay $H \to t \bar t$  almost fully saturates the total width of $H$, while for $M_H \gtrsim  600 \, {\rm GeV}$ the decay mode $H \to AZ$ becomes important quickly and even dominant for $M_H \gtrsim  800 \, {\rm GeV}$. A similar picture arises in the case of the $A$ with $A \to t \bar t$ and $A \to HZ$ representing the two dominant decay modes for $M_A \gtrsim 600 \, {\rm GeV}$.  This feature is illustrated in the  upper right panel in Figure~\ref{fig:branchingratios}. 

Notice  that  to obtain the latter plots we have fixed $M_{H^\pm} = M_H$ and $M_{H^\pm} = M_A$, respectively. These choices are  well-motivated, because only in these two cases~\cite{Haber:1992py,Pomarol:1993mu,Gerard:2007kn,Grzadkowski:2010dj,Haber:2010bw} can the $H$ or the $A$ have a sizeable mass splitting from the rest of the non-SM Higgses without being in conflict with EW precision measurements. The left (right) panel shown in the lower row of~Figure~\ref{fig:branchingratios} illustrate how the decay pattern of $H$ ($A$) changes if the charged Higgs mass is instead  set equal to the mass of the heavy CP-odd (CP-even) Higgs. One observes that  for such parameter choices besides $H \to t \bar t$ and  $H \to AZ$ ($A \to t \bar t$ and $A \to HZ$) also the channel  $H \to H^\pm W^\mp$ ($A \to H^\pm W^\mp$) is important  at high $M_H$ ($M_A$). This  feature is expected because $H$ ($A$) decays to a charged Higgs and a $W$~boson are  kinematically allowed if $M_{H} > M_{H^\pm} + M_W$  ($M_{A} > M_{H^\pm} + M_W$) and unsuppressed in the alignment limit $\big($see (\ref{eq:ghHAZ})$\big)$. From the lower left panel one furthermore sees that for a non-zero value of $c_{\beta-\alpha}$ the branching ratio of $H \to hh$ exceed the few-percent level  for $M_H \gtrsim  600 \, {\rm GeV}$, making it the fourth largest branching ratio for heavy CP-even Higgses $H$.  We add that the results shown in the latter panel correspond  to the choice $\lambda_3 = 3$, where $\lambda_3$ is the quartic coupling that multiplies the term $|H_1|^2 \, |H_2|^2$ in the 2HDM scalar potential, and $H_1$ and $H_2$ denote the two Higgs doublets in the~$Z_2$ basis. 

\section{Anatomy of the $\bm{t \bar t Z}$ signature}
\label{sec:anatomyttZ}

\begin{figure}[!t]
\begin{center}
\includegraphics[width=0.75\textwidth]{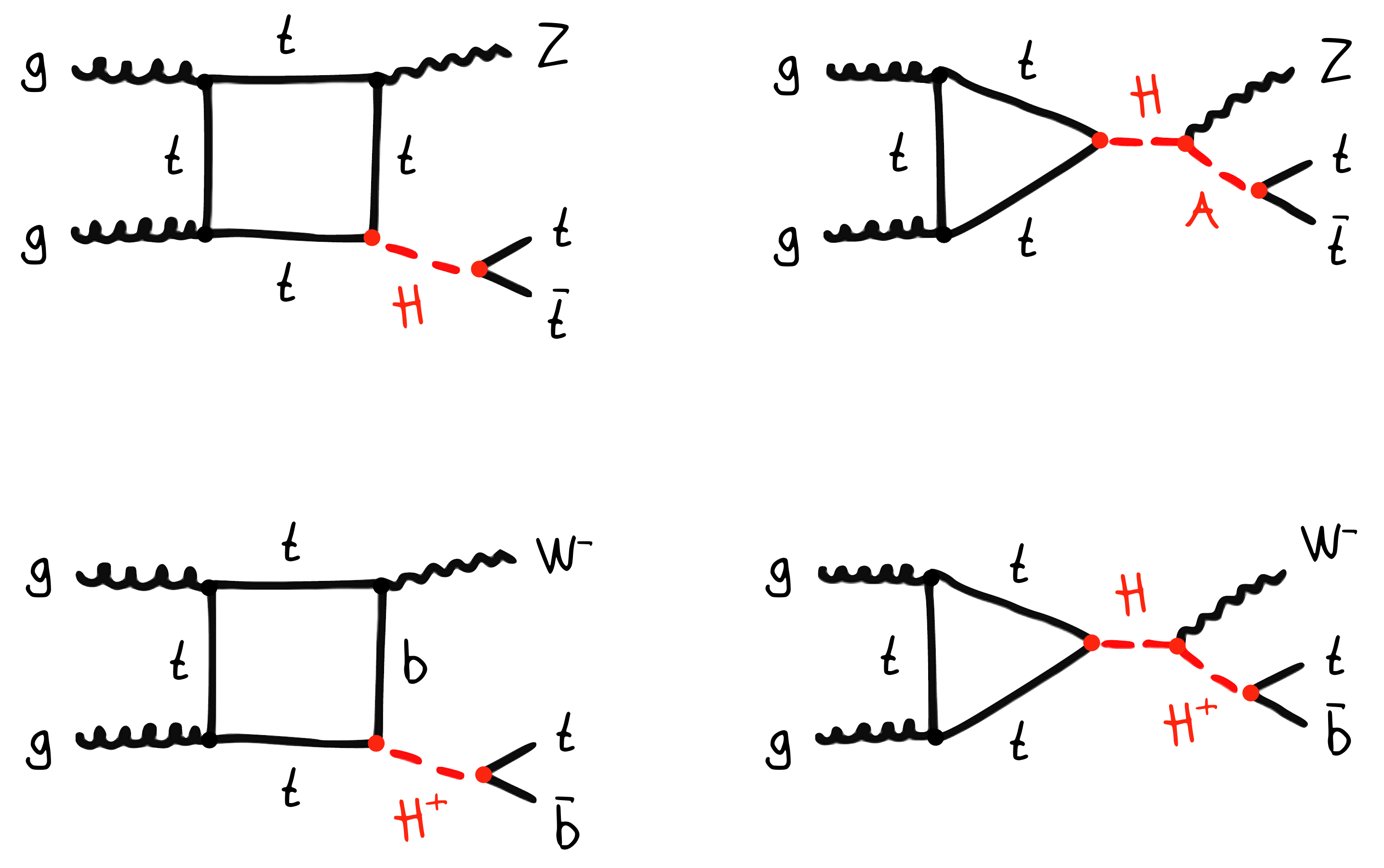}  
\vspace{4mm}
\caption{\label{fig:diagrams} Examples of Feynman diagrams that involve  the exchange of non-SM Higgses and contribute to the process $gg \to t \bar t Z$ (upper row) and $gg \to t \bar b W^-$ (lower row). See text for further explanations.}
\end{center}
\end{figure}

The discussion in the last section singles out the  $t \bar t Z$ and $tbW$ final states as promising  to search for the presence of heavy Higgs particles.  Prototypes of Feynman diagrams that  lead to the former signal in 2HDMs are shown in the upper row of Figure~\ref{fig:diagrams}. In the graph on the left-hand side a  $H$ is produced in association with a $Z$ boson from a top-quark box, while in the right diagram the $H$ is emitted from a top-quark triangle and then decays via $H \to AZ \to t \bar t Z$.  Graphs where the role of the neutral  Higgses $H$ and $A$ is interchanged also contribute to the $t \bar t Z$ signature in 2HDMs but are not explicitly shown in the figure. 

\begin{figure}[!t]
\begin{center}
\includegraphics[width=\textwidth]{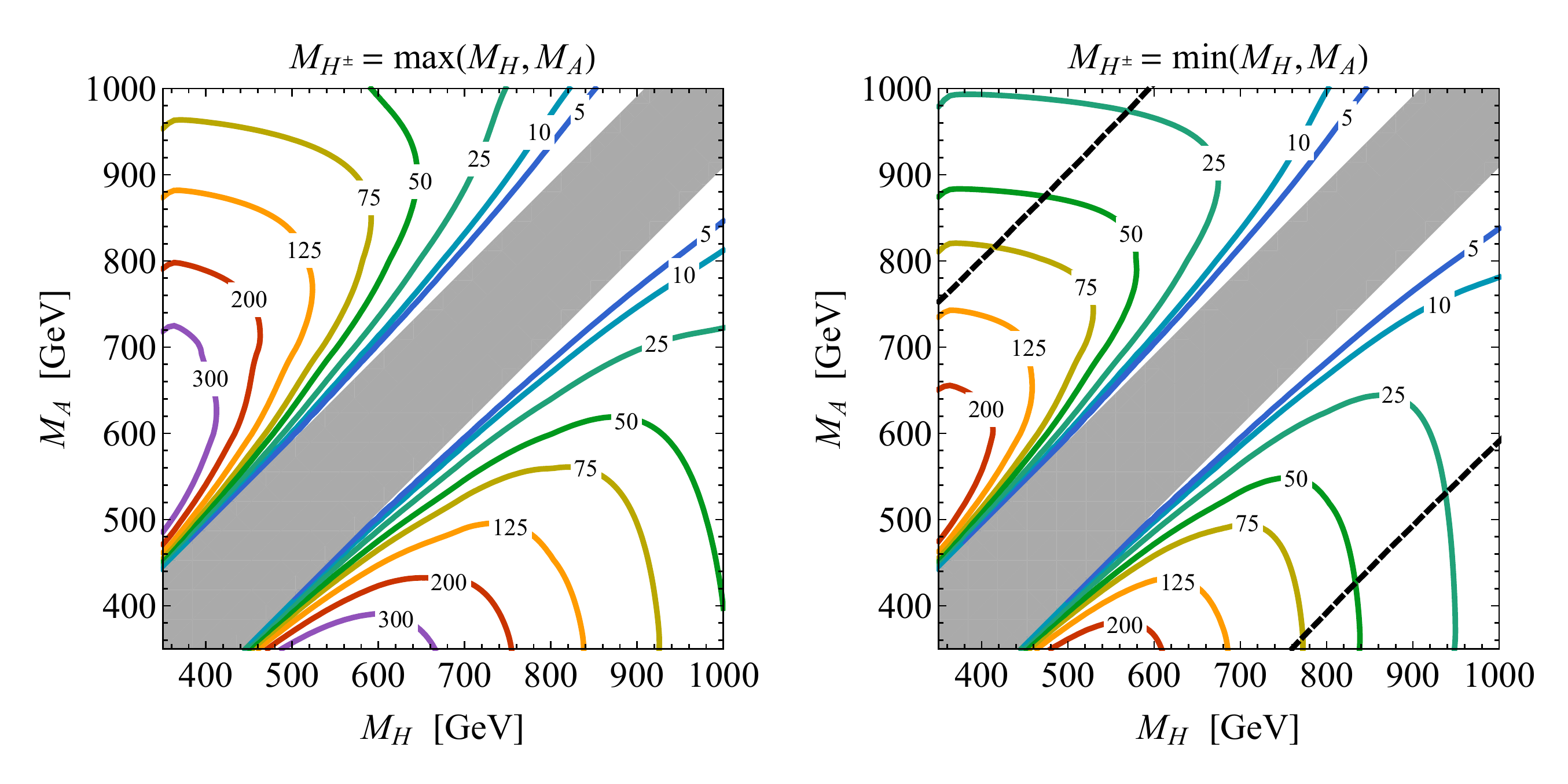} 
\vspace{-6mm}
\caption{\label{fig:ttZsignalstrengths} Approximate signal strengths  for $pp \to t \bar t Z$ production in units of $\rm fb$. The~left~(right) plot is obtained by fixing the charged Higgs mass to $M_{H^\pm} = {\rm max} \left (M_H , M_A \right )$ $\big($$M_{H^\pm} = {\rm min} \left (M_H , M_A \right )$$\big)$ in the scan. Both panels correspond to the type-II 2HDM and employ $\sqrt{s} = 14 \, {\rm TeV}$, $c_{\beta-\alpha} = 0$ and $t_\beta=1$. The grey bands indicate the values of $M_H$ and $M_A$ that are kinematically inaccessible in $pp \to H/A \to A/H Z \to t \bar t Z$.  In the upper left and lower right corner  of the right panel the total decay widths of the Higgses  become sizeable. The dashed black lines correspond to the contour  ${\rm max} \left ( \Gamma_H/M_H, \Gamma_A/M_A, \Gamma_{H^\pm} /M_{H^\pm}  \right ) = 30\%$. } 
\end{center}
\end{figure}

In order to understand the anatomy of the $t \bar t Z$ signal in the 2HDM context, one first has to notice that the upper right Feynman diagram in Figure~\ref{fig:diagrams} allows for resonant $t \bar t Z$ production if the two conditions $M_H > M_A + M_Z$ and $M_A > 2  m_t$ are satisfied. Once the channels $H \to A Z$ and $A \to t \bar t$ are kinematically accessible the triangle graph therefore always dominates over the box contribution displayed on the upper left  of the latter figure. In fact, the dominance of the triangle contribution allows one to estimate the signal strength $s \left (pp \to t \bar t Z \right )$. Since  in the narrow-width approximation~(NWA) the signal strength factorises into production and decay and given that ${\rm BR} \left (A \to t \bar t \right ) \simeq 100\%$ for the parameters of interest, one obtains in the  case of $pp \to H \to A Z \to t \bar t Z$ the following approximate result 
\begin{equation} \label{eq:signalstrength}
\sigma \left (pp \to t \bar t Z \right ) \simeq  \sigma \left ( p p \to H \right ) {\rm BR} \left (H \to AZ \right ) \,.
\end{equation} 
If  the $t \bar tZ$ signature arises instead through $A \to HZ$, the role of $H$ and $A$ has to simply be interchanged. The total $H,A$ production cross sections appearing in~(\ref{eq:signalstrength}) are easy to calculate at  leading~order~(LO). In the exact alignment limit and assuming that $t_\beta$ is not too large, we obtain at $\sqrt{s} = 14 \, {\rm TeV}$ the following expressions
\begin{equation} \label{eq:sigmaHA}
\begin{split}
\sigma \left ( p p \to H \right )  \simeq \frac{1}{t_\beta^2} \, \left ( \frac{570 \, {\rm GeV}}{M_H} \right )^{4.6} \, {\rm pb} \,, \qquad 
\sigma \left ( p p \to A \right )  \simeq \frac{1.7}{t_\beta^2} \, \left ( \frac{570 \, {\rm GeV}}{M_A} \right )^{5.2} \, {\rm pb} \,. 
\end{split}
\end{equation}
These approximations work to better than 20\% for Higgs masses in the range of $[400, 1000] \, {\rm GeV}$. They imply  that the total production cross section of a $A$ is always larger than that of a $H$ if these particles have the same mass. For $M_{H} = M_A  = 600 \, {\rm GeV}$ the relations (\ref{eq:sigmaHA}) predict for example an enhancement factor of around 1.6. We emphasise that the formulas given in (\ref{eq:sigmaHA}) serve mostly an illustrative purpose and have only been used to obtain the approximate signal strengths  for $pp \to t \bar t Z$ and $pp \to t b W$ production as shown in Figures~\ref{fig:ttZsignalstrengths} and~\ref{fig:tbWsignalstrengths}.  Our numerical results presented in Section~\ref{sec:results} and Appendices~\ref{app:pTZfit} and \ref{app:discovery} instead do not use the approximations (\ref{eq:sigmaHA}).

Figure~\ref{fig:ttZsignalstrengths} displays  the $t \bar t Z$  signal strengths  at $\sqrt{s} = 14 \, {\rm TeV}$ as a function of $M_H$ and $M_A$ in the type-II 2HDM. The shown results are obtained by treating the process $pp \to H/A \to A/H Z \to t \bar t Z$ in the NWA. The kinematically inaccessible region in the $M_H\hspace{0.5mm}$--$\hspace{0.5mm} M_A$ plane that separates  $pp \to H \to AZ \to t \bar t Z$ (lower right corners) from  $pp \to A \to HZ \to t \bar t Z$ (upper left corners) are indicated in grey. Both panels employ $c_{\beta - \alpha} = 0$ and $t_\beta = 1$. The left plot illustrates the choice $M_{H^\pm} = {\rm max} \left ( M_H, M_A \right )$ meaning that the decay channels $H/A \to H^\pm W^\mp$  are closed. One sees that in this case the $t \bar t Z$ signal strength can reach and even exceed $300 \, {\rm fb}$ for $M_H \simeq 600 \, {\rm GeV}$ and $M_A \simeq 375 \, {\rm GeV}$ or vice versa. This number should be compared to the LO result for the SM $t \bar tZ$ production cross section at $\sqrt{s} = 14 \, {\rm TeV}$ which amounts to $\sigma \left ( p p \to t \bar t Z \right )_{\rm SM} \simeq 700 \, {\rm fb}$. Notice that in accordance with (\ref{eq:sigmaHA}) the signal strengths for $pp \to A \to HZ \to t \bar t Z$ are always slightly larger than those for  $pp \to H \to AZ \to t \bar t Z$. From the panel on the right-hand side of~Figure~\ref{fig:ttZsignalstrengths} one moreover observes that the signal-over-background ratio is less favourable for the choice $M_{H^\pm} = {\rm min} \left ( M_H, M_A \right )$, because in this case the heavy neutral Higgses can  decay to a charged Higgs and a $W$ boson. Despite this suppression, the signal strength can reach up to around $200 \, {\rm fb}$, meaning that it still constitute a non-negligible fraction of the total SM $t \bar t Z$ production cross section. Notice that in the upper right (lower left) corner the width of the $A$ ($H$) becomes large because of $A \to H^{\pm} W^{\mp}$ ($H \to H^{\pm} W^{\mp}$) decays. To indicate this feature we have included in the figure dashed black contour lines that correspond to parameter choices leading to ${\rm max} \left ( \Gamma_H/M_H, \Gamma_A/M_A, \Gamma_{H^\pm} /M_{H^\pm}  \right ) = 30\%$. For relative decay widths below the quoted value the NWA should be applicable.  

The resonant contributions not only enhance the $t \bar t Z$ signal cross section, but also lead to interesting kinematic features that  one can harness to discriminate signal from background. Firstly, since both heavy Higgses tend to be on-shell in the production chain $p p \to H \to A Z \to t \bar t Z$, the invariant masses $m_{t \bar t Z}$ and $m_{t \bar t}$ of the $t \bar t Z$ and $t \bar t$ systems show characteristic Breit-Wigner peaks at 
\begin{equation} \label{eq:invariantmasses}
m_{t \bar t Z} \simeq M_H \,, \qquad m_{t \bar t} \simeq M_A \,.
\end{equation}
The difference $\Delta m$ between $m_{t \bar t Z}$ and $m_{t \bar t}$ can therefore be used to determine the mass splitting of the heavy Higgses.  In the considered case, the $\Delta m$ distribution of the $t \bar t Z$ signal will for instance be peaked at 
\begin{equation} \label{eq:Deltam}
\Delta m \equiv m_{t \bar t Z} - m_{t \bar t} \simeq M_H - M_A \,. 
\end{equation}

\begin{figure}[!t]
\begin{center}
\includegraphics[width=\textwidth]{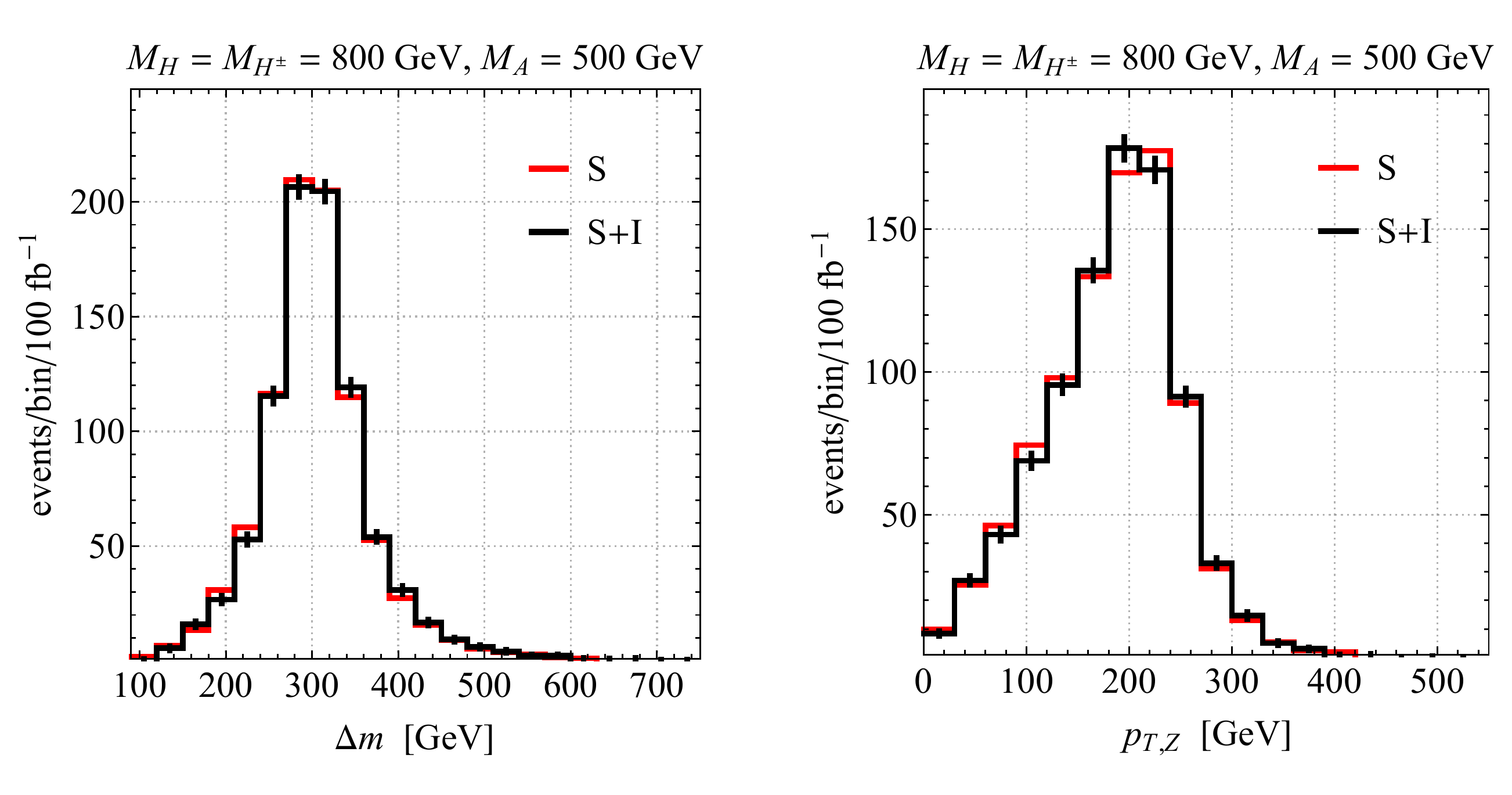} 
\vspace{-6mm}
\caption{\label{fig:distributionsttZ} Distributions of $\Delta m$ (left panel) and $p_{T,Z}$ (right panel) of the $t \bar t Z$ signal. The shown results correspond to the type-II 2HDM and employ  the parameters $M_H = M_{H^\pm} = 800 \, {\rm GeV}$, $M_A  = 500 \, {\rm GeV}$, $c_{\beta - \alpha} = 0$ and $t_\beta = 1$. The red and black curves correspond to the $t \bar t Z$ signal at $\sqrt{s} = 14 \, {\rm TeV}$ ignoring ($\text{S}$) and including ($\text{S}\!+\!\text{I}$) the interference with the SM~background, respectively. The black error bars represent statistical uncertainties.}
\end{center}
\end{figure}

Second, since the four-momenta of the decay products $A$ and $Z$ that enter $H \to A Z$ are fixed by~$H$ being preferentially on-shell,  also  the $p_{T,Z}$ spectrum will have a characteristic shape.   In fact, it is straightforward to show that the $p_{T,Z}$ distribution of the resulting $t \bar t Z$ signal is a steeply rising function of~$p_{T,Z}$  with a  cut-off at
\begin{equation} \label{eq:pTZ}
p_{T,Z}^{\rm max}  \simeq \frac{1}{2 M_H} \sqrt{ \left (M_H^2 - M_A^2 - M_Z^2 \right )^2 - 4 M_A^2 M_Z^2} \,,
\end{equation}
that is smeared by the total decay width $\Gamma_H$ of the heavy Higgs $H$. Needless to say, that the same line of reasoning and formulas similar to (\ref{eq:invariantmasses}), (\ref{eq:Deltam}) and (\ref{eq:pTZ}) apply when one considers  the process $pp \to A \to HZ \to t \bar t Z$ instead of $p p \to H \to A Z \to t \bar t Z$. 

Figure~\ref{fig:distributionsttZ} shows the $\Delta m$ and $p_{T,Z}$ distribution that we obtain from a {\tt MadGraph5\_aMCNLO}~\cite{Alwall:2014hca} simulation of the new-physics contribution to the $t \bar t Z$ final state. The displayed results have been obtained in the context of the type-II 2HDM  by employing a modified {\tt UFO} implementation~\cite{Degrande:2011ua} of the 2HDM discussed in~\cite{Bauer:2017ota}. The chosen parameters are $M_H = M_{H^\pm} = 800 \, {\rm GeV}$, $M_A  = 500 \, {\rm GeV}$, $c_{\beta - \alpha} = 0$ and $t_\beta = 1$.  The red predictions correspond to the pure new-physics signal ($\text{S}$), while the black  distributions  take into account  the interference between the signal process  and the background from SM $t \bar t Z$ production ($\text{S}\!+\!\text{I}$). All relevant box and triangle diagrams have been included in our simulation. The distinctive kinematic features of the signal discussed earlier  are clearly visible in the two panels with the $\Delta m$ distribution peaked at about $300 \, {\rm GeV}$ and an edge in the $p_{T,Z}$ spectrum at around $230 \, {\rm GeV}$. One also observes that, in contrast to the case of $t \bar t$~production~\cite{Gaemers:1984sj,Dicus:1994bm,Bernreuther:1997gs,Frederix:2007gi,Hespel:2016qaf}, signal-background interference leads only to minor distortions of the shapes of the most interesting~$t \bar t Z$ distributions. Although the interference effects are observed  to be small (roughly of the size of the statistical uncertainties in the shown example), we will include them in Section~\ref{sec:results} when determining the sensitivity of the~$t \bar t Z$ signature in constraining the parameter space of 2HDMs. 

\section{Anatomy of the $\bm{t bW}$ signature}
\label{sec:anatomytbW}

Two example diagrams that gives rise to a $t bW$ signal through the exchange of a charged Higgs boson are displayed in the lower row of~Figure~\ref{fig:diagrams}.   In the left graph a $H^+$ and a $W^-$ are radiated off a box diagram with internal top and bottom quarks, while in the diagram on the right-hand side a~$H$ is emitted from a top-quark triangle which then decays via $H \to H^+ W^-$. In both cases the charged~Higgs boson decays to a $t \bar b$ pair.  Notice that diagrams with $H^-$ or $A$ exchange also lead to a $tbW$ signal. These contributions while not explicitly shown in the lower row of~Figure~\ref{fig:diagrams} are all included in our analysis. 

\begin{figure}[!t]
\begin{center}
\includegraphics[width=0.475\textwidth]{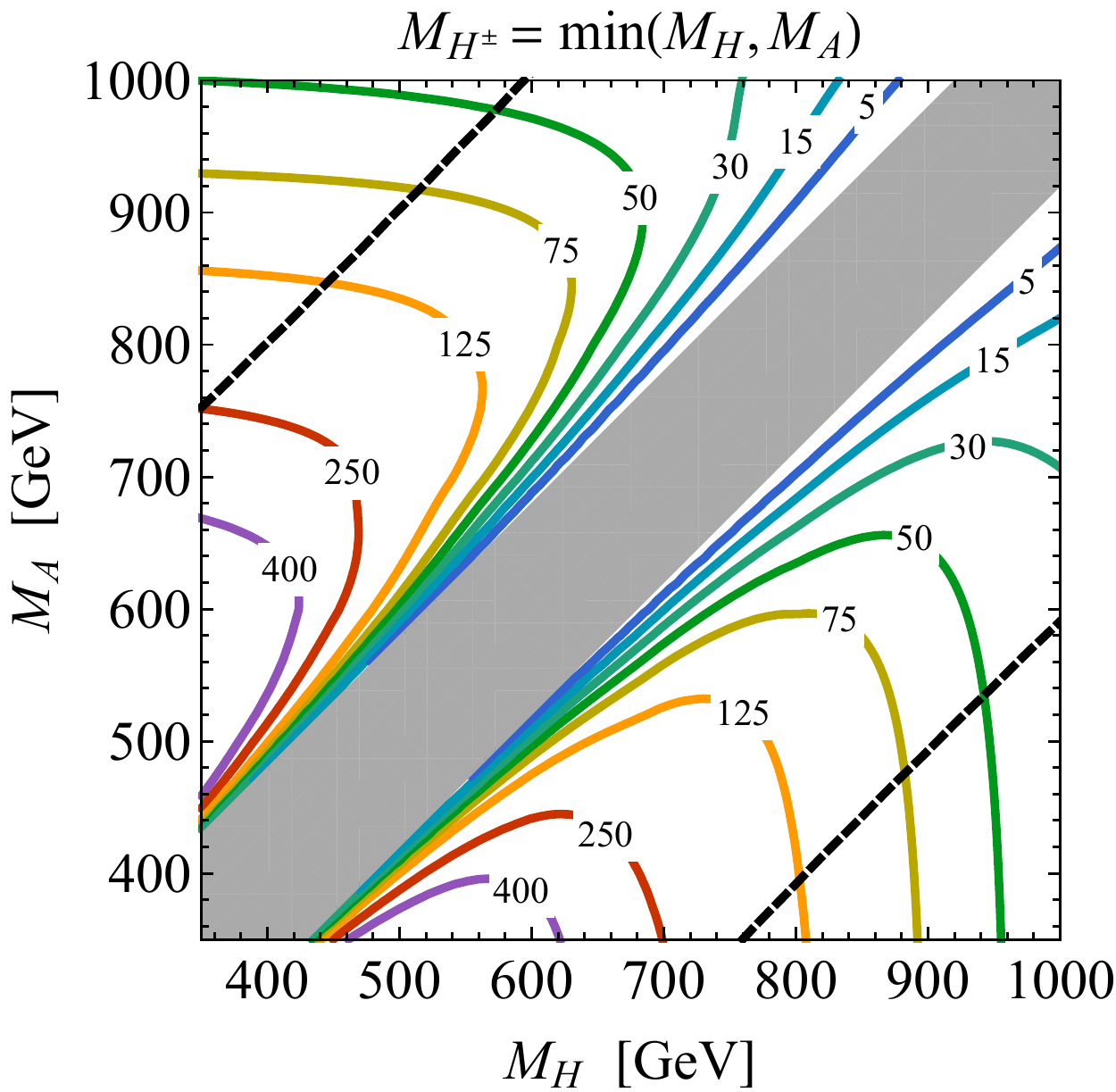} 
\vspace{3mm}
\caption{\label{fig:tbWsignalstrengths} Approximate signal strengths  at $\sqrt{s} = 14 \, {\rm TeV}$ for $pp \to t b W$ production in units of $\rm fb$. The given results correspond to the type-II 2HDM  and use $c_{\beta-\alpha} = 0$, $t_\beta=1$ and  $M_{H^\pm} = {\rm min} \left (M_H , M_A \right )$. The region in the   $M_H\hspace{0.5mm}$--$\hspace{0.5mm} M_A$ plane that is kinematically inaccessible through $pp \to H/A \to H^\pm W^\mp \to t b W$ is coloured  grey.  The plot corners that are enclosed by dashed black lines indicate the parameter space where  ${\rm max} \left ( \Gamma_H/M_H, \Gamma_A/M_A, \Gamma_{H^\pm} /M_{H^\pm}  \right ) > 30\%$. } 
\end{center}
\end{figure}

The~$tbW$ final state can be resonantly produced via $pp \to H \to H^\pm W^\mp$  ($pp \to A \to H^\pm W^\mp$) followed by~the decay $H^\pm \to tb$ if the two conditions $M_H > M_{H^\pm} + M_W$ ($M_A > M_{H^\pm} + M_W$)  and $M_{H^\pm} > m_t + m_b$ are  fulfilled. In such a case triangle diagrams   provide the leading contribution to the $t b W$ signal strength. In Figure~\ref{fig:tbWsignalstrengths} we show $s \left ( p p \to tbW \right)$ at $\sqrt{s} = 14 \, {\rm TeV}$ in the $M_H\hspace{0.5mm}$--$\hspace{0.5mm} M_A$ plane, treating the process $pp \to H/A \to H^\pm W^\mp \to t b W$ in the NWA. The depicted results correspond to the type-II 2HDM and $c_{\beta-\alpha} = 0$, $t_\beta=1$ and  $M_{H^\pm} = {\rm min} \left (M_H , M_A \right )$. The regions of parameter space in which the new-physics signal arises from $pp \to H \to H^\pm W^\mp \to t b W$~(lower right corner) or from  $pp \to A \to H^\pm W^\mp \to t b W$~(upper left corner) are divided by a grey stripe that masks Higgs masses satisfying $M_{H,A} < M_{H^\pm} + M_W$. From the figure one observes that the $t bW$ signal strength can  be as large as $400 \, {\rm fb}$ (or even larger) for $M_{H,A} \simeq 600 \, {\rm GeV}$ and $M_{A,H} \simeq 400 \, {\rm GeV}$. Since  for $M_H = M_A$ the total production cross section  $\sigma \left ( pp \to A \right)$  is bigger than $\sigma \left (pp \to H \right) $, one again notices a small asymmetry between the signal strengths $pp \to A \to H^\pm W^\mp \to t b W$ and $pp \to H \to H^\pm W^\mp \to t b W$ with the former being always slightly larger than the latter. 

\begin{figure}[!t]
\begin{center}
\includegraphics[width=\textwidth]{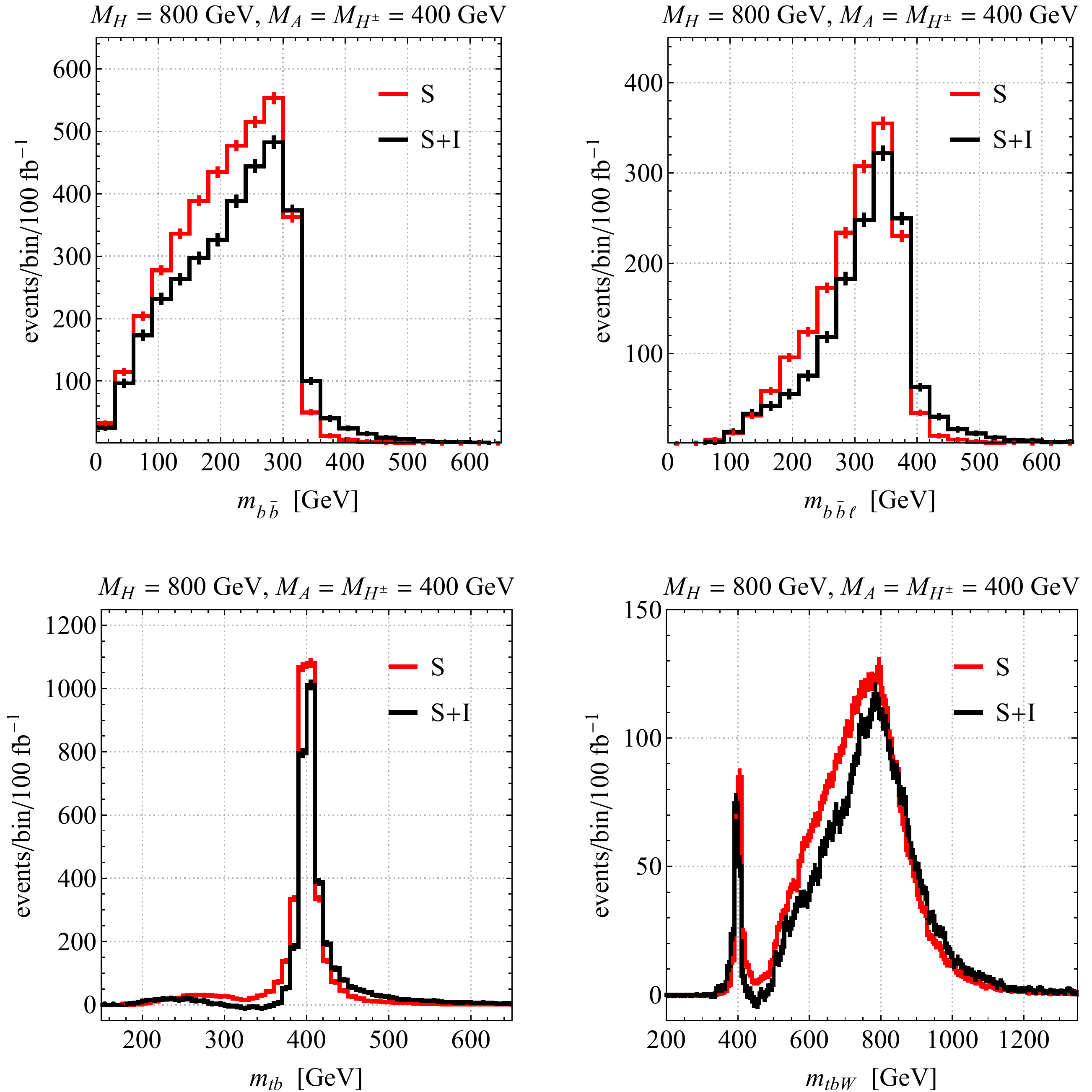} 
\vspace{-2mm}
\caption{\label{fig:distributionstbW}  Invariant mass distributions of the $t bW$ signal:  $m_{b \bar b}$ (upper left panel), $m_{b \bar b \ell}$ (upper right panel), $m_{t b}$ (lower left panel) and $m_{tbW}$ (lower right panel),. The displayed predictions have been obtained in the type-II~2HDM using $M_H = 800 \, {\rm GeV}$, $M_A  = M_{H^\pm} = 400 \, {\rm GeV}$, $c_{\beta - \alpha} = 0$ and $t_\beta = 1$. The red and black curves correspond to the results at $\sqrt{s} = 14 \, {\rm TeV}$ ignoring ($\text{S}$) and including ($\text{S}\!+\!\text{I}$) the interference with the SM~background. The error bars represent statistical uncertainties.}
\end{center}
\end{figure}

Like the case of the $t \bar t Z$ signal, also the kinematic distributions of the  $t b W$ signature have distinctive features that can be exploited to tame SM~backgrounds.   Figure~\ref{fig:distributionstbW} shows an assortment of invariant mass distributions that can serve this purpose. The displayed results have been obtained in the type-II 2HDM using {\tt MadGraph5\_aMCNLO}. The choice of parameters is $M_H =  800 \, {\rm GeV}$, $M_A  = M_{H^\pm} = 400 \, {\rm GeV}$, $c_{\beta - \alpha} = 0$ and $t_\beta = 1$. The left panel in the upper row of the figure depicts the invariant mass $m_{b \bar b}$ of the $b \bar b$ system in $p p \to t b W \to b \bar b W^+ W^-$. One sees that the $m_{b \bar b}$ distribution has a sharp edge at around~$320 \, {\rm GeV}$, which corresponds to the  kinematic endpoint~\cite{Allanach:2000kt,Lester:705139}
\begin{equation} \label{eq:mbbmax}
 m_{b \bar b}^{\rm max}  \simeq  \frac{1}{m_t} \sqrt{\left (M_{H^\pm}^2 - m_t^2 \right ) \left ( m_t^2 - M_W^2 \right )} \,.
\end{equation}
Similarly, also the invariant mass $m_{b \bar b \ell}$ of the $b \bar b \ell$ final-state configuration that appears in the $tbW$ channel from the sequential decay $H^\pm \to t b \to b\bar b W \to b \bar b \ell \nu$  has a kinematic endpoint. It is located at~\cite{Allanach:2000kt,Lester:705139}  
\begin{equation} \label{eq:mbblmax}
m_{b \bar b \ell}^{\rm max}  \simeq  \sqrt{M_{H^\pm}^2 - M_W^2} \,.
\end{equation}
The associated edge in the $m_{b \bar b \ell}$  spectrum arises at about $390 \, {\rm GeV}$, a feature that is evident in the upper right panel of Figure~\ref{fig:distributionstbW}. Notice that $b \bar b \ell$ final states also arise from the leptonic decay of the $W$ bosons involved in $H/A \to H^\pm W^\mp$. The corresponding invariant mass $m_{\ell b \bar b}$ has a very soft endpoint at $m_{\ell b \bar b} \simeq M_{H/A} - M_W$ and no edge because the lepton is not emitted directly from the backbone of the whole decay chain. Since experimentally one can  separate the two cases (see Section~\ref{sec:strategytbW}), an example of a $m_{\ell b \bar b}$ signal distribution has not been depicted in Figure~\ref{fig:distributionstbW}.  

In the lower left panel the invariant mass $m_{tb}$ of the~$tb$ system is  depicted. As expected, this distribution shows a Breit-Wigner peak at
\begin{equation} \label{eq:mtb}
m_{tb} \simeq M_{H^\pm} \,. 
\end{equation}
The invariant mass $m_{tbW}$ of the $tbW$ final state is displayed in the lower right panel of Figure~\ref{fig:distributionstbW}. The two mass peaks at 
\begin{equation} \label{eq:mtbW}
m_{tbW} \simeq M_{H} \,, \qquad m_{tbW} \simeq M_{A} \,, 
\end{equation}
resulting from  $pp  \to H \to H^\pm W^\mp \to tbW$ and $pp  \to A \to H^\pm W^\mp \to tbW$, respectively, are clearly visible in the figure. Notice that the peak at approximately $800 \, {\rm GeV}$  is smeared by the total decay width $\Gamma_H$ of the heavy Higgs $H$ which in the case at hand amounts to $\Gamma_H/M_H \simeq 30\%$. The resonance peak centred at $400 \, {\rm GeV}$  is on the other hand  narrow since $\Gamma_A/M_A \simeq 3\%$. 

Realise that not only the process $gg \to H^\pm W^\mp \to tbW$ (example diagrams are shown in the lower row of Figure~\ref{fig:diagrams}) but also  graphs corresponding to $gg \to H/A \to t \bar t \to tbW$ contribute to the $tbW$ signature in 2HDMs. To separate the charged Higgs contributions to the $tbW$ channel from the neutral Higgs contributions associated to $t \bar t$ production, we employ the so-called diagram removal~(DR)~procedure~\cite{Frixione:2008yi}. In this scheme the $tbW$ final state is defined by removing from the $tbW$ scattering amplitude all doubly resonant diagrams, i.e.~graphs in which  the intermediate top quarks can be on-shell. Singly resonant contributions are on the other hand kept. The DR procedure is also applied to the SM amplitudes, and as a result only $tb$-fusion (but no top-fragmentation) diagrams contribute at LO in QCD to the $tbW$ final state. 

Based on the DR definition of the $tbW$ final state, we have studied the impact of signal-background interference. The red curves in~Figure~\ref{fig:distributionstbW} correspond to the pure new-physics signal~($\text{S}$), while the black  distributions  in the four panels take into account  the interference between the signal process  and the background from $t b W$~production within the SM ($\text{S}\!+\!\text{I}$). Comparing the two sets of histograms, we observe that the kinematic features in the  $m_{b \bar b}$, $m_{b \bar b \ell}$, $m_{tb}$ and $m_{tbW}$ distributions are always less pronounced for the $\text{S}\!+\!\text{I}$ predictions than the $\text{S}$ results. Since the size of the signal-background interference typically exceeds the statistical uncertainties expected in future LHC runs, a rigorous assessment of the prospects of the $tbW$ final state to search for heavy Higgses should be based on MC simulations that include interference effects between the new-physics signal and the SM~background.

\section{MC generation and detector simulation}
\label{sec:montecarlo}

In our study  we consider throughout $pp$ collisions at $\sqrt{s} = 14 \, {\rm TeV}$. We generate the  signal samples using a modified  version of the {\tt Pseudoscalar\_2HDM} UFO together with  {\tt  MadGraph5\_aMC@NLO} and {\tt NNPDF23\_lo\_as\_0130} parton distribution functions~\cite{Ball:2012cx}. Compared to the UFO presented in~\cite{Bauer:2017ota} our new implementation is able to calculate the interference between the loop-induced $t \bar t Z$ and $t bW$ signals and the corresponding tree-level SM~backgrounds. The obtained parton-level events are then decayed and showered with {\tt PYTHIA~8.2}~\cite{Sjostrand:2014zea}  which allows us to study the fully interfered signals and backgrounds at the detector level. 

Our $t \bar t Z$ analysis will address the three-lepton final state, with two opposite-sign same-flavour leptons from the $Z$-boson decay and one lepton from the semileptonic decay of one of the two top quarks. For the description of the SM~backgrounds to this final state, SM processes involving at least three leptons coming from the decay of EW gauge bosons are simulated. Most of the backgrounds are generated at LO with {\tt  MadGraph5\_aMC@NLO}. The dominant irreducible background is $t \bar t Z$ which is generated with up to an additional jet. The $t b W$ background is instead generated with up to two additional jets. The dominant diboson background,~i.e.~$WZ$, is simulated with up to three additional jets. The minor backgrounds considered  are $tZ$ and $tWZ$ both of which are obtained using {\tt   MadGraph5\_aMC@NLO}. In each case the decay of the top quarks and the EW gauge bosons is performed by {\tt MadGraph5\_aMC@NLO}.  The reducible $ZZ$ background is generated at next-to-leading order (NLO) with {\tt POWHEG~BOX}~\cite{Alioli:2010xd}. 

A potential significant background in the case of the $t \bar t Z$ channel arises from processes where two leptons are produced in the decays of EW gauge bosons, and a third lepton is either the result of a misidentification in the detector or the decay of a $B$ meson. The latter is experimentally strongly suppressed by requiring the leptons to be isolated. The estimate of these backgrounds requires a profound understanding of the detector performance, and indeed the ATLAS and CMS use data-driven techniques rather than MC simulations to determine them.  A recently published search for~$t \bar t Z$ by the ATLAS experiment~\cite{Aad:2015eua} shows that the requirement of having at least four jets of which two  are identified  as coming from the fragmentation of  bottom quarks reduces the background from misidentified leptons to a level well below the other backgrounds. 

In the case of the  $tbW$ analysis, the background evaluation is performed through the generation of SM processes involving at least two leptons coming from the decays of EW gauge bosons. The backgrounds from $t \bar t$~\cite{Campbell:2014kua}, $tW$~\cite{Re:2010bp}, $WW$, $WZ$ and $ZZ$ production~\cite{Melia:2011tj,Nason:2013ydw} were all generated at NLO with {\tt POWHEG~BOX}. The $Z + {\rm jets}$  sample is generated at~LO with {\tt  MadGraph5\_aMC@NLO}, considering up to four jets for the matrix element calculation. The latter MC code is also used to simulate the $t \bar t V$ backgrounds with $V = W,Z$ at~LO with a multiplicity of up to two jets, and the $tZ$ and $tWZ$ backgrounds at LO. As for the $t \bar t Z$ analysis, we do not consider final states where one or both of the leptons are either fake electrons from jet misidentification or real non-isolated leptons  from the decay of heavy flavours. 

All partonic events are showered with {\tt PYTHIA~8.2} and the SM~backgrounds are normalised to their NLO cross section calculated either with {\tt MadGraph5\_aMC@NLO} or with {\tt POWHEG~BOX} where relevant. The simulated analyses are performed on experimentally identified electrons, muons,  photons, jets and $E_{T, \rm miss}$ which are constructed from the stable particles in the generator output. Jets are constructed by clustering the true momenta of all the particles interacting in the calorimeters, with the exception of muons. An anti-$k_t$ algorithm~\cite{Cacciari:2008gp} with a parameter $R=0.4$ is used, as implemented in  {\tt FastJet}~\cite{Cacciari:2011ma}. Jets originating from the hadronisation of bottom quarks ($b$-jets) are experimentally tagged in the detector ($b$-tagged). The variable $p_{T,\rm miss}$ with magnitude $E_{T, \rm miss}$ is defined at truth level,~i.e.~before applying detector effects, as the vector sum of the transverse momenta of all the invisible particles (neutrinos in our case). The effect of the detector on the kinematic quantities used in the analysis is simulated by applying a Gaussian smearing to the momenta of  the different reconstructed objects and reconstruction and tagging efficiency factors. The parametrisation of the smearing and of the reconstruction and tagging efficiencies is tuned to mimic the performance of the  ATLAS detector~\cite{Aad:2008zzm,Aad:2009wy} and is applied as a function of the momentum and the pseudorapidity  of the physical objects. The discrimination of the signal from the background is significantly affected  by the experimental smearing assumed for $E_{T, \rm miss}$. To simulate this effect,  the transverse momenta of unsmeared electrons, muons and jets are subtracted from the truth $E_{T, \rm miss}$  and replaced by the corresponding smeared quantities.  The residual truth imbalance  is then smeared as a function of the scalar sum of the transverse momenta of the particles not assigned to electrons or jets. The same techniques have also been employed in~\cite{Haisch:2016gry,Pani:2017qyd}. 

\section{Analysis strategy for the $\bm{t\bar tZ}$ signature}
\label{sec:strategyttZ}

In the case of the $t \bar t Z$ channel the generated signal and background events are preselected by requiring exactly three charged leptons (electrons or muons) with a pseudorapidity of $|\eta_\ell|<2.5$.  The leading lepton must have $p_{T,\ell}>25 \, {\rm GeV}$, while the other two are required to satisfy  $p_{T,\ell}>20 \, {\rm GeV}$.  At least one pair of leptons of opposite charge and same flavour must be present, and the invariant mass $m_{\ell \ell}$ of this pair must meet the requirement $|m_{\ell \ell} - M_Z| < 15 \, {\rm GeV}$. In case the event includes more than one such  lepton pair, the pair with the invariant mass closest to the nominal value of~$M_Z$ is selected as the $Z$-boson candidate. All events  furthermore need to contain four jets with $p_{T,j} > 20 \, {\rm GeV}$ and  $|\eta_j|<2.5$, of which two must be tagged as bottom-quark jets ($b$-tagged).

Notice that the large jet multiplicity and the requirement of having two $b$-tagged jets  leads, on the one hand, to a strong reduction of the $WZ$ and $tWZ$   SM~backgrounds, and on the other  selects all  objects needed for a full reconstruction of the event. The leptonically decaying $W$ boson is reconstructed from the  charged lepton not assigned to $Z \to \ell^+ \ell^-$ and the amount of $E_{T, \rm miss}$ by solving the $W$-boson  mass constraint. If the solution for the longitudinal momentum of the neutrino is imaginary, the real part of $p_{z, \nu}$ is taken. The reconstructed jets are assigned to the different $W$ and top decays, by choosing the assignment which gives the best compatibility with the decay of two top quarks in terms of reconstructed masses. If in the process two solutions are found for the  reconstructed leptonic $W$ decay, the one giving the best mass compatibility  is selected. 

\begin{figure}[!t]
\begin{center}
\includegraphics[width=\textwidth]{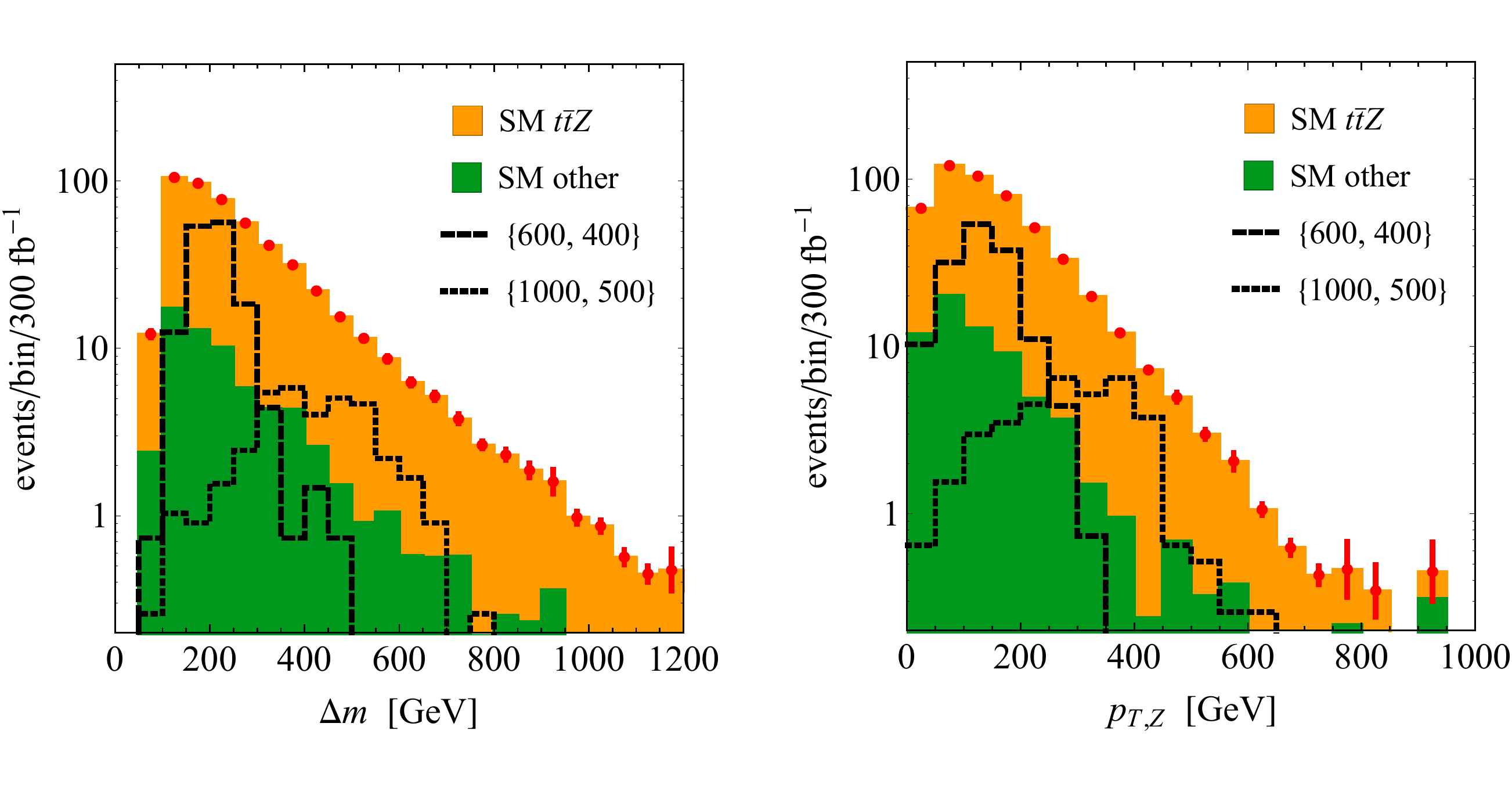} 
\vspace{-8mm}
\caption{\label{fig:SBdistributionsttZ} 
Distributions of $\Delta m$ (left panel) and $p_{T,Z}$ (right panel) after imposing the experimental selection requirements as detailed in the text. The coloured histograms are stacked and represent the SM~backgrounds with the label ``SM other'' referring to the contributions from $tbW$, $WZ$, $tZ$ and $tWZ$.  The shown error bars represent the statistical uncertainties of the sum of the SM~backgrounds. The signal predictions in the type-II 2HDM corresponding to $M_H = M_{H^\pm} = 600 \, {\rm GeV}$, $M_A  = 400 \, {\rm GeV}$ ($M_H = M_{H^\pm} = 1000 \, {\rm GeV}$, $M_A  = 500 \, {\rm GeV}$), $c_{\beta - \alpha} = 0$ and $t_\beta = 1$ are superimposed as dashed (dotted) black lines. All predictions are obtained at $\sqrt{s} = 14 \, {\rm TeV}$.}
\end{center}
\end{figure}

In Figure~\ref{fig:SBdistributionsttZ} the distributions of $\Delta m$ and $p_{T,Z}$ for the SM~backgrounds and two type-II 2HDM benchmark models after applying  the selections described above are displayed. Our benchmarks correspond to $M_H = M_{H^\pm} = 600 \, {\rm GeV}$, $M_A  = 400 \, {\rm GeV}$ ($M_H = M_{H^\pm} = 1000 \, {\rm GeV}$, $M_A  = 500 \, {\rm GeV}$), $c_{\beta - \alpha} = 0$ and $t_\beta = 1$ and are indicated by the dashed (dotted) black lines. In the case of the~$\Delta m$ spectrum, one observes from the left panel that the sum of the SM~backgrounds is a steeply falling distribution, while both new-physics $t \bar t Z$ signals exhibit a Breit-Wigner peak. In fact,   as expected from~(\ref{eq:Deltam}) the   peaks  are located at around $200 \, {\rm GeV}$ and $500 \, {\rm GeV}$.  Our results for the~$p_{T,Z}$ distributions are presented in the right panel of the latter figure.  In agreement with~(\ref{eq:pTZ}) the two 2HDM benchmark models lead to spectra that show distinctive Jacobian peaks with edges at roughly  $150 \, {\rm GeV}$ and $370 \, {\rm GeV}$. The SM~backgrounds are in contrast again smoothly falling  and featureless. Notice that a measurement of the $p_{T,Z}$ distribution in $t \bar t Z$ production does, unlike a measurement of the difference $\Delta m$ of invariant masses, not require the full  reconstruction of the final state. As a result,  $p_{T,Z}$ is  less  subject to experimental  uncertainties  than $\Delta m$. In order to stress the experimental robustness of the proposed $t \bar t Z$ signature, we will in our sensitivity study consider both the $\Delta m$ and the~$p_{T,Z}$ distribution as final discriminants.

\section{Analysis strategy for the $\bm{tbW}$ signature}
\label{sec:strategytbW}

In the case of the $pp \to  H^\pm W^{\mp} \to tbW \to b \bar b W W$ signal the dominant QCD  backgrounds are~$t \bar t$~production and $tW$ production in association with a $b$-jet.  By vetoing events where the observed $W$ bosons and $b$-jets are kinematically compatible with the decay of two top quarks the  overwhelming $t \bar t$ background can be reduced by approximately two orders of magnitude, making it comparable to the $tW$ background in size. After this selection the signal is however still two orders of magnitude smaller than the background. Notice that this is in contrast to the $t \bar t Z$ channel where the signal and the background are of the same size after background suppression. To improve the signal-over-background ratio in the case of the $tbW$ signal, one needs to exploit the decay kinematics of the heavy Higgses by identifying the decay products  of the top quark in the signal events. The invariant mass of the top quark with the additional $b$-jet will be peaked  at $M_{H^\pm}$, while the invariant mass of the two $b$-jets and the two $W$ bosons equals $M_H$ or $M_A$ depending on which mass is larger.  Experimentally the  signal can be looked for in events with two, one or zero isolated  charged leptons  resulting from $W \to \ell \nu$. In the following we will sketch a possible analysis procedure for the two-lepton final state. Given the small signal-to-background ratio for the irreducible backgrounds, we however expect that our conclusions will be valid for the one-lepton final state as well. 

For the two-lepton case, the reconstruction of mass peaks is not possible due to the presence of two neutrinos in each event. However, given the presence of multi-step sequential decays leading to undetected neutrinos, the invariant mass distributions of the visible decay products are bounded from above~\cite{Hinchliffe:1996iu}. In the case of top decays, the invariant mass $m_{b \ell}$ of the resulting $b$-quark and lepton  must be lower than $m_{b\ell}^{\rm max} = \sqrt{m_t^2 - M_W^2} \simeq 153 \, {\rm GeV}$.  Thus exactly two opposite-sign leptons ($\ell_1$ and $\ell_2$) with  $p_{T,\ell_1} > 30 \, {\rm GeV}$ and $p_{T,\ell_2} > 25 \, {\rm GeV}$ and exactly two $b$-tagged jets ($b_1$ and $b_2$) with $p_{T,b} > 30 \, {\rm GeV}$ are required in the event, and events are selected in which none of the two possible pairings among $b$-jets and leptons is compatible with the decay of two top quarks. A convenient way of rejecting events compatible with two top decays consists in introducing the observable 
\begin{equation}
m_{b \ell}^t =\mathrm{min} \hspace{0.5mm} \Big  (\mathrm{max} \hspace{0.5mm}  \big ( m_{l_1 j_a}, m_{l_2 j_b} \big  ) \Big ) \,,
\end{equation}
where the minimisation runs over all  pairs $\{j_a, j_b\}$ of distinct jets inside a predefined set of test jets. Based on the number of $b$-tagged jets in the event, the set of test jets is defined as follows. If the event includes one or two $b$-tagged jets, an additional test jet is considered, chosen as the non-$b$-tagged jet with the highest $b$-tagging weight and $p_{T,j} >25 \, {\rm GeV}$. If three $b$-tagged jets are found, they are all taken as test jets. The requirement $m_{b \ell}^t  >180 \, {\rm GeV}$ suppresses the $t \bar t$ background by approximately two orders of magnitude. A dangerous background is also also due to the production of a $Z$ boson in association with $b$-jets. Vetoing  same-flavour lepton pairs compatible with a $Z$-boson decay and requiring  some $E_{T, \rm miss}$ reduces this background to roughly  the level of the signal, albeit with large uncertainties.  To avoid this possible issue we completely remove the latter  and all other backgrounds including a real $Z$ boson by requiring that the two selected leptons have different flavours. After these selections the remaining background consists of approximately one half of $t \bar t$ and one half of $tW$ events.

\begin{figure}[!t]
\begin{center}
\includegraphics[width=\textwidth]{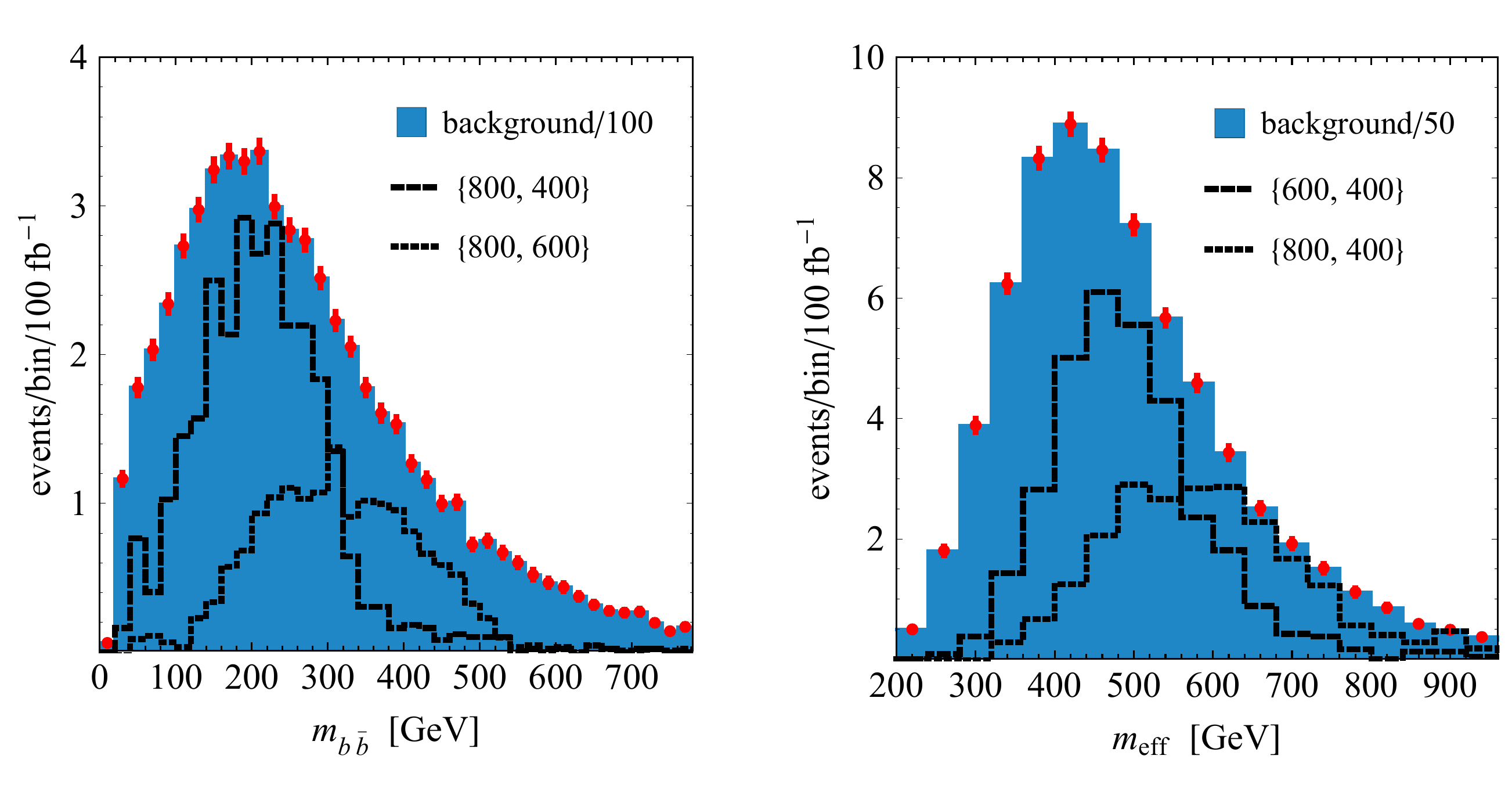} 
\vspace{-6mm}
\caption{\label{fig:mbbmeff} Left: The coloured histogram depicts the background $m_{b \bar b}$ distribution scaled down by a factor of~100. The predictions in the type-II 2HDM corresponding to $M_H = 800 \, {\rm GeV}$, $M_A  = M_{H^\pm} = 400 \, {\rm GeV}$ ($M_H = 800 \, {\rm GeV}$, $M_A  = M_{H^\pm} = 600 \, {\rm GeV}$), $c_{\beta - \alpha} = 0$ and $t_\beta = 1$ are superimposed as dashed (dotted) black lines. Right: The coloured histogram corresponds to the  $m_{\rm eff}$ distribution of the background reduced by a factor of 50. The signal prediction in the type-II 2HDM has been obtained for $M_H = 600 \, {\rm GeV}$, $M_A  = M_{H^\pm} = 400 \, {\rm GeV}$ ($M_H = 800 \, {\rm GeV}$, $M_A  = M_{H^\pm} = 400 \, {\rm GeV}$), $c_{\beta - \alpha} = 0$ and $t_\beta = 1$ and is represented by  a dashed (dotted) black line.  All predictions are obtained at $\sqrt{s} = 14 \, {\rm TeV}$ and take into account the selection requirements that are detailed in the text. The shown error bars represent the statistical uncertainties of the entire SM~background.} 
\end{center}
\end{figure}

A further separation of signal from background can be achieved by exploiting the fact that in the case of the signal the invariant mass $m_{b \bar b}$ of the two $b$-tagged jets as well as the  invariant mass~$m_{b\bar b \ell}$ of the two $b$-tagged jets with the lepton from a top decay are bounded from above. See~(\ref{eq:mbbmax}) and~(\ref{eq:mbblmax}). In order to  illustrate this point we show in the left panel of  Figure~\ref{fig:mbbmeff}  the distribution of~$m_{b \bar b}$ for two signal samples with $M_A  = M_{H^\pm} = 400 \, {\rm GeV}$ (dashed black line) and $M_A  = M_{H^\pm} = 600 \, {\rm GeV}$ (dotted black line), respectively. The remaining 2HDM parameters are set to $M_H = 800 \, {\rm GeV}$, $c_{\beta - \alpha} = 0$ and $t_\beta = 1$  and the  background has been scaled down by a factor of 100 for better visibility. Upper cuts on $m_{b \bar b}$ matching the kinematic endpoint of the signal for different values of $M_{H^{\pm}}$ will improve the signal-over-background ratio, bringing it to a level of at most 3\% over the parameter space relevant for this analysis. The variable $m_{b \bar b \ell}$ is less effective as it has a less sharp edge, and suffers from an ambiguity in the choice of the lepton.

The final experimental handle is the fact that the invariant mass of the $tbW$ system will peak at the mass of the $H/A$ bosons for the bulk of the signal (cf.~Figure~\ref{fig:distributionstbW}) whereas for the background $m_{tbW}$ has a broad distribution centred at around $400 \, {\rm GeV}$. However, if the charged Higgs mass   is not known, the observable $m_{tbW}$ cannot  be reconstructed because of the two undetected neutrinos. A large literature on the reconstruction of the mass of new particles in events with two invisible particles in the final state is available (see~e.g.~\cite{Gripaios:2011na}  for a review).  In our exploratory study, we employ the variable $m_{\rm eff}$ as an estimator of $m_{tbW}$, which is defined as~\cite{Paige:1996nx,Hinchliffe:1996iu}
\begin{equation} \label{eq:meff}
m_{\rm eff}\equiv \sum_{a=\ell_1,\ell_2, b_1, b_2} p_{T, a} + E_{T, \rm miss} \,.
\end{equation} 
In the right panel of Figure~\ref{fig:mbbmeff} we show the $m_{\rm eff}$ distribution for $M_{H} = 600 \, {\rm GeV}$~(dashed black line) and $M_{H} =800  \, {\rm GeV}$~(dotted black line), fixing the other parameters to $M_A = M_{H^{\pm}}=400 \, {\rm GeV}$, $c_{\beta - \alpha} =0$ and  $t_\beta=1$. For better visibility the background has been scaled down by a factor 50 after applying the cut $m_{b \bar b}<280 \, {\rm GeV}$. The significance of the signal can  be extracted from a shape fit to the $m_{\rm eff}$ spectrum for signal and background. Given the difference in shape between signal  and background, and the large number of kinematic handles, it should be possible to extract a significant signal for a signal sample corresponding to 3~ab$^{-1}$ of integrated luminosity if the shape of the background can be experimentally controlled to a level below $2\%$. In these conditions a reliable evaluation  of the coverage in parameter space can only be performed by the experimental collaborations. It is however worth noting that, due to  the shape of the $m_{b \bar b}$ background distribution, the maximum  sensitivity of the $tbW$ analysis is expected to arise for $M_A = M_{H^{\pm}} \lesssim 400 \, {\rm GeV}$,  making the $tbW$ coverage complementary to that of the $t \bar t Z$ search. 

\section{Numerical results}
\label{sec:results}

Based on the search strategy outlined in Section~\ref{sec:strategyttZ}, we now study the sensitivity of future LHC runs to the $t \bar t Z$ signature. To evaluate the upper limit on the ratio of the signal yield to that predicted in the 2HDM framework, a profiled likelihood ratio test statistic applied to the shapes of the~$\Delta m$ and $p_{T,Z}$ distributions is used. The CLs method~\cite{Read:2002hq} is employed to derive exclusion limits at 95\% confidence level~(CL). The statistical analysis has been performed by employing the {\tt RooStat} toolkit~\cite{Moneta:2010pm}.  A systematic uncertainty on the absolute normalisation of the SM~background  (signal) of 15\% (5\%) is assumed. This choice of uncertainties is in accordance with the uncertainties obtained by ATLAS and CMS for existing searches in similar final states. For the signal, the main uncertainty is generated by the impact on the selection efficiency of uncertainties on the measurement of quantities such as~e.g.~the energy scale and resolution for jets and $E_T^{\rm miss}$. In the case of the background there is in addition an important contribution to the total uncertainty that is associated with the procedure used to obtain the background estimate, which is typically achieved through a mixture of MC and data-driven techniques. Since we perform a shape analysis, the obtained fit results have reduced sensitivity to the absolute normalisation uncertainties, and are essentially determined by the uncertainties on the prediction of the shape of the distribution of the fitted variable for the SM~background. The magnitude of these uncertainties is difficult to forecast, as they include different factors, such as the shape distortion from uncertainties on energy and efficiency determinations, or theoretical uncertainties associated to the simulation of the background. A variety of techniques are used by the experiments to control  shape uncertainties, including the usage of appropriate control regions and the profiling of experimental uncertainties. In the case of the $t \bar t Z$ final state, shape uncertainties of a few percent seem to be an achievable goal, and we will determine the LHC reach, assuming a representative value of $5\%$ for the latter~uncertainty. 

The results  given in the following are for integrated luminosities of $300 \, {\rm fb}^{-1}$ and $3 \, {\rm ab}^{-1}$, corresponding to the LHC~Run-3  phase and  the high-luminosity option of the LHC (HL-LHC), respectively.  As the  LHC experimental community is still working on the detailed assessment of  the impact of the high pileup on the detector performance in the HL-LHC phase, we assume for simplicity the same detector performance for the two benchmark luminosities.  The analysis based on the $\Delta m$ variable, relying on an accurate measurement of $E_T^{\rm miss}$ and the momenta of jets will likely be affected by pileup. In contrast, we expect only a minor impact on the $p_{T,Z}$ variable, built from two high-$p_T$ leptons. Under the assumptions presented above, we find that a shape analysis using $\Delta m$ leads to  only marginally better results than a fit to $p_{T,Z}$. We are therefore convinced that the conclusions of this study  are valid also in the presence of a much higher pileup than the one experienced in the ongoing LHC run.  In  this section,  we  will only  show  the  results   of  our $\Delta m$ shape~fit. A comparison of the performance of the $\Delta m$ and $p_{T,Z}$ fits is provided in Appendix~\ref{app:pTZfit}.

\begin{figure}[!t]
\begin{center}
\includegraphics[width=\textwidth]{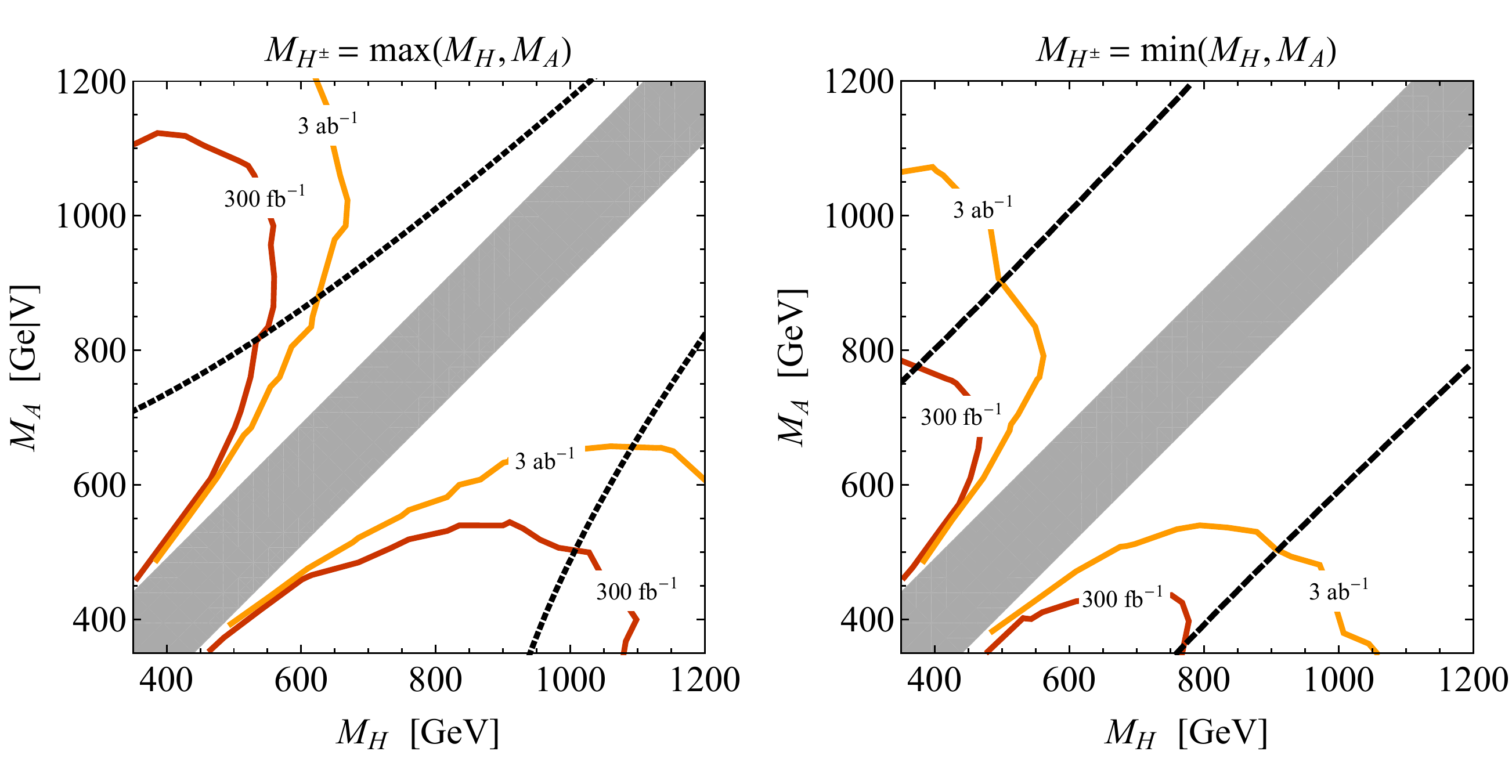} 
\vspace{-4mm}
\caption{\label{fig:lpplots} Hypothetical constraints in the $M_H\hspace{0.5mm}$--$\hspace{0.5mm} M_A$ plane  arising  from the proposed $t\bar t Z$ analysis. The given results correspond to the type-II 2HDM  and adopt  $c_{\beta-\alpha} = 0$, $t_\beta=1$, $M_{H^\pm} = {\rm max} \left (M_H , M_A \right )$~(left panel) and $M_{H^\pm} = {\rm min} \left (M_H , M_A \right )$~(right panel). The parameter space below and the left of the red (yellow) contours are excluded at 95\%~CL assuming  $300 \, {\rm fb}^{-1}$ $\big ( 3 \ {\rm ab}^{-1} \big)$ of~$14 \, {\rm TeV}$~LHC data. The regions in the   $M_H\hspace{0.5mm}$--$\hspace{0.5mm} M_A$ plane that are kinematically inaccessible through $pp \to H/A \to A/HZ \to t \bar t Z$ are depicted in  grey. The dotted~(dashed) black curves in the left (right) panel represents the  combined constraint from perturbativity and vacuum stability (the parameter region where ${\rm max} \left ( \Gamma_H/M_H, \Gamma_A/M_A, \Gamma_{H^\pm} /M_{H^\pm}  \right ) > 30\%$). See text for further explanations.}
\end{center}
\end{figure}

\begin{figure}[!t]
\begin{center}
\includegraphics[width=\textwidth]{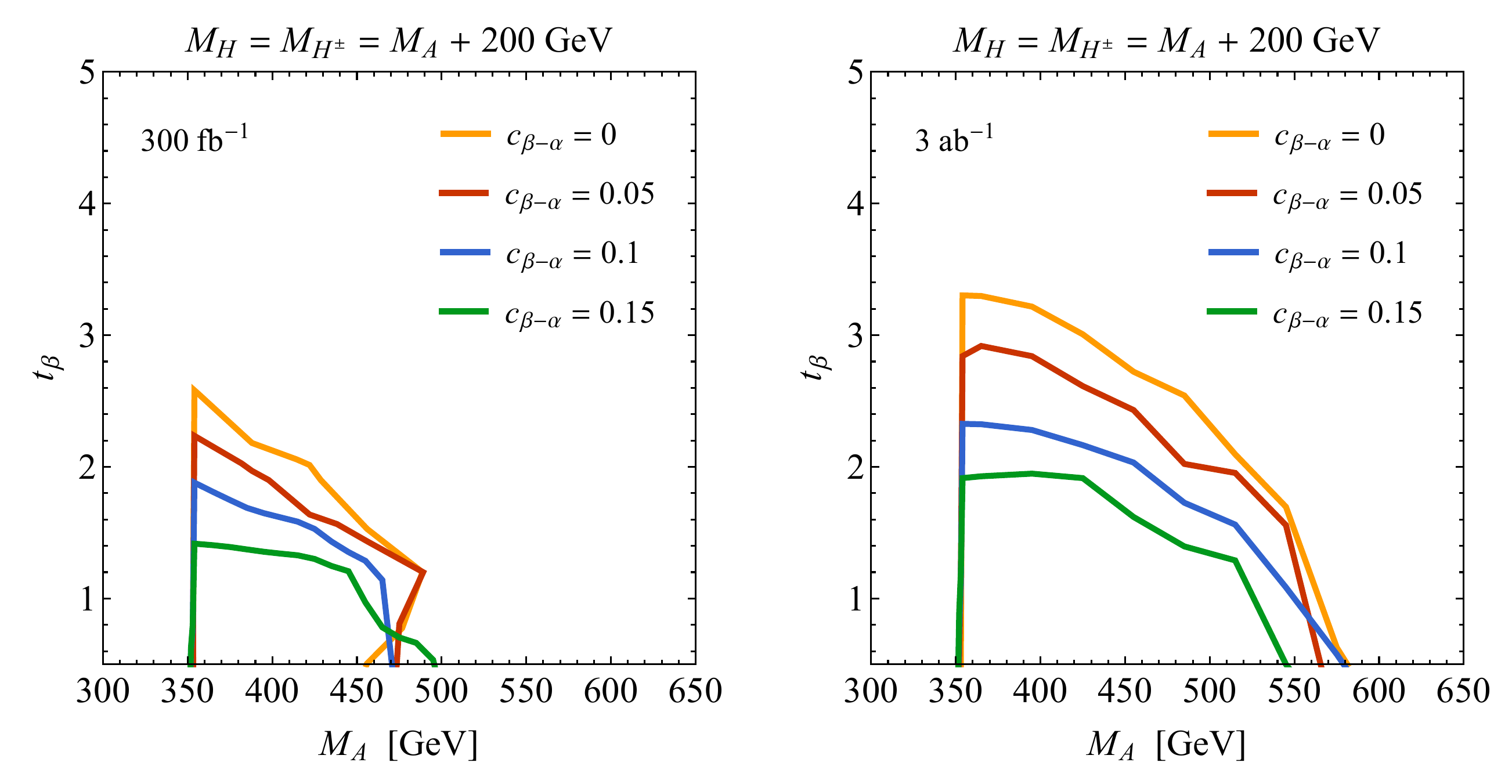} 
\vspace{-4mm}
\caption{\label{fig:mtbplots} 95\%~CL limits on $t_\beta$ in the type-II 2HDM resulting from a hypothetical $t\bar t Z$ search. The results shown on the left (right) are based on $300 \, {\rm fb}^{-1}$ $\big(3 \, {\rm  ab}^{-1}\big)$ of LHC data taken at $\sqrt{s} = 14 \, {\rm TeV}$. They employ  $M_H = M_{H^\pm} = M_A + 200 \, {\rm GeV}$,  $\lambda_3 = 6$ and assume  four different values of $c_{\beta - \alpha}$ as indicated in the legend of each plot. The  regions below the coloured contours represent  the excluded parameter space.}
\end{center}
\end{figure}

The results of our sensitivity study are displayed in Figures~\ref{fig:lpplots} and \ref{fig:mtbplots}. In the two panels of the first figure, we show the 95\%~CL exclusion limits  in the $M_H\hspace{0.5mm}$--$\hspace{0.5mm} M_A$ plane  that derive  from a shape~fit to the $\Delta m$ observable introduced in (\ref{eq:Deltam}). The red (yellow) contours illustrate  the constraints that follow from $300 \, {\rm fb}^{-1}$~$\big ( 3 \ {\rm ab}^{-1} \big)$ of  data collected at $\sqrt{s} = 14 \, {\rm TeV}$. They  are obtained in the type-II 2HDM employing $c_{\beta-\alpha} = 0$, $t_\beta=1$, $M_{H^\pm} = {\rm max} \left (M_H , M_A \right )$~(left panel) and $M_{H^\pm} = {\rm min} \left (M_H , M_A \right )$~(right~panel). The region in the   $M_H\hspace{0.5mm}$--$\hspace{0.5mm} M_A$ plane that is kinematically inaccessible  is indicated in  grey. From the red contours in the left  plot, one sees that if the intermediate $H/A$ can only decay to the $A/HZ$ final state but not to $H^\pm W^\mp$, based on the entire LHC~Run-3 data set it should be possible to exclude masses $M_{H/A}$ in the range of approximately $[450, 1150] \, {\rm GeV}$ ($[350, 500] \, {\rm GeV}$) for $M_{A/H} = 350 \, {\rm GeV}$ ($M_{A/H} = 1000 \, {\rm GeV}$). If, on the other hand, the decay channels $H/A \to H^\pm W^\mp$ are open, the exclusion reduces to $[450, 750] \, {\rm GeV}$  for $M_{A/H} = 350 \, {\rm GeV}$ as illustrated by the red contour lines in the right panel. It is also evident from the two panels, that with  $3 \, {\rm ab}^{-1}$ of data that  the HL-LHC is expected to collect, it may be possible to improve the LHC Run-3  sensitivity by up to a factor of 1.5. The corresponding contours are coloured yellow in Figure~\ref{fig:lpplots}. The improvements are more pronounced for $M_{H^\pm} = {\rm min} \left (M_H , M_A \right )$ than for $M_{H^\pm} = {\rm max} \left (M_H , M_A \right )$, and numerically largest for mass hierarchies $|M_H - M_A| \gg M_Z$. Notice that in these cases the signal strengths are small and in consequence the proposed $t \bar t Z$ search is statistics limited at LHC Run-3. The $5 \sigma$ discovery reach corresponding to Figure~\ref{fig:lpplots} can be found  in~Appendix~\ref{app:discovery}.

At this point, one should mention that  large mass splittings between the heavy Higgses are in general constrained by  theoretical arguments such as perturbativity and vacuum stability. In~order to illustrate this point, we depict in the left panel of Figure~\ref{fig:lpplots} the parameter space that is disfavoured by requiring  simultaneously the quartic coupling $\lambda_3$ to be perturbative,~i.e.~$\lambda_3 < 4 \pi$, and the simplest 2HDM scalar potential to be bounded from below~\cite{Gunion:2002zf}. The displayed constraints can be relaxed in more general 2HDMs containing additional quartic couplings  like for example $\lambda_6 \hspace{0.5mm} \big ( |H_1|^2 H_1^\dagger H_2 + {\rm h.c} \big )$, and  the shown  contours should   therefore only  be considered as indicative, having the mere purpose to identify  theoretically (dis)favoured parameter regions (see~\cite{Bauer:2017fsw} for a more detailed discussion of this point). In the right plot in Figure~\ref{fig:lpplots}, perturbativity and vacuum stability arguments instead do not lead to any restriction on the shown parameter~space. As discussed before, in this case the total decay width of $A$ ($H$) however becomes large because the $A \to H^{\pm} W^{\mp}$ ($H \to H^{\pm} W^{\mp}$) channel is open. The family of $\{M_H, M_A\}$ values that leads to ${\rm max} \left ( \Gamma_H/M_H, \Gamma_A/M_A, \Gamma_{H^\pm} /M_{H^\pm}  \right ) = 30\%$ is indicated by the dashed black lines in the right panel. Although our $t \bar t Z$ analysis is performed keeping effects due to off-shell $H/A$ production and decay, and due to the interference with the SM~background  (see Section~\ref{sec:anatomyttZ}), it ignores possible modifications of the $H/A$ line shape~\cite{Seymour:1995np,Goria:2011wa,Passarino:2010qk}. The latter effects have been studied in~\cite{Anastasiou:2011pi,Anastasiou:2012hx}, where it was found that for a heavy Higgs boson different treatments of its propagator can lead to notable changes in the inclusive production cross sections compared to the case of a Breit-Wigner with a fixed width, as  used  in our work. In consequence, the exclusion limits in the upper left and lower right corner of the right plot in  Figure~\ref{fig:lpplots} carry some (hard to quantify) model dependence related to the precise  treatment of the  $H/A$ propagators.  

In Figure~\ref{fig:mtbplots} we show furthermore the 95\%~CL exclusion contours in the $M_A\hspace{0.5mm}$--$t_\beta$ plane for the type-II~2HDM scenarios with $M_H = M_{H^\pm} = M_A + 200 \, {\rm GeV}$ and $\lambda_3 = 6$. In both panels the results of our $\Delta m$ shape fit are given for  four different values of $c_{\beta-\alpha}$. Notice that for the chosen parameters there are no  issues with perturbativity and vacuum stability, and that the $A$ is sufficiently narrow for the NWA to hold. From the results shown on the left-hand side one observes that with $300 \, {\rm fb}^{-1}$ of $\sqrt{s} = 14 \, {\rm TeV}$ data, all values of $t_\beta \lesssim 2.5$ can be excluded for $A$ masses close to the top threshold  in the exact alignment limit,~i.e.~$c_{\beta - \alpha} = 0$.  If the Higgs sector is not perfectly aligned, the branching ratio $H \to A Z$ is reduced  $\big ($see (\ref{eq:ghHAZ})$\big)$, and as a result the bounds in  the $M_A\hspace{0.5mm}$--$t_\beta$ plane become weaker. For instance, for the choice $c_{\beta - \alpha} = 0.15$, we find that the reach is decreased by roughly a factor of 1.5 compared to the case of full alignment. One also sees that  $M_A$ values up to around $450 \, {\rm GeV}$ can be excluded with $300 \, {\rm fb}^{-1}$ for $c_{\beta - \alpha} \leq 0.15$ and $t_\beta = {\cal O} (1)$. With $3 \, {\rm ab}^{-1}$ of integrated luminosity it turns out that the obtained limits can be pushed to values of $M_A$ and $t_\beta$ that are higher by approximately  $30 \%$ than the corresponding $300 \, {\rm fb}^{-1}$ bounds. This statement is illustrated by the coloured contours that are displayed in the right panel of Figure~\ref{fig:mtbplots}. The discovery reach corresponding to the latter figure are provided  in~Appendix~\ref{app:discovery}.

The constraints on the type-II~2HDM parameter space presented  in this section should be compared to the  bounds that have been derived in~\cite{Craig:2015jba,Hajer:2015gka,Gori:2016zto}. These analyses have considered the  $pp \to H/A \to t \bar t$, $pp \to t \bar t H/A \to t \bar t t \bar t$ and $pp \to b \bar b H/A \to b \bar b t \bar t$ channels, and there seems to be a consensus that future searches for $pp \to t \bar t H/A \to t \bar t t \bar t$  should provide the best sensitivity to neutral Higgses with masses $M_{H,A} > 2 m_t$ at low values of $t_\beta$. While a one-to-one comparison with the exclusions obtained in~\cite{Craig:2015jba,Hajer:2015gka,Gori:2016zto}  is not possible, we note that the limits derived in our work  appear to be more stringent than the bounds reported in the latter articles. In this context it is important to realise that the reach of the $pp \to t \bar t H/A \to t \bar t t \bar t$ searches has found to be strongly dependent on the systematic uncertainty of the normalisation of the $t \bar t$ background. The $t \bar t Z$ analysis strategy proposed by us does in contrast not rely on knowing the absolute  size of the relevant backgrounds to the level of a few percent, since the search gains its discriminating power from shape differences. We therefore expect future $t \bar t Z$ searches to lead to the most robust coverage of the 2HDM parameter space with $M_{H,A} > 350 \, {\rm GeV}$, $|M_H - M_A| > M_Z$ and $t_\beta = {\cal O} (1)$. 

\section{Conclusions}
\label{sec:conclusions}

In this article, we have proposed to use the $t \bar t Z$ and $t bW$ final states to search for heavy Higgs bosons at the LHC. These final states are interesting, because in the 2HDM context they  can arise resonantly from $pp \to H/A \to A/H Z \to t \bar t Z$ or $pp \to H/A \to H^\pm W^\mp \to t b W$,  if the requirements $M_{H/A} > M_{A/H} + M_Z$ and $M_{A/H} > 2 m_t$ or $M_{H/A} > M_{H^\pm} + M_W$ and $M_{H^\pm} > m_t + m_b$ are satisfied. In~fact,~the  involved couplings $g_{HAZ}$, $g_{Ht\bar t}$, $g_{At \bar t}$, $g_{HH^\pm W^\mp}$, $g_{AH^\pm W^\mp}$ and $g_{H^\pm tb}$  are all  non-vanishing for $c_{\beta-\alpha} \ll 1$ $\big($see (\ref{eq:ghHAZ}) and (\ref{eq:ghHAtt})$\big)$ which corresponds to the so-called alignment limit that is experimentally favoured by the agreement of the LHC Higgs measurements with SM predictions. As a result, appreciable $t\bar tZ$ and $tbW$ rates associated to $H/A$ production turn out to be a rather generic prediction in 2HDMs that feature a SM-like $125 \, {\rm GeV}$ scalar and non-SM Higges  that are heavier than about $350 \, {\rm GeV}$ with some of their masses split by around $100 \, {\rm GeV}$ or more. 

By analysing the anatomy of the $t \bar t Z$ and $t bW$ signatures in 2HDMs, we have demonstrated that many of the resulting final-state distributions show peaks and/or edges that are characteristic for the on-shell production of a resonance followed by its sequential decay into visible and invisible particles. These kinematic features can be used to disentangle the new-physics signal from the SM~background. In the case of the $t \bar t Z$ final state, we found that the difference $\Delta m \equiv m_{t \bar t Z} - m_{t \bar t}$ between the masses of the $t \bar t Z$ and $t \bar t$ systems and the transverse momentum~$p_{T,Z}$ of the $Z$ boson are powerful discriminants, while in the case of the $t bW$ final state the invariant masses $m_{b \bar b}$, $m_{b \bar b \ell}$, $m_{tb}$ and $m_{tbW}$ can be exploited for a signal-background separation. Our MC simulations have furthermore shown that the discussed observables can all be reconstructed and well measured under realistic experimental conditions through either a dedicated three-lepton~($\Delta m$ and $p_{T,Z}$) or  two-lepton~($m_{b \bar b}$ and $m_{b \bar b \ell}$) analysis strategy (see Sections~\ref{sec:strategyttZ} and \ref{sec:strategytbW}). 

Applying our three-lepton analysis strategy to simulated 14 TeV LHC data, we have then presented a comprehensive sensitivity study of the $t \bar t Z$ signature in the 2HDM framework. We have derived various 95\%~CL exclusion limits on the parameter space of the type-II 2HDM that follow from a shape fit to the $\Delta m$ (see Section~\ref{sec:results}) and $p_{T,Z}$ (see Appendix~\ref{app:pTZfit})  distributions. Our analysis shows that for the parameter choices $c_{\beta-\alpha}=0$, $t_\beta = 1$, $M_{H^\pm} = {\rm max} \left ( M_H , M_A\right )$ and assuming  $300 \, {\rm fb}^{-1}$ of integrated luminosity, it should be possible to exclude all mass combinations $\{M_{H/A}, M_{A/H}\}$ inside a roughly triangular region spanned by the points $\{ 450, 350 \} \, {\rm GeV}$, $\{1000, 500\}  \, {\rm GeV}$ and $\{1150, 350\}  \, {\rm GeV}$. In the case of the mass hierarchy $M_{H^\pm} = {\rm min} \left ( M_H , M_A\right )$ the decays $H/A \to H^\pm W^\mp$ are open and we instead find that the exclusions only reach up to around $\{700, 450\}  \, {\rm GeV}$ and $\{750, 350\}  \, {\rm GeV}$ (see~Figure~\ref{fig:lpplots}).    For the scenarios $M_H = M_{H^\pm} = M_A + 200 \, {\rm GeV}$, we have also derived the  95\%~CL exclusion limits in the  $M_A\hspace{0.5mm}$--$t_\beta$ plane for four different values of $c_{\beta -\alpha}$ (see Figure~\ref{fig:mtbplots}). For the choice $c_{\beta - \alpha} = 0.1$, we found for instance that it should be possible to exclude $t_\beta$ values up to almost 2 for $M_{A} = 350 \, {\rm GeV}$, assuming $300 \, {\rm fb}^{-1}$ of data. The HL-LHC  is expected to  improve the quoted LHC Run-3 limits noticeably. The $5 \sigma$~discovery reach in the $M_H\hspace{0.5mm}$--$\hspace{0.5mm} M_A$ and  $M_A\hspace{0.5mm}$--$t_\beta$  plane can be found in Appendix~\ref{app:discovery}. The constraints obtained in our work are complementary to and in many cases stronger than the exclusions that future LHC searches for the processes $pp \to H/A \to t \bar t$, $pp \to t \bar t H/A \to t \bar t t \bar t$ and $pp \to b \bar b H/A \to b \bar b t \bar t$ are expected to be able to provide (cf.~\cite{Craig:2015jba,Hajer:2015gka,Gori:2016zto}) on 2HDMs with neutral Higgses with $M_{H,A} > 2 m_t$ and small values of  $t_\beta$. 

In the case of the $tbW$ final state, we have found that for both the two-lepton and one-lepton analysis the signal-over-background ratios do not exceed the level of a few percent. As a result,  a reliable evaluation of the coverage of the 2HDM parameter space would require to make strong assumptions about the systematic uncertainties that plague  the normalisation and shape of the $t \bar t$~background at future LHC searches. Since we feel that it would be premature to make these assumptions, we hope that the ATLAS and CMS collaborations will explore the $tbW$ signature further. This is a worthwhile exercise, because we expect that at the HL-LHC with $3 \, {\rm ab}^{-1}$ of integrated luminosity,  this channel should also allow to probe parts of the 2HDM parameter space that feature heavy non-SM Higgses and $t_\beta$~values of the order of a few. In fact, the maximum sensitivity of our~$tbW$ analysis  arises for $M_A = M_{H^\pm} \lesssim 400 \, {\rm GeV}$, making the $tbW$ coverage complementary to that of the proposed $t \bar t Z$ search.

\acknowledgments 
We are grateful to  Stefania~Gori and Priscilla~Pani for useful conversations,  and would like to thank Valentin~Hirschi and Eleni~Vryonidou for help with  {\tt MadGraph5\_aMCNLO}. UH acknowledges the continued hospitality  and support of the CERN Theoretical Physics Department.

\begin{appendix}

\section{Shape analysis using $\bm{p_{T,Z}}$ }
\label{app:pTZfit}

\begin{figure}[!t]
\begin{center}
\includegraphics[width=\textwidth]{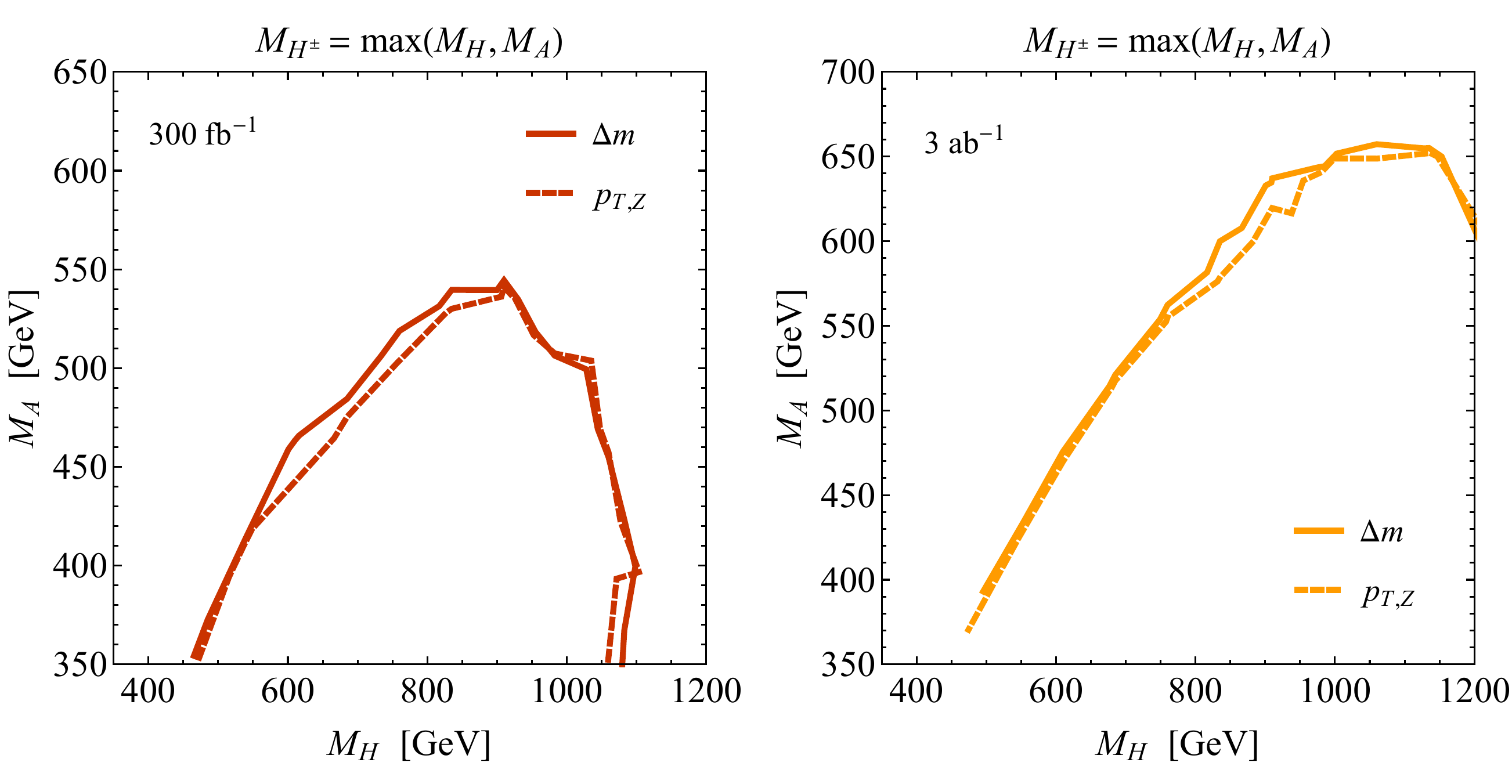} 
\vspace{-4mm}
\caption{\label{fig:ptZfit}  Comparison of the performance of the $\Delta m$ (solid contours) and $p_{T,Z}$ (dashed contours) shape fit. The left (right) panel shows the 95\%~CL exclusions obtained using $300 \, {\rm fb}^{-1}$ ($3 \, {\rm ab}^{-1}$) of $\sqrt{s} = 14 \, {\rm TeV}$ data. The used parameters resemble those employed in Figure~\ref{fig:lpplots}. See text for further details. }
\end{center}
\end{figure}

The 95\%~CL exclusions shown in Section~\ref{sec:results} have been obtained from a shape analysis of the $\Delta m$ variable~(\ref{eq:Deltam}). Since the measurement of $\Delta m$ relies on accurate measurements of $E_T^{\rm miss}$ and the momenta of jets it will likely be affected by the large pileup present in the HL-LHC phase. In contrast, pileup is expected to have only a minor impact on  $p_{T,Z}$,  because this observable can be reconstructed from the measurement of two charged leptons with high transverse momentum. To corroborate the statement made in Section~\ref{sec:results} that our $t \bar t Z$ analysis strategy is robust with respect to pileup, we compare in Figure~\ref{fig:ptZfit} the performance of the proposed $\Delta m$ and $p_{T,Z}$ shape analyses. The given  results are obtained in the type-II 2HDM employing $c_{\beta - \alpha} = 0$, $t_\beta = 1$, $M_{H^\pm} = {\rm max} \left (M_H , M_A \right )$ and only the parameter space with $M_H > M_A$ is shown. The assumptions about the uncertainties entering our analyses are specified in Section~\ref{sec:results}. One observes that a shape analysis based  on $\Delta m$~(solid~contours)  leads to  only marginally better 95\%~CL exclusions than a fit using $p_{T,Z}$~(dashed~contours) at both $300 \, {\rm fb}^{-1}$~(left panel) and $3 \, {\rm ab}^{-1}$~(right panel). This observation makes us confident that the main conclusions of this work  also hold in the presence of the large pileup expected at the~HL-LHC. 

\section{Discovery reach}
\label{app:discovery}

\begin{figure}[!t]
\begin{center}
\includegraphics[width=\textwidth]{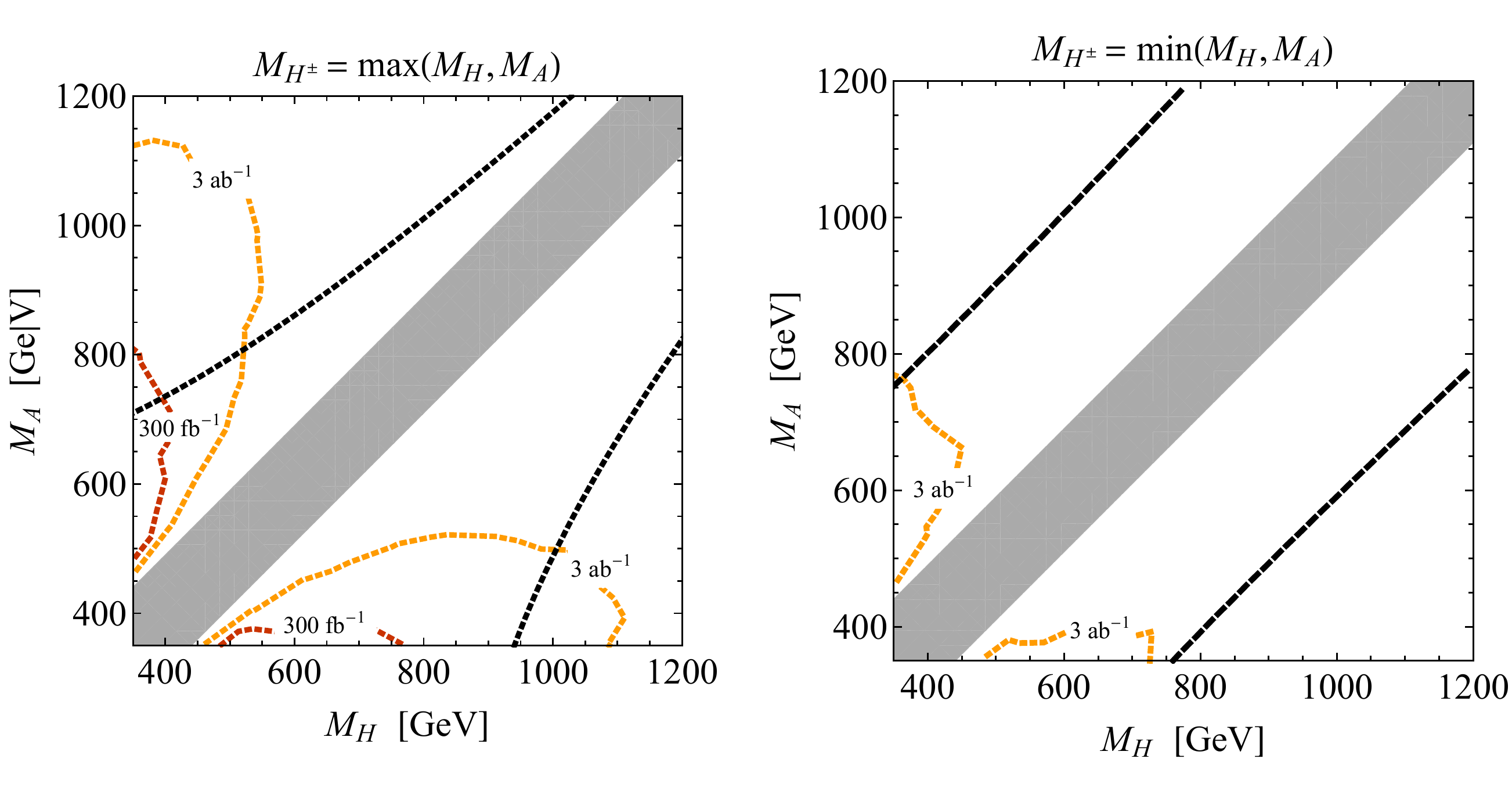} 
\vspace{-4mm}
\caption{\label{fig:discovery1} Hypothetical $5 \sigma$ discovery reach in the $M_H\hspace{0.5mm}$--$\hspace{0.5mm} M_A$ plane  arising  from a shape~fit to the $\Delta m$ observable introduced in (\ref{eq:Deltam}). The used input parameters and the meaning of the different elements that are shown in the two panels are identical to Figure~\ref{fig:lpplots}.}
\end{center}
\end{figure}

\begin{figure}[!t]
\begin{center}
\includegraphics[width=\textwidth]{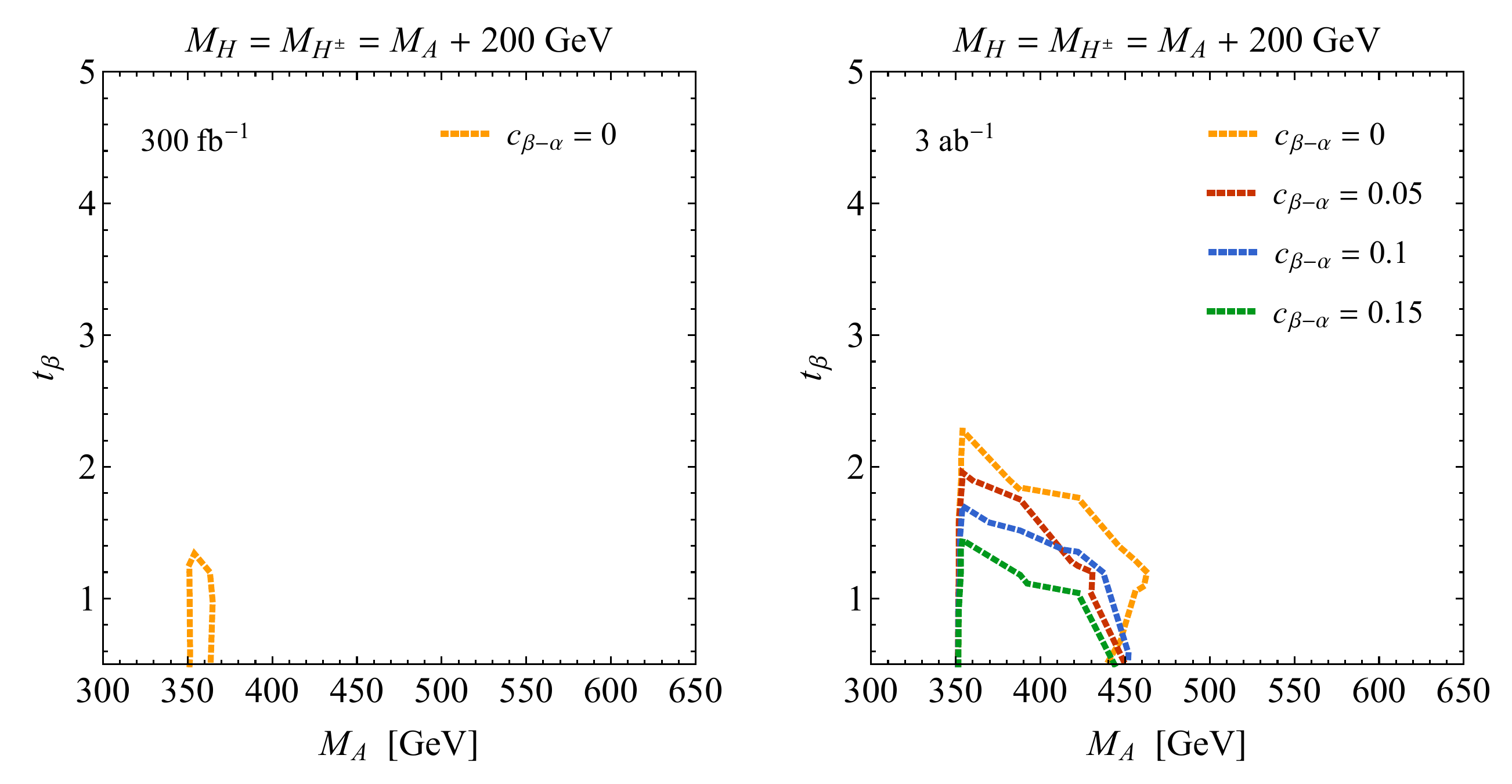} 
\vspace{-4mm}
\caption{\label{fig:discovery2}  The $5 \sigma$ discovery reach on $t_\beta$ in the type-II 2HDM resulting from a $\Delta m$ shape~fit. The used integrated luminosities and the choice of parameters is identical to that of  Figure~\ref{fig:mtbplots}.}
\end{center}
\end{figure}

In this appendix we extend the numerical study performed in Section~\ref{sec:results} by presenting the $5 \sigma$ discovery reach corresponding to Figures~\ref{fig:lpplots} and \ref{fig:mtbplots}. The limits  in the $M_H\hspace{0.5mm}$--$\hspace{0.5mm} M_A$ plane  that stem  from our $\Delta m$ shape~fit  are shown in Figure~\ref{fig:discovery1}.   The dotted red (dotted yellow) contours correspond to $300 \, {\rm fb}^{-1}$~$\big (3 \ {\rm ab}^{-1} \big)$ of  integrated luminosity at $\sqrt{s} = 14 \, {\rm TeV}$. The displayed limits  are obtained in the type-II 2HDM using $c_{\beta-\alpha} = 0$, $t_\beta=1$, $M_{H^\pm} = {\rm max} \left (M_H , M_A \right )$~(left panel) and $M_{H^\pm} = {\rm min} \left (M_H , M_A \right )$~(right~panel). The part in the   $M_H\hspace{0.5mm}$--$\hspace{0.5mm} M_A$ plane that is kinematically inaccessible  is shaded grey. One observes that the full LHC Run-3  has  a quite limited discovery reach, as it can achieve $5 \sigma$ significance only for masses $M_{H/A}$ in the range of around $[500, 800] \, {\rm GeV}$ assuming that  $M_{A/H} = 350 \, {\rm GeV}$ and $M_{H^\pm} = {\rm max} \left (M_H , M_A \right )$.  Furthermore, in the case of $M_{H^\pm} = {\rm min} \left (M_H , M_A \right )$ no discovery seems  possible with $300 \, {\rm fb}^{-1}$ of data. The situation is however expected to improve significantly  with $3 \, {\rm ab}^{-1}$ of integrated luminosity. In~the case of $M_{H^\pm} = {\rm max} \left (M_H , M_A \right )$, we find that the HL-LHC  may be able to discover all mass combinations $\{M_{H/A}, M_{A/H}\}$ inside a region spanned by the points $\{ 450, 350 \} \, {\rm GeV}$, $\{1000, 500\}  \, {\rm GeV}$ and $\{1150, 350\}  \, {\rm GeV}$, while for $M_{H^\pm} = {\rm min} \left (M_H , M_A \right )$ only heavy neutral Higgs bosons with masses in the range of  $[450, 750] \, {\rm GeV}$ can potentially be discovered if $M_{A/H} = 350 \, {\rm GeV}$. Notice that the quoted $5 \sigma$ HL-LHC   limits are similar to the 95\%~CL exclusions obtained  in Section~\ref{sec:results} for LHC~Run-3. 

Figure~\ref{fig:discovery2} displays in addition the discovery reach   in the $M_A\hspace{0.5mm}$--$t_\beta$ plane for the type-II~2HDM scenarios with $M_H = M_{H^\pm} = M_A + 200 \, {\rm GeV}$, $\lambda_3 = 6$ and four different values of $c_{\beta-\alpha}$. From  the results obtained for  $300 \, {\rm fb}^{-1}$ of $\sqrt{s} = 14 \, {\rm TeV}$ data (left panel) one can see that a discovery seems only possible in the exact alignment limit $c_{\beta-\alpha} = 0$ for $t_\beta \lesssim 1.5$ and $M_A$~masses  close to the top threshold. The discovery reach is again significantly improved at the HL-LHC with $3 \, {\rm ab}^{-1}$ of integrated luminosity (right panel), which should be able to achieve a significance of $5 \sigma$ for all scenarios with $M_A$ in the range of approximately $[350, 450] \, {\rm GeV}$, $\tan \beta = {\cal O} (1)$ and $c_{\beta - \alpha} \leq 0.15$. 

\end{appendix}


\begin{thebibliography}{89}%
\makeatletter
\providecommand \@ifxundefined [1]{%
 \@ifx{#1\undefined}
}%
\providecommand \@ifnum [1]{%
 \ifnum #1\expandafter \@firstoftwo
 \else \expandafter \@secondoftwo
 \fi
}%
\providecommand \@ifx [1]{%
 \ifx #1\expandafter \@firstoftwo
 \else \expandafter \@secondoftwo
 \fi
}%
\providecommand \natexlab [1]{#1}%
\providecommand \enquote  [1]{``#1''}%
\providecommand \bibnamefont  [1]{#1}%
\providecommand \bibfnamefont [1]{#1}%
\providecommand \citenamefont [1]{#1}%
\providecommand \href@noop [0]{\@secondoftwo}%
\providecommand \href [0]{\begingroup \@sanitize@url \@href}%
\providecommand \@href[1]{\@@startlink{#1}\@@href}%
\providecommand \@@href[1]{\endgroup#1\@@endlink}%
\providecommand \@sanitize@url [0]{\catcode `\\12\catcode `\$12\catcode
  `\&12\catcode `\#12\catcode `\^12\catcode `\_12\catcode `\%12\relax}%
\providecommand \@@startlink[1]{}%
\providecommand \@@endlink[0]{}%
\providecommand \url  [0]{\begingroup\@sanitize@url \@url }%
\providecommand \@url [1]{\endgroup\@href {#1}{\urlprefix }}%
\providecommand \urlprefix  [0]{URL }%
\providecommand \Eprint [0]{\href }%
\providecommand \doibase [0]{http://dx.doi.org/}%
\providecommand \selectlanguage [0]{\@gobble}%
\providecommand \bibinfo  [0]{\@secondoftwo}%
\providecommand \bibfield  [0]{\@secondoftwo}%
\providecommand \translation [1]{[#1]}%
\providecommand \BibitemOpen [0]{}%
\providecommand \bibitemStop [0]{}%
\providecommand \bibitemNoStop [0]{.\EOS\space}%
\providecommand \EOS [0]{\spacefactor3000\relax}%
\providecommand \BibitemShut  [1]{\csname bibitem#1\endcsname}%
\let\auto@bib@innerbib\@empty
\bibitem [{\citenamefont {Aad}\ \emph {et~al.}(2012)\citenamefont {Aad} \emph
  {et~al.}}]{Aad:2012tfa}%
  \BibitemOpen
  \bibfield  {author} {\bibinfo {author} {\bibfnamefont {G.}~\bibnamefont
  {Aad}} \emph {et~al.} (\bibinfo {collaboration} {ATLAS}),\ }\href {\doibase
  10.1016/j.physletb.2012.08.020} {\bibfield  {journal} {\bibinfo  {journal}
  {Phys. Lett.}\ }\textbf {\bibinfo {volume} {B716}},\ \bibinfo {pages} {1}
  (\bibinfo {year} {2012})},\ \Eprint {http://arxiv.org/abs/1207.7214}
  {arXiv:1207.7214 [hep-ex]} \BibitemShut {NoStop}%
\bibitem [{\citenamefont {Chatrchyan}\ \emph {et~al.}(2012)\citenamefont
  {Chatrchyan} \emph {et~al.}}]{Chatrchyan:2012xdj}%
  \BibitemOpen
  \bibfield  {author} {\bibinfo {author} {\bibfnamefont {S.}~\bibnamefont
  {Chatrchyan}} \emph {et~al.} (\bibinfo {collaboration} {CMS}),\ }\href
  {\doibase 10.1016/j.physletb.2012.08.021} {\bibfield  {journal} {\bibinfo
  {journal} {Phys. Lett.}\ }\textbf {\bibinfo {volume} {B716}},\ \bibinfo
  {pages} {30} (\bibinfo {year} {2012})},\ \Eprint
  {http://arxiv.org/abs/1207.7235} {1207.7235} \BibitemShut {NoStop}%
\bibitem [{\citenamefont {Sirunyan}\ \emph
  {et~al.}(2017{\natexlab{a}})\citenamefont {Sirunyan} \emph
  {et~al.}}]{Sirunyan:2017exp}%
  \BibitemOpen
  \bibfield  {author} {\bibinfo {author} {\bibfnamefont {A.~M.}\ \bibnamefont
  {Sirunyan}} \emph {et~al.} (\bibinfo {collaboration} {CMS}),\ }\href
  {\doibase 10.1007/JHEP11(2017)047} {\bibfield  {journal} {\bibinfo  {journal}
  {JHEP}\ }\textbf {\bibinfo {volume} {11}},\ \bibinfo {pages} {047} (\bibinfo
  {year} {2017}{\natexlab{a}})},\ \Eprint {http://arxiv.org/abs/1706.09936}
  {arXiv:1706.09936 [hep-ex]} \BibitemShut {NoStop}%
\bibitem [{\citenamefont {Aaboud}\ \emph
  {et~al.}(2017{\natexlab{a}})\citenamefont {Aaboud} \emph
  {et~al.}}]{Aaboud:2017oem}%
  \BibitemOpen
  \bibfield  {author} {\bibinfo {author} {\bibfnamefont {M.}~\bibnamefont
  {Aaboud}} \emph {et~al.} (\bibinfo {collaboration} {ATLAS}),\ }\href
  {\doibase 10.1007/JHEP10(2017)132} {\bibfield  {journal} {\bibinfo  {journal}
  {JHEP}\ }\textbf {\bibinfo {volume} {10}},\ \bibinfo {pages} {132} (\bibinfo
  {year} {2017}{\natexlab{a}})},\ \Eprint {http://arxiv.org/abs/1708.02810}
  {arXiv:1708.02810 [hep-ex]} \BibitemShut {NoStop}%
\bibitem [{\citenamefont {Aaboud}\ \emph
  {et~al.}(2018{\natexlab{a}})\citenamefont {Aaboud} \emph
  {et~al.}}]{Aaboud:2018xdt}%
  \BibitemOpen
  \bibfield  {author} {\bibinfo {author} {\bibfnamefont {M.}~\bibnamefont
  {Aaboud}} \emph {et~al.} (\bibinfo {collaboration} {ATLAS}),\ }\href@noop {}
  {\  (\bibinfo {year} {2018}{\natexlab{a}})},\ \Eprint
  {http://arxiv.org/abs/1802.04146} {arXiv:1802.04146 [hep-ex]} \BibitemShut
  {NoStop}%
\bibitem [{\citenamefont {Sirunyan}\ \emph
  {et~al.}(2018{\natexlab{a}})\citenamefont {Sirunyan} \emph
  {et~al.}}]{Sirunyan:2018ouh}%
  \BibitemOpen
  \bibfield  {author} {\bibinfo {author} {\bibfnamefont {A.~M.}\ \bibnamefont
  {Sirunyan}} \emph {et~al.} (\bibinfo {collaboration} {CMS}),\ }\href@noop {}
  {\  (\bibinfo {year} {2018}{\natexlab{a}})},\ \Eprint
  {http://arxiv.org/abs/1804.02716} {arXiv:1804.02716 [hep-ex]} \BibitemShut
  {NoStop}%
\bibitem [{\citenamefont {Aad}\ \emph {et~al.}(2016)\citenamefont {Aad} \emph
  {et~al.}}]{Khachatryan:2016vau}%
  \BibitemOpen
  \bibfield  {author} {\bibinfo {author} {\bibfnamefont {G.}~\bibnamefont
  {Aad}} \emph {et~al.} (\bibinfo {collaboration} {ATLAS, CMS}),\ }\href
  {\doibase 10.1007/JHEP08(2016)045} {\bibfield  {journal} {\bibinfo  {journal}
  {JHEP}\ }\textbf {\bibinfo {volume} {08}},\ \bibinfo {pages} {045} (\bibinfo
  {year} {2016})},\ \Eprint {http://arxiv.org/abs/1606.02266} {arXiv:1606.02266
  [hep-ex]} \BibitemShut {NoStop}%
\bibitem [{\citenamefont {Aad}\ \emph {et~al.}(2015{\natexlab{a}})\citenamefont
  {Aad} \emph {et~al.}}]{Aad:2015pla}%
  \BibitemOpen
  \bibfield  {author} {\bibinfo {author} {\bibfnamefont {G.}~\bibnamefont
  {Aad}} \emph {et~al.} (\bibinfo {collaboration} {ATLAS}),\ }\href {\doibase
  10.1007/JHEP11(2015)206} {\bibfield  {journal} {\bibinfo  {journal} {JHEP}\
  }\textbf {\bibinfo {volume} {11}},\ \bibinfo {pages} {206} (\bibinfo {year}
  {2015}{\natexlab{a}})},\ \Eprint {http://arxiv.org/abs/1509.00672}
  {arXiv:1509.00672 [hep-ex]} \BibitemShut {NoStop}%
\bibitem [{CMS(2016{\natexlab{a}})}]{CMS-PAS-HIG-16-007}%
  \BibitemOpen
  \href {https://cds.cern.ch/record/2142432} {\emph {\bibinfo {title} {{Summary
  results of high mass BSM Higgs searches using CMS run-I data}}}},\ \bibinfo
  {type} {Tech. Rep.}\ \bibinfo {number} {CMS-PAS-HIG-16-007}\ (\bibinfo
  {institution} {CERN},\ \bibinfo {address} {Geneva},\ \bibinfo {year}
  {2016})\BibitemShut {NoStop}%
\bibitem [{\citenamefont {Aaboud}\ \emph
  {et~al.}(2018{\natexlab{b}})\citenamefont {Aaboud} \emph
  {et~al.}}]{Aaboud:2017sjh}%
  \BibitemOpen
  \bibfield  {author} {\bibinfo {author} {\bibfnamefont {M.}~\bibnamefont
  {Aaboud}} \emph {et~al.} (\bibinfo {collaboration} {ATLAS}),\ }\href
  {\doibase 10.1007/JHEP01(2018)055} {\bibfield  {journal} {\bibinfo  {journal}
  {JHEP}\ }\textbf {\bibinfo {volume} {01}},\ \bibinfo {pages} {055} (\bibinfo
  {year} {2018}{\natexlab{b}})},\ \Eprint {http://arxiv.org/abs/1709.07242}
  {arXiv:1709.07242 [hep-ex]} \BibitemShut {NoStop}%
\bibitem [{\citenamefont {Sirunyan}\ \emph
  {et~al.}(2018{\natexlab{b}})\citenamefont {Sirunyan} \emph
  {et~al.}}]{Sirunyan:2018zut}%
  \BibitemOpen
  \bibfield  {author} {\bibinfo {author} {\bibfnamefont {A.~M.}\ \bibnamefont
  {Sirunyan}} \emph {et~al.} (\bibinfo {collaboration} {CMS}),\ }\href
  {\doibase 10.1007/JHEP09(2018)007} {\bibfield  {journal} {\bibinfo  {journal}
  {JHEP}\ }\textbf {\bibinfo {volume} {09}},\ \bibinfo {pages} {007} (\bibinfo
  {year} {2018}{\natexlab{b}})},\ \Eprint {http://arxiv.org/abs/1803.06553}
  {arXiv:1803.06553 [hep-ex]} \BibitemShut {NoStop}%
\bibitem [{\citenamefont {Khachatryan}\ \emph
  {et~al.}(2015{\natexlab{a}})\citenamefont {Khachatryan} \emph
  {et~al.}}]{Khachatryan:2015tra}%
  \BibitemOpen
  \bibfield  {author} {\bibinfo {author} {\bibfnamefont {V.}~\bibnamefont
  {Khachatryan}} \emph {et~al.} (\bibinfo {collaboration} {CMS}),\ }\href
  {\doibase 10.1007/JHEP11(2015)071} {\bibfield  {journal} {\bibinfo  {journal}
  {JHEP}\ }\textbf {\bibinfo {volume} {11}},\ \bibinfo {pages} {071} (\bibinfo
  {year} {2015}{\natexlab{a}})},\ \Eprint {http://arxiv.org/abs/1506.08329}
  {arXiv:1506.08329 [hep-ex]} \BibitemShut {NoStop}%
\bibitem [{CMS(2016{\natexlab{b}})}]{CMS-PAS-HIG-16-025}%
  \BibitemOpen
  \href {https://cds.cern.ch/record/2204928} {\emph {\bibinfo {title} {{Search
  for a narrow heavy decaying to bottom quark pairs in the 13~TeV data
  sample}}}},\ \bibinfo {type} {Tech. Rep.}\ \bibinfo {number}
  {CMS-PAS-HIG-16-025}\ (\bibinfo  {institution} {CERN},\ \bibinfo {address}
  {Geneva},\ \bibinfo {year} {2016})\BibitemShut {NoStop}%
\bibitem [{\citenamefont {Sirunyan}\ \emph
  {et~al.}(2018{\natexlab{c}})\citenamefont {Sirunyan} \emph
  {et~al.}}]{Sirunyan:2018taj}%
  \BibitemOpen
  \bibfield  {author} {\bibinfo {author} {\bibfnamefont {A.~M.}\ \bibnamefont
  {Sirunyan}} \emph {et~al.} (\bibinfo {collaboration} {CMS}),\ }\href
  {\doibase 10.1007/JHEP08(2018)113} {\bibfield  {journal} {\bibinfo  {journal}
  {JHEP}\ }\textbf {\bibinfo {volume} {08}},\ \bibinfo {pages} {113} (\bibinfo
  {year} {2018}{\natexlab{c}})},\ \Eprint {http://arxiv.org/abs/1805.12191}
  {arXiv:1805.12191 [hep-ex]} \BibitemShut {NoStop}%
\bibitem [{\citenamefont {Khachatryan}\ \emph
  {et~al.}(2015{\natexlab{b}})\citenamefont {Khachatryan} \emph
  {et~al.}}]{Khachatryan:2015cwa}%
  \BibitemOpen
  \bibfield  {author} {\bibinfo {author} {\bibfnamefont {V.}~\bibnamefont
  {Khachatryan}} \emph {et~al.} (\bibinfo {collaboration} {CMS}),\ }\href
  {\doibase 10.1007/JHEP10(2015)144} {\bibfield  {journal} {\bibinfo  {journal}
  {JHEP}\ }\textbf {\bibinfo {volume} {10}},\ \bibinfo {pages} {144} (\bibinfo
  {year} {2015}{\natexlab{b}})},\ \Eprint {http://arxiv.org/abs/1504.00936}
  {arXiv:1504.00936 [hep-ex]} \BibitemShut {NoStop}%
\bibitem [{\citenamefont {Aaboud}\ \emph
  {et~al.}(2018{\natexlab{c}})\citenamefont {Aaboud} \emph
  {et~al.}}]{Aaboud:2017gsl}%
  \BibitemOpen
  \bibfield  {author} {\bibinfo {author} {\bibfnamefont {M.}~\bibnamefont
  {Aaboud}} \emph {et~al.} (\bibinfo {collaboration} {ATLAS}),\ }\href
  {\doibase 10.1140/epjc/s10052-017-5491-4} {\bibfield  {journal} {\bibinfo
  {journal} {Eur. Phys. J.}\ }\textbf {\bibinfo {volume} {C78}},\ \bibinfo
  {pages} {24} (\bibinfo {year} {2018}{\natexlab{c}})},\ \Eprint
  {http://arxiv.org/abs/1710.01123} {arXiv:1710.01123 [hep-ex]} \BibitemShut
  {NoStop}%
\bibitem [{\citenamefont {Aaboud}\ \emph
  {et~al.}(2018{\natexlab{d}})\citenamefont {Aaboud} \emph
  {et~al.}}]{Aaboud:2017itg}%
  \BibitemOpen
  \bibfield  {author} {\bibinfo {author} {\bibfnamefont {M.}~\bibnamefont
  {Aaboud}} \emph {et~al.} (\bibinfo {collaboration} {ATLAS}),\ }\href
  {\doibase 10.1007/JHEP03(2018)009} {\bibfield  {journal} {\bibinfo  {journal}
  {JHEP}\ }\textbf {\bibinfo {volume} {03}},\ \bibinfo {pages} {009} (\bibinfo
  {year} {2018}{\natexlab{d}})},\ \Eprint {http://arxiv.org/abs/1708.09638}
  {arXiv:1708.09638 [hep-ex]} \BibitemShut {NoStop}%
\bibitem [{\citenamefont {Aaboud}\ \emph
  {et~al.}(2018{\natexlab{e}})\citenamefont {Aaboud} \emph
  {et~al.}}]{Aaboud:2017rel}%
  \BibitemOpen
  \bibfield  {author} {\bibinfo {author} {\bibfnamefont {M.}~\bibnamefont
  {Aaboud}} \emph {et~al.} (\bibinfo {collaboration} {ATLAS}),\ }\href
  {\doibase 10.1140/epjc/s10052-018-5686-3} {\bibfield  {journal} {\bibinfo
  {journal} {Eur. Phys. J.}\ }\textbf {\bibinfo {volume} {C78}},\ \bibinfo
  {pages} {293} (\bibinfo {year} {2018}{\natexlab{e}})},\ \Eprint
  {http://arxiv.org/abs/1712.06386} {arXiv:1712.06386 [hep-ex]} \BibitemShut
  {NoStop}%
\bibitem [{\citenamefont {Sirunyan}\ \emph
  {et~al.}(2018{\natexlab{d}})\citenamefont {Sirunyan} \emph
  {et~al.}}]{Sirunyan:2018qlb}%
  \BibitemOpen
  \bibfield  {author} {\bibinfo {author} {\bibfnamefont {A.~M.}\ \bibnamefont
  {Sirunyan}} \emph {et~al.} (\bibinfo {collaboration} {CMS}),\ }\href
  {\doibase 10.1007/JHEP06(2018)127} {\bibfield  {journal} {\bibinfo  {journal}
  {JHEP}\ }\textbf {\bibinfo {volume} {06}},\ \bibinfo {pages} {127} (\bibinfo
  {year} {2018}{\natexlab{d}})},\ \Eprint {http://arxiv.org/abs/1804.01939}
  {arXiv:1804.01939 [hep-ex]} \BibitemShut {NoStop}%
\bibitem [{\citenamefont {Khachatryan}\ \emph
  {et~al.}(2016{\natexlab{a}})\citenamefont {Khachatryan} \emph
  {et~al.}}]{Khachatryan:2016are}%
  \BibitemOpen
  \bibfield  {author} {\bibinfo {author} {\bibfnamefont {V.}~\bibnamefont
  {Khachatryan}} \emph {et~al.} (\bibinfo {collaboration} {CMS}),\ }\href
  {\doibase 10.1016/j.physletb.2016.05.087} {\bibfield  {journal} {\bibinfo
  {journal} {Phys. Lett.}\ }\textbf {\bibinfo {volume} {B759}},\ \bibinfo
  {pages} {369} (\bibinfo {year} {2016}{\natexlab{a}})},\ \Eprint
  {http://arxiv.org/abs/1603.02991} {arXiv:1603.02991 [hep-ex]} \BibitemShut
  {NoStop}%
\bibitem [{CMS(2016{\natexlab{c}})}]{CMS-PAS-HIG-16-010}%
  \BibitemOpen
  \href {https://cds.cern.ch/record/2140613} {\emph {\bibinfo {title} {{Search
  for $H \to Z( \ell^+ \ell^-)+A(b \bar b)$ with 2015 data}}}},\ \bibinfo
  {type} {Tech. Rep.}\ \bibinfo {number} {CMS-PAS-HIG-16-010}\ (\bibinfo
  {institution} {CERN},\ \bibinfo {address} {Geneva},\ \bibinfo {year}
  {2016})\BibitemShut {NoStop}%
\bibitem [{\citenamefont {Aaboud}\ \emph
  {et~al.}(2018{\natexlab{f}})\citenamefont {Aaboud} \emph
  {et~al.}}]{Aaboud:2018eoy}%
  \BibitemOpen
  \bibfield  {author} {\bibinfo {author} {\bibfnamefont {M.}~\bibnamefont
  {Aaboud}} \emph {et~al.} (\bibinfo {collaboration} {ATLAS}),\ }\href
  {\doibase 10.1016/j.physletb.2018.07.006} {\bibfield  {journal} {\bibinfo
  {journal} {Phys. Lett.}\ }\textbf {\bibinfo {volume} {B783}},\ \bibinfo
  {pages} {392} (\bibinfo {year} {2018}{\natexlab{f}})},\ \Eprint
  {http://arxiv.org/abs/1804.01126} {arXiv:1804.01126 [hep-ex]} \BibitemShut
  {NoStop}%
\bibitem [{\citenamefont {Aaboud}\ \emph
  {et~al.}(2018{\natexlab{g}})\citenamefont {Aaboud} \emph
  {et~al.}}]{Aaboud:2017cxo}%
  \BibitemOpen
  \bibfield  {author} {\bibinfo {author} {\bibfnamefont {M.}~\bibnamefont
  {Aaboud}} \emph {et~al.} (\bibinfo {collaboration} {ATLAS}),\ }\href
  {\doibase 10.1007/JHEP03(2018)174} {\bibfield  {journal} {\bibinfo  {journal}
  {JHEP}\ }\textbf {\bibinfo {volume} {03}},\ \bibinfo {pages} {174} (\bibinfo
  {year} {2018}{\natexlab{g}})},\ \Eprint {http://arxiv.org/abs/1712.06518}
  {arXiv:1712.06518 [hep-ex]} \BibitemShut {NoStop}%
\bibitem [{\citenamefont {Khachatryan}\ \emph
  {et~al.}(2016{\natexlab{b}})\citenamefont {Khachatryan} \emph
  {et~al.}}]{Khachatryan:2015tha}%
  \BibitemOpen
  \bibfield  {author} {\bibinfo {author} {\bibfnamefont {V.}~\bibnamefont
  {Khachatryan}} \emph {et~al.} (\bibinfo {collaboration} {CMS}),\ }\href
  {\doibase 10.1016/j.physletb.2016.01.056} {\bibfield  {journal} {\bibinfo
  {journal} {Phys. Lett.}\ }\textbf {\bibinfo {volume} {B755}},\ \bibinfo
  {pages} {217} (\bibinfo {year} {2016}{\natexlab{b}})},\ \Eprint
  {http://arxiv.org/abs/1510.01181} {arXiv:1510.01181 [hep-ex]} \BibitemShut
  {NoStop}%
\bibitem [{ATL(2016)}]{ATLAS-CONF-2016-071}%
  \BibitemOpen
  \href {https://cds.cern.ch/record/2206222} {\emph {\bibinfo {title} {{Search
  for Higgs boson pair production in the final state of $\gamma\gamma
  WW^*$($\rightarrow l\nu jj$) using 13.3 fb$^{-1}$ of $pp$ collision data
  recorded at $\sqrt{s}= $ 13 TeV with the ATLAS detector}}}},\ \bibinfo {type}
  {Tech. Rep.}\ \bibinfo {number} {ATLAS-CONF-2016-071}\ (\bibinfo
  {institution} {CERN},\ \bibinfo {address} {Geneva},\ \bibinfo {year}
  {2016})\BibitemShut {NoStop}%
\bibitem [{\citenamefont {Sirunyan}\ \emph
  {et~al.}(2018{\natexlab{e}})\citenamefont {Sirunyan} \emph
  {et~al.}}]{Sirunyan:2018zkk}%
  \BibitemOpen
  \bibfield  {author} {\bibinfo {author} {\bibfnamefont {A.~M.}\ \bibnamefont
  {Sirunyan}} \emph {et~al.} (\bibinfo {collaboration} {CMS}),\ }\href
  {\doibase 10.1007/JHEP08(2018)152} {\bibfield  {journal} {\bibinfo  {journal}
  {JHEP}\ }\textbf {\bibinfo {volume} {08}},\ \bibinfo {pages} {152} (\bibinfo
  {year} {2018}{\natexlab{e}})},\ \Eprint {http://arxiv.org/abs/1806.03548}
  {arXiv:1806.03548 [hep-ex]} \BibitemShut {NoStop}%
\bibitem [{\citenamefont {Khachatryan}\ \emph {et~al.}(2017)\citenamefont
  {Khachatryan} \emph {et~al.}}]{Khachatryan:2016yec}%
  \BibitemOpen
  \bibfield  {author} {\bibinfo {author} {\bibfnamefont {V.}~\bibnamefont
  {Khachatryan}} \emph {et~al.} (\bibinfo {collaboration} {CMS}),\ }\href
  {\doibase 10.1016/j.physletb.2017.01.027} {\bibfield  {journal} {\bibinfo
  {journal} {Phys. Lett.}\ }\textbf {\bibinfo {volume} {B767}},\ \bibinfo
  {pages} {147} (\bibinfo {year} {2017})},\ \Eprint
  {http://arxiv.org/abs/1609.02507} {arXiv:1609.02507 [hep-ex]} \BibitemShut
  {NoStop}%
\bibitem [{\citenamefont {Aaboud}\ \emph
  {et~al.}(2017{\natexlab{b}})\citenamefont {Aaboud} \emph
  {et~al.}}]{Aaboud:2017yyg}%
  \BibitemOpen
  \bibfield  {author} {\bibinfo {author} {\bibfnamefont {M.}~\bibnamefont
  {Aaboud}} \emph {et~al.} (\bibinfo {collaboration} {ATLAS}),\ }\href
  {\doibase 10.1016/j.physletb.2017.10.039} {\bibfield  {journal} {\bibinfo
  {journal} {Phys. Lett.}\ }\textbf {\bibinfo {volume} {B775}},\ \bibinfo
  {pages} {105} (\bibinfo {year} {2017}{\natexlab{b}})},\ \Eprint
  {http://arxiv.org/abs/1707.04147} {arXiv:1707.04147 [hep-ex]} \BibitemShut
  {NoStop}%
\bibitem [{\citenamefont {Aaboud}\ \emph
  {et~al.}(2017{\natexlab{c}})\citenamefont {Aaboud} \emph
  {et~al.}}]{Aaboud:2017uhw}%
  \BibitemOpen
  \bibfield  {author} {\bibinfo {author} {\bibfnamefont {M.}~\bibnamefont
  {Aaboud}} \emph {et~al.} (\bibinfo {collaboration} {ATLAS}),\ }\href
  {\doibase 10.1007/JHEP10(2017)112} {\bibfield  {journal} {\bibinfo  {journal}
  {JHEP}\ }\textbf {\bibinfo {volume} {10}},\ \bibinfo {pages} {112} (\bibinfo
  {year} {2017}{\natexlab{c}})},\ \Eprint {http://arxiv.org/abs/1708.00212}
  {arXiv:1708.00212 [hep-ex]} \BibitemShut {NoStop}%
\bibitem [{\citenamefont {Sirunyan}\ \emph
  {et~al.}(2017{\natexlab{b}})\citenamefont {Sirunyan} \emph
  {et~al.}}]{Sirunyan:2017hsb}%
  \BibitemOpen
  \bibfield  {author} {\bibinfo {author} {\bibfnamefont {A.~M.}\ \bibnamefont
  {Sirunyan}} \emph {et~al.} (\bibinfo {collaboration} {CMS}),\ }\href@noop {}
  {\  (\bibinfo {year} {2017}{\natexlab{b}})},\ \Eprint
  {http://arxiv.org/abs/1712.03143} {arXiv:1712.03143 [hep-ex]} \BibitemShut
  {NoStop}%
\bibitem [{\citenamefont {Djouadi}\ \emph {et~al.}(2015)\citenamefont
  {Djouadi}, \citenamefont {Maiani}, \citenamefont {Polosa}, \citenamefont
  {Quevillon},\ and\ \citenamefont {Riquer}}]{Djouadi:2015jea}%
  \BibitemOpen
  \bibfield  {author} {\bibinfo {author} {\bibfnamefont {A.}~\bibnamefont
  {Djouadi}}, \bibinfo {author} {\bibfnamefont {L.}~\bibnamefont {Maiani}},
  \bibinfo {author} {\bibfnamefont {A.}~\bibnamefont {Polosa}}, \bibinfo
  {author} {\bibfnamefont {J.}~\bibnamefont {Quevillon}}, \ and\ \bibinfo
  {author} {\bibfnamefont {V.}~\bibnamefont {Riquer}},\ }\href {\doibase
  10.1007/JHEP06(2015)168} {\bibfield  {journal} {\bibinfo  {journal} {JHEP}\
  }\textbf {\bibinfo {volume} {06}},\ \bibinfo {pages} {168} (\bibinfo {year}
  {2015})},\ \Eprint {http://arxiv.org/abs/1502.05653} {arXiv:1502.05653
  [hep-ph]} \BibitemShut {NoStop}%
\bibitem [{\citenamefont {Craig}\ \emph {et~al.}(2015)\citenamefont {Craig},
  \citenamefont {D'Eramo}, \citenamefont {Draper}, \citenamefont {Thomas},\
  and\ \citenamefont {Zhang}}]{Craig:2015jba}%
  \BibitemOpen
  \bibfield  {author} {\bibinfo {author} {\bibfnamefont {N.}~\bibnamefont
  {Craig}}, \bibinfo {author} {\bibfnamefont {F.}~\bibnamefont {D'Eramo}},
  \bibinfo {author} {\bibfnamefont {P.}~\bibnamefont {Draper}}, \bibinfo
  {author} {\bibfnamefont {S.}~\bibnamefont {Thomas}}, \ and\ \bibinfo {author}
  {\bibfnamefont {H.}~\bibnamefont {Zhang}},\ }\href {\doibase
  10.1007/JHEP06(2015)137} {\bibfield  {journal} {\bibinfo  {journal} {JHEP}\
  }\textbf {\bibinfo {volume} {06}},\ \bibinfo {pages} {137} (\bibinfo {year}
  {2015})},\ \Eprint {http://arxiv.org/abs/1504.04630} {arXiv:1504.04630
  [hep-ph]} \BibitemShut {NoStop}%
\bibitem [{\citenamefont {Hajer}\ \emph {et~al.}(2015)\citenamefont {Hajer},
  \citenamefont {Li}, \citenamefont {Liu},\ and\ \citenamefont
  {Shiu}}]{Hajer:2015gka}%
  \BibitemOpen
  \bibfield  {author} {\bibinfo {author} {\bibfnamefont {J.}~\bibnamefont
  {Hajer}}, \bibinfo {author} {\bibfnamefont {Y.-Y.}\ \bibnamefont {Li}},
  \bibinfo {author} {\bibfnamefont {T.}~\bibnamefont {Liu}}, \ and\ \bibinfo
  {author} {\bibfnamefont {J.~F.~H.}\ \bibnamefont {Shiu}},\ }\href {\doibase
  10.1007/JHEP11(2015)124} {\bibfield  {journal} {\bibinfo  {journal} {JHEP}\
  }\textbf {\bibinfo {volume} {11}},\ \bibinfo {pages} {124} (\bibinfo {year}
  {2015})},\ \Eprint {http://arxiv.org/abs/1504.07617} {arXiv:1504.07617
  [hep-ph]} \BibitemShut {NoStop}%
\bibitem [{\citenamefont {Gori}\ \emph {et~al.}(2016)\citenamefont {Gori},
  \citenamefont {Kim}, \citenamefont {Shah},\ and\ \citenamefont
  {Zurek}}]{Gori:2016zto}%
  \BibitemOpen
  \bibfield  {author} {\bibinfo {author} {\bibfnamefont {S.}~\bibnamefont
  {Gori}}, \bibinfo {author} {\bibfnamefont {I.-W.}\ \bibnamefont {Kim}},
  \bibinfo {author} {\bibfnamefont {N.~R.}\ \bibnamefont {Shah}}, \ and\
  \bibinfo {author} {\bibfnamefont {K.~M.}\ \bibnamefont {Zurek}},\ }\href
  {\doibase 10.1103/PhysRevD.93.075038} {\bibfield  {journal} {\bibinfo
  {journal} {Phys. Rev.}\ }\textbf {\bibinfo {volume} {D93}},\ \bibinfo {pages}
  {075038} (\bibinfo {year} {2016})},\ \Eprint
  {http://arxiv.org/abs/1602.02782} {arXiv:1602.02782 [hep-ph]} \BibitemShut
  {NoStop}%
\bibitem [{\citenamefont {Alvarez}\ \emph {et~al.}(2017)\citenamefont
  {Alvarez}, \citenamefont {Faroughy}, \citenamefont {Kamenik}, \citenamefont
  {Morales},\ and\ \citenamefont {Szynkman}}]{Alvarez:2016nrz}%
  \BibitemOpen
  \bibfield  {author} {\bibinfo {author} {\bibfnamefont {E.}~\bibnamefont
  {Alvarez}}, \bibinfo {author} {\bibfnamefont {D.~A.}\ \bibnamefont
  {Faroughy}}, \bibinfo {author} {\bibfnamefont {J.~F.}\ \bibnamefont
  {Kamenik}}, \bibinfo {author} {\bibfnamefont {R.}~\bibnamefont {Morales}}, \
  and\ \bibinfo {author} {\bibfnamefont {A.}~\bibnamefont {Szynkman}},\ }\href
  {\doibase 10.1016/j.nuclphysb.2016.11.024} {\bibfield  {journal} {\bibinfo
  {journal} {Nucl. Phys.}\ }\textbf {\bibinfo {volume} {B915}},\ \bibinfo
  {pages} {19} (\bibinfo {year} {2017})},\ \Eprint
  {http://arxiv.org/abs/1611.05032} {arXiv:1611.05032 [hep-ph]} \BibitemShut
  {NoStop}%
\bibitem [{\citenamefont {Aaboud}\ \emph
  {et~al.}(2017{\natexlab{d}})\citenamefont {Aaboud} \emph
  {et~al.}}]{Aaboud:2017hnm}%
  \BibitemOpen
  \bibfield  {author} {\bibinfo {author} {\bibfnamefont {M.}~\bibnamefont
  {Aaboud}} \emph {et~al.} (\bibinfo {collaboration} {ATLAS}),\ }\href
  {\doibase 10.1103/PhysRevLett.119.191803} {\bibfield  {journal} {\bibinfo
  {journal} {Phys. Rev. Lett.}\ }\textbf {\bibinfo {volume} {119}},\ \bibinfo
  {pages} {191803} (\bibinfo {year} {2017}{\natexlab{d}})},\ \Eprint
  {http://arxiv.org/abs/1707.06025} {arXiv:1707.06025 [hep-ex]} \BibitemShut
  {NoStop}%
\bibitem [{\citenamefont {Sirunyan}\ \emph
  {et~al.}(2018{\natexlab{f}})\citenamefont {Sirunyan} \emph
  {et~al.}}]{Sirunyan:2017roi}%
  \BibitemOpen
  \bibfield  {author} {\bibinfo {author} {\bibfnamefont {A.~M.}\ \bibnamefont
  {Sirunyan}} \emph {et~al.} (\bibinfo {collaboration} {CMS}),\ }\href
  {\doibase 10.1140/epjc/s10052-018-5607-5} {\bibfield  {journal} {\bibinfo
  {journal} {Eur. Phys. J.}\ }\textbf {\bibinfo {volume} {C78}},\ \bibinfo
  {pages} {140} (\bibinfo {year} {2018}{\natexlab{f}})},\ \Eprint
  {http://arxiv.org/abs/1710.10614} {arXiv:1710.10614 [hep-ex]} \BibitemShut
  {NoStop}%
\bibitem [{\citenamefont {Aaboud}\ \emph
  {et~al.}(2018{\natexlab{h}})\citenamefont {Aaboud} \emph
  {et~al.}}]{Aaboud:2018xuw}%
  \BibitemOpen
  \bibfield  {author} {\bibinfo {author} {\bibfnamefont {M.}~\bibnamefont
  {Aaboud}} \emph {et~al.} (\bibinfo {collaboration} {ATLAS}),\ }\href@noop {}
  {\  (\bibinfo {year} {2018}{\natexlab{h}})},\ \Eprint
  {http://arxiv.org/abs/1803.09678} {arXiv:1803.09678 [hep-ex]} \BibitemShut
  {NoStop}%
\bibitem [{\citenamefont {Gaemers}\ and\ \citenamefont
  {Hoogeveen}(1984)}]{Gaemers:1984sj}%
  \BibitemOpen
  \bibfield  {author} {\bibinfo {author} {\bibfnamefont {K.~J.~F.}\
  \bibnamefont {Gaemers}}\ and\ \bibinfo {author} {\bibfnamefont
  {F.}~\bibnamefont {Hoogeveen}},\ }\href {\doibase
  10.1016/0370-2693(84)91711-8} {\bibfield  {journal} {\bibinfo  {journal}
  {Phys. Lett.}\ }\textbf {\bibinfo {volume} {146B}},\ \bibinfo {pages} {347}
  (\bibinfo {year} {1984})}\BibitemShut {NoStop}%
\bibitem [{\citenamefont {Dicus}\ \emph {et~al.}(1994)\citenamefont {Dicus},
  \citenamefont {Stange},\ and\ \citenamefont {Willenbrock}}]{Dicus:1994bm}%
  \BibitemOpen
  \bibfield  {author} {\bibinfo {author} {\bibfnamefont {D.}~\bibnamefont
  {Dicus}}, \bibinfo {author} {\bibfnamefont {A.}~\bibnamefont {Stange}}, \
  and\ \bibinfo {author} {\bibfnamefont {S.}~\bibnamefont {Willenbrock}},\
  }\href {\doibase 10.1016/0370-2693(94)91017-0} {\bibfield  {journal}
  {\bibinfo  {journal} {Phys. Lett.}\ }\textbf {\bibinfo {volume} {B333}},\
  \bibinfo {pages} {126} (\bibinfo {year} {1994})},\ \Eprint
  {http://arxiv.org/abs/hep-ph/9404359} {arXiv:hep-ph/9404359 [hep-ph]}
  \BibitemShut {NoStop}%
\bibitem [{\citenamefont {Bernreuther}\ \emph {et~al.}(1998)\citenamefont
  {Bernreuther}, \citenamefont {Flesch},\ and\ \citenamefont
  {Haberl}}]{Bernreuther:1997gs}%
  \BibitemOpen
  \bibfield  {author} {\bibinfo {author} {\bibfnamefont {W.}~\bibnamefont
  {Bernreuther}}, \bibinfo {author} {\bibfnamefont {M.}~\bibnamefont {Flesch}},
  \ and\ \bibinfo {author} {\bibfnamefont {P.}~\bibnamefont {Haberl}},\ }\href
  {\doibase 10.1103/PhysRevD.58.114031} {\bibfield  {journal} {\bibinfo
  {journal} {Phys. Rev.}\ }\textbf {\bibinfo {volume} {D58}},\ \bibinfo {pages}
  {114031} (\bibinfo {year} {1998})},\ \Eprint
  {http://arxiv.org/abs/hep-ph/9709284} {arXiv:hep-ph/9709284 [hep-ph]}
  \BibitemShut {NoStop}%
\bibitem [{\citenamefont {Frederix}\ and\ \citenamefont
  {Maltoni}(2009)}]{Frederix:2007gi}%
  \BibitemOpen
  \bibfield  {author} {\bibinfo {author} {\bibfnamefont {R.}~\bibnamefont
  {Frederix}}\ and\ \bibinfo {author} {\bibfnamefont {F.}~\bibnamefont
  {Maltoni}},\ }\href {\doibase 10.1088/1126-6708/2009/01/047} {\bibfield
  {journal} {\bibinfo  {journal} {JHEP}\ }\textbf {\bibinfo {volume} {01}},\
  \bibinfo {pages} {047} (\bibinfo {year} {2009})},\ \Eprint
  {http://arxiv.org/abs/0712.2355} {arXiv:0712.2355 [hep-ph]} \BibitemShut
  {NoStop}%
\bibitem [{\citenamefont {Hespel}\ \emph {et~al.}(2016)\citenamefont {Hespel},
  \citenamefont {Maltoni},\ and\ \citenamefont {Vryonidou}}]{Hespel:2016qaf}%
  \BibitemOpen
  \bibfield  {author} {\bibinfo {author} {\bibfnamefont {B.}~\bibnamefont
  {Hespel}}, \bibinfo {author} {\bibfnamefont {F.}~\bibnamefont {Maltoni}}, \
  and\ \bibinfo {author} {\bibfnamefont {E.}~\bibnamefont {Vryonidou}},\ }\href
  {\doibase 10.1007/JHEP10(2016)016} {\bibfield  {journal} {\bibinfo  {journal}
  {JHEP}\ }\textbf {\bibinfo {volume} {10}},\ \bibinfo {pages} {016} (\bibinfo
  {year} {2016})},\ \Eprint {http://arxiv.org/abs/1606.04149} {arXiv:1606.04149
  [hep-ph]} \BibitemShut {NoStop}%
\bibitem [{\citenamefont {Glashow}\ and\ \citenamefont
  {Weinberg}(1977)}]{Glashow:1976nt}%
  \BibitemOpen
  \bibfield  {author} {\bibinfo {author} {\bibfnamefont {S.~L.}\ \bibnamefont
  {Glashow}}\ and\ \bibinfo {author} {\bibfnamefont {S.}~\bibnamefont
  {Weinberg}},\ }\href {\doibase 10.1103/PhysRevD.15.1958} {\bibfield
  {journal} {\bibinfo  {journal} {Phys. Rev.}\ }\textbf {\bibinfo {volume}
  {D15}},\ \bibinfo {pages} {1958} (\bibinfo {year} {1977})}\BibitemShut
  {NoStop}%
\bibitem [{\citenamefont {Paschos}(1977)}]{Paschos:1976ay}%
  \BibitemOpen
  \bibfield  {author} {\bibinfo {author} {\bibfnamefont {E.~A.}\ \bibnamefont
  {Paschos}},\ }\href {\doibase 10.1103/PhysRevD.15.1966} {\bibfield  {journal}
  {\bibinfo  {journal} {Phys. Rev.}\ }\textbf {\bibinfo {volume} {D15}},\
  \bibinfo {pages} {1966} (\bibinfo {year} {1977})}\BibitemShut {NoStop}%
\bibitem [{\citenamefont {Craig}\ \emph {et~al.}(2013)\citenamefont {Craig},
  \citenamefont {Galloway},\ and\ \citenamefont {Thomas}}]{Craig:2013hca}%
  \BibitemOpen
  \bibfield  {author} {\bibinfo {author} {\bibfnamefont {N.}~\bibnamefont
  {Craig}}, \bibinfo {author} {\bibfnamefont {J.}~\bibnamefont {Galloway}}, \
  and\ \bibinfo {author} {\bibfnamefont {S.}~\bibnamefont {Thomas}},\
  }\href@noop {} {\  (\bibinfo {year} {2013})},\ \Eprint
  {http://arxiv.org/abs/1305.2424} {arXiv:1305.2424 [hep-ph]} \BibitemShut
  {NoStop}%
\bibitem [{\citenamefont {Djouadi}\ \emph {et~al.}(1996)\citenamefont
  {Djouadi}, \citenamefont {Kalinowski},\ and\ \citenamefont
  {Zerwas}}]{Djouadi:1995gv}%
  \BibitemOpen
  \bibfield  {author} {\bibinfo {author} {\bibfnamefont {A.}~\bibnamefont
  {Djouadi}}, \bibinfo {author} {\bibfnamefont {J.}~\bibnamefont {Kalinowski}},
  \ and\ \bibinfo {author} {\bibfnamefont {P.~M.}\ \bibnamefont {Zerwas}},\
  }\href {\doibase 10.1007/s002880050121} {\bibfield  {journal} {\bibinfo
  {journal} {Z. Phys.}\ }\textbf {\bibinfo {volume} {C70}},\ \bibinfo {pages}
  {435} (\bibinfo {year} {1996})},\ \Eprint
  {http://arxiv.org/abs/hep-ph/9511342} {arXiv:hep-ph/9511342 [hep-ph]}
  \BibitemShut {NoStop}%
\bibitem [{\citenamefont {Djouadi}(2008{\natexlab{a}})}]{Djouadi:2005gi}%
  \BibitemOpen
  \bibfield  {author} {\bibinfo {author} {\bibfnamefont {A.}~\bibnamefont
  {Djouadi}},\ }\href {\doibase 10.1016/j.physrep.2007.10.004} {\bibfield
  {journal} {\bibinfo  {journal} {Phys. Rept.}\ }\textbf {\bibinfo {volume}
  {457}},\ \bibinfo {pages} {1} (\bibinfo {year} {2008}{\natexlab{a}})},\
  \Eprint {http://arxiv.org/abs/hep-ph/0503172} {arXiv:hep-ph/0503172 [hep-ph]}
  \BibitemShut {NoStop}%
\bibitem [{\citenamefont {Djouadi}(2008{\natexlab{b}})}]{Djouadi:2005gj}%
  \BibitemOpen
  \bibfield  {author} {\bibinfo {author} {\bibfnamefont {A.}~\bibnamefont
  {Djouadi}},\ }\href {\doibase 10.1016/j.physrep.2007.10.005} {\bibfield
  {journal} {\bibinfo  {journal} {Phys. Rept.}\ }\textbf {\bibinfo {volume}
  {459}},\ \bibinfo {pages} {1} (\bibinfo {year} {2008}{\natexlab{b}})},\
  \Eprint {http://arxiv.org/abs/hep-ph/0503173} {arXiv:hep-ph/0503173 [hep-ph]}
  \BibitemShut {NoStop}%
\bibitem [{\citenamefont {Alves}\ \emph {et~al.}(2017)\citenamefont {Alves},
  \citenamefont {El~Hedri}, \citenamefont {Taki},\ and\ \citenamefont
  {Weiner}}]{Alves:2017snd}%
  \BibitemOpen
  \bibfield  {author} {\bibinfo {author} {\bibfnamefont {D.~S.~M.}\
  \bibnamefont {Alves}}, \bibinfo {author} {\bibfnamefont {S.}~\bibnamefont
  {El~Hedri}}, \bibinfo {author} {\bibfnamefont {A.~M.}\ \bibnamefont {Taki}},
  \ and\ \bibinfo {author} {\bibfnamefont {N.}~\bibnamefont {Weiner}},\ }\href
  {\doibase 10.1103/PhysRevD.96.075032} {\bibfield  {journal} {\bibinfo
  {journal} {Phys. Rev.}\ }\textbf {\bibinfo {volume} {D96}},\ \bibinfo {pages}
  {075032} (\bibinfo {year} {2017})},\ \Eprint
  {http://arxiv.org/abs/1703.06834} {arXiv:1703.06834 [hep-ph]} \BibitemShut
  {NoStop}%
\bibitem [{\citenamefont {Haisch}\ and\ \citenamefont
  {Malinauskas}(2018)}]{Haisch:2017gql}%
  \BibitemOpen
  \bibfield  {author} {\bibinfo {author} {\bibfnamefont {U.}~\bibnamefont
  {Haisch}}\ and\ \bibinfo {author} {\bibfnamefont {A.}~\bibnamefont
  {Malinauskas}},\ }\href {\doibase 10.1007/JHEP03(2018)135} {\bibfield
  {journal} {\bibinfo  {journal} {JHEP}\ }\textbf {\bibinfo {volume} {03}},\
  \bibinfo {pages} {135} (\bibinfo {year} {2018})},\ \Eprint
  {http://arxiv.org/abs/1712.06599} {arXiv:1712.06599 [hep-ph]} \BibitemShut
  {NoStop}%
\bibitem [{\citenamefont {Haisch}\ \emph {et~al.}(2018)\citenamefont {Haisch},
  \citenamefont {Kamenik}, \citenamefont {Malinauskas},\ and\ \citenamefont
  {Spira}}]{Haisch:2018kqx}%
  \BibitemOpen
  \bibfield  {author} {\bibinfo {author} {\bibfnamefont {U.}~\bibnamefont
  {Haisch}}, \bibinfo {author} {\bibfnamefont {J.~F.}\ \bibnamefont {Kamenik}},
  \bibinfo {author} {\bibfnamefont {A.}~\bibnamefont {Malinauskas}}, \ and\
  \bibinfo {author} {\bibfnamefont {M.}~\bibnamefont {Spira}},\ }\href
  {\doibase 10.1007/JHEP03(2018)178} {\bibfield  {journal} {\bibinfo  {journal}
  {JHEP}\ }\textbf {\bibinfo {volume} {03}},\ \bibinfo {pages} {178} (\bibinfo
  {year} {2018})},\ \Eprint {http://arxiv.org/abs/1802.02156} {arXiv:1802.02156
  [hep-ph]} \BibitemShut {NoStop}%
\bibitem [{\citenamefont {Haber}\ and\ \citenamefont
  {Pomarol}(1993)}]{Haber:1992py}%
  \BibitemOpen
  \bibfield  {author} {\bibinfo {author} {\bibfnamefont {H.~E.}\ \bibnamefont
  {Haber}}\ and\ \bibinfo {author} {\bibfnamefont {A.}~\bibnamefont
  {Pomarol}},\ }\href {\doibase 10.1016/0370-2693(93)90423-F} {\bibfield
  {journal} {\bibinfo  {journal} {Phys. Lett.}\ }\textbf {\bibinfo {volume}
  {B302}},\ \bibinfo {pages} {435} (\bibinfo {year} {1993})},\ \Eprint
  {http://arxiv.org/abs/hep-ph/9207267} {arXiv:hep-ph/9207267 [hep-ph]}
  \BibitemShut {NoStop}%
\bibitem [{\citenamefont {Pomarol}\ and\ \citenamefont
  {Vega}(1994)}]{Pomarol:1993mu}%
  \BibitemOpen
  \bibfield  {author} {\bibinfo {author} {\bibfnamefont {A.}~\bibnamefont
  {Pomarol}}\ and\ \bibinfo {author} {\bibfnamefont {R.}~\bibnamefont {Vega}},\
  }\href {\doibase 10.1016/0550-3213(94)90611-4} {\bibfield  {journal}
  {\bibinfo  {journal} {Nucl. Phys.}\ }\textbf {\bibinfo {volume} {B413}},\
  \bibinfo {pages} {3} (\bibinfo {year} {1994})},\ \Eprint
  {http://arxiv.org/abs/hep-ph/9305272} {arXiv:hep-ph/9305272 [hep-ph]}
  \BibitemShut {NoStop}%
\bibitem [{\citenamefont {Gerard}\ and\ \citenamefont
  {Herquet}(2007)}]{Gerard:2007kn}%
  \BibitemOpen
  \bibfield  {author} {\bibinfo {author} {\bibfnamefont {J.~M.}\ \bibnamefont
  {Gerard}}\ and\ \bibinfo {author} {\bibfnamefont {M.}~\bibnamefont
  {Herquet}},\ }\href {\doibase 10.1103/PhysRevLett.98.251802} {\bibfield
  {journal} {\bibinfo  {journal} {Phys. Rev. Lett.}\ }\textbf {\bibinfo
  {volume} {98}},\ \bibinfo {pages} {251802} (\bibinfo {year} {2007})},\
  \Eprint {http://arxiv.org/abs/hep-ph/0703051} {arXiv:hep-ph/0703051 [HEP-PH]}
  \BibitemShut {NoStop}%
\bibitem [{\citenamefont {Grzadkowski}\ \emph {et~al.}(2011)\citenamefont
  {Grzadkowski}, \citenamefont {Maniatis},\ and\ \citenamefont
  {Wudka}}]{Grzadkowski:2010dj}%
  \BibitemOpen
  \bibfield  {author} {\bibinfo {author} {\bibfnamefont {B.}~\bibnamefont
  {Grzadkowski}}, \bibinfo {author} {\bibfnamefont {M.}~\bibnamefont
  {Maniatis}}, \ and\ \bibinfo {author} {\bibfnamefont {J.}~\bibnamefont
  {Wudka}},\ }\href {\doibase 10.1007/JHEP11(2011)030} {\bibfield  {journal}
  {\bibinfo  {journal} {JHEP}\ }\textbf {\bibinfo {volume} {11}},\ \bibinfo
  {pages} {030} (\bibinfo {year} {2011})},\ \Eprint
  {http://arxiv.org/abs/1011.5228} {arXiv:1011.5228 [hep-ph]} \BibitemShut
  {NoStop}%
\bibitem [{\citenamefont {Haber}\ and\ \citenamefont
  {O'Neil}(2011)}]{Haber:2010bw}%
  \BibitemOpen
  \bibfield  {author} {\bibinfo {author} {\bibfnamefont {H.~E.}\ \bibnamefont
  {Haber}}\ and\ \bibinfo {author} {\bibfnamefont {D.}~\bibnamefont {O'Neil}},\
  }\href {\doibase 10.1103/PhysRevD.83.055017} {\bibfield  {journal} {\bibinfo
  {journal} {Phys. Rev.}\ }\textbf {\bibinfo {volume} {D83}},\ \bibinfo {pages}
  {055017} (\bibinfo {year} {2011})},\ \Eprint {http://arxiv.org/abs/1011.6188}
  {arXiv:1011.6188 [hep-ph]} \BibitemShut {NoStop}%
\bibitem [{\citenamefont {Alwall}\ \emph {et~al.}(2014)\citenamefont {Alwall},
  \citenamefont {Frederix}, \citenamefont {Frixione}, \citenamefont {Hirschi},
  \citenamefont {Maltoni}, \citenamefont {Mattelaer}, \citenamefont {Shao},
  \citenamefont {Stelzer}, \citenamefont {Torrielli},\ and\ \citenamefont
  {Zaro}}]{Alwall:2014hca}%
  \BibitemOpen
  \bibfield  {author} {\bibinfo {author} {\bibfnamefont {J.}~\bibnamefont
  {Alwall}}, \bibinfo {author} {\bibfnamefont {R.}~\bibnamefont {Frederix}},
  \bibinfo {author} {\bibfnamefont {S.}~\bibnamefont {Frixione}}, \bibinfo
  {author} {\bibfnamefont {V.}~\bibnamefont {Hirschi}}, \bibinfo {author}
  {\bibfnamefont {F.}~\bibnamefont {Maltoni}}, \bibinfo {author} {\bibfnamefont
  {O.}~\bibnamefont {Mattelaer}}, \bibinfo {author} {\bibfnamefont {H.~S.}\
  \bibnamefont {Shao}}, \bibinfo {author} {\bibfnamefont {T.}~\bibnamefont
  {Stelzer}}, \bibinfo {author} {\bibfnamefont {P.}~\bibnamefont {Torrielli}},
  \ and\ \bibinfo {author} {\bibfnamefont {M.}~\bibnamefont {Zaro}},\ }\href
  {\doibase 10.1007/JHEP07(2014)079} {\bibfield  {journal} {\bibinfo  {journal}
  {JHEP}\ }\textbf {\bibinfo {volume} {07}},\ \bibinfo {pages} {079} (\bibinfo
  {year} {2014})},\ \Eprint {http://arxiv.org/abs/1405.0301} {arXiv:1405.0301
  [hep-ph]} \BibitemShut {NoStop}%
\bibitem [{\citenamefont {Degrande}\ \emph {et~al.}(2012)\citenamefont
  {Degrande}, \citenamefont {Duhr}, \citenamefont {Fuks}, \citenamefont
  {Grellscheid}, \citenamefont {Mattelaer},\ and\ \citenamefont
  {Reiter}}]{Degrande:2011ua}%
  \BibitemOpen
  \bibfield  {author} {\bibinfo {author} {\bibfnamefont {C.}~\bibnamefont
  {Degrande}}, \bibinfo {author} {\bibfnamefont {C.}~\bibnamefont {Duhr}},
  \bibinfo {author} {\bibfnamefont {B.}~\bibnamefont {Fuks}}, \bibinfo {author}
  {\bibfnamefont {D.}~\bibnamefont {Grellscheid}}, \bibinfo {author}
  {\bibfnamefont {O.}~\bibnamefont {Mattelaer}}, \ and\ \bibinfo {author}
  {\bibfnamefont {T.}~\bibnamefont {Reiter}},\ }\href {\doibase
  10.1016/j.cpc.2012.01.022} {\bibfield  {journal} {\bibinfo  {journal}
  {Comput. Phys. Commun.}\ }\textbf {\bibinfo {volume} {183}},\ \bibinfo
  {pages} {1201} (\bibinfo {year} {2012})},\ \Eprint
  {http://arxiv.org/abs/1108.2040} {arXiv:1108.2040 [hep-ph]} \BibitemShut
  {NoStop}%
\bibitem [{\citenamefont {Bauer}\ \emph {et~al.}(2017)\citenamefont {Bauer},
  \citenamefont {Haisch},\ and\ \citenamefont {Kahlhoefer}}]{Bauer:2017ota}%
  \BibitemOpen
  \bibfield  {author} {\bibinfo {author} {\bibfnamefont {M.}~\bibnamefont
  {Bauer}}, \bibinfo {author} {\bibfnamefont {U.}~\bibnamefont {Haisch}}, \
  and\ \bibinfo {author} {\bibfnamefont {F.}~\bibnamefont {Kahlhoefer}},\
  }\href {\doibase 10.1007/JHEP05(2017)138} {\bibfield  {journal} {\bibinfo
  {journal} {JHEP}\ }\textbf {\bibinfo {volume} {05}},\ \bibinfo {pages} {138}
  (\bibinfo {year} {2017})},\ \Eprint {http://arxiv.org/abs/1701.07427}
  {arXiv:1701.07427 [hep-ph]} \BibitemShut {NoStop}%
\bibitem [{\citenamefont {Allanach}\ \emph {et~al.}(2000)\citenamefont
  {Allanach}, \citenamefont {Lester}, \citenamefont {Parker},\ and\
  \citenamefont {Webber}}]{Allanach:2000kt}%
  \BibitemOpen
  \bibfield  {author} {\bibinfo {author} {\bibfnamefont {B.~C.}\ \bibnamefont
  {Allanach}}, \bibinfo {author} {\bibfnamefont {C.~G.}\ \bibnamefont
  {Lester}}, \bibinfo {author} {\bibfnamefont {M.~A.}\ \bibnamefont {Parker}},
  \ and\ \bibinfo {author} {\bibfnamefont {B.~R.}\ \bibnamefont {Webber}},\
  }\href {\doibase 10.1088/1126-6708/2000/09/004} {\bibfield  {journal}
  {\bibinfo  {journal} {JHEP}\ }\textbf {\bibinfo {volume} {09}},\ \bibinfo
  {pages} {004} (\bibinfo {year} {2000})},\ \Eprint
  {http://arxiv.org/abs/hep-ph/0007009} {arXiv:hep-ph/0007009 [hep-ph]}
  \BibitemShut {NoStop}%
\bibitem [{\citenamefont {Lester}\ and\ \citenamefont
  {Parker}(2001)}]{Lester:705139}%
  \BibitemOpen
  \bibfield  {author} {\bibinfo {author} {\bibfnamefont {C.~G.}\ \bibnamefont
  {Lester}}\ and\ \bibinfo {author} {\bibfnamefont {M.~A.}\ \bibnamefont
  {Parker}},\ }\href {https://cds.cern.ch/record/705139} {\enquote {\bibinfo
  {title} {{Model independent sparticle mass measurements at ATLAS}},}\ }
  (\bibinfo {year} {2001}),\ \bibinfo {note} {presented on 12 Dec
  2001}\BibitemShut {NoStop}%
\bibitem [{\citenamefont {Frixione}\ \emph {et~al.}(2008)\citenamefont
  {Frixione}, \citenamefont {Laenen}, \citenamefont {Motylinski}, \citenamefont
  {Webber},\ and\ \citenamefont {White}}]{Frixione:2008yi}%
  \BibitemOpen
  \bibfield  {author} {\bibinfo {author} {\bibfnamefont {S.}~\bibnamefont
  {Frixione}}, \bibinfo {author} {\bibfnamefont {E.}~\bibnamefont {Laenen}},
  \bibinfo {author} {\bibfnamefont {P.}~\bibnamefont {Motylinski}}, \bibinfo
  {author} {\bibfnamefont {B.~R.}\ \bibnamefont {Webber}}, \ and\ \bibinfo
  {author} {\bibfnamefont {C.~D.}\ \bibnamefont {White}},\ }\href {\doibase
  10.1088/1126-6708/2008/07/029} {\bibfield  {journal} {\bibinfo  {journal}
  {JHEP}\ }\textbf {\bibinfo {volume} {07}},\ \bibinfo {pages} {029} (\bibinfo
  {year} {2008})},\ \Eprint {http://arxiv.org/abs/0805.3067} {arXiv:0805.3067
  [hep-ph]} \BibitemShut {NoStop}%
\bibitem [{\citenamefont {Ball}\ \emph {et~al.}(2013)\citenamefont {Ball} \emph
  {et~al.}}]{Ball:2012cx}%
  \BibitemOpen
  \bibfield  {author} {\bibinfo {author} {\bibfnamefont {R.~D.}\ \bibnamefont
  {Ball}} \emph {et~al.},\ }\href {\doibase 10.1016/j.nuclphysb.2012.10.003}
  {\bibfield  {journal} {\bibinfo  {journal} {Nucl. Phys.}\ }\textbf {\bibinfo
  {volume} {B867}},\ \bibinfo {pages} {244} (\bibinfo {year} {2013})},\ \Eprint
  {http://arxiv.org/abs/1207.1303} {arXiv:1207.1303 [hep-ph]} \BibitemShut
  {NoStop}%
\bibitem [{\citenamefont {Sj{\"o}strand}\ \emph {et~al.}(2015)\citenamefont
  {Sj{\"o}strand}, \citenamefont {Ask}, \citenamefont {Christiansen},
  \citenamefont {Corke}, \citenamefont {Desai}, \citenamefont {Ilten},
  \citenamefont {Mrenna}, \citenamefont {Prestel}, \citenamefont {Rasmussen},\
  and\ \citenamefont {Skands}}]{Sjostrand:2014zea}%
  \BibitemOpen
  \bibfield  {author} {\bibinfo {author} {\bibfnamefont {T.}~\bibnamefont
  {Sj{\"o}strand}}, \bibinfo {author} {\bibfnamefont {S.}~\bibnamefont {Ask}},
  \bibinfo {author} {\bibfnamefont {J.~R.}\ \bibnamefont {Christiansen}},
  \bibinfo {author} {\bibfnamefont {R.}~\bibnamefont {Corke}}, \bibinfo
  {author} {\bibfnamefont {N.}~\bibnamefont {Desai}}, \bibinfo {author}
  {\bibfnamefont {P.}~\bibnamefont {Ilten}}, \bibinfo {author} {\bibfnamefont
  {S.}~\bibnamefont {Mrenna}}, \bibinfo {author} {\bibfnamefont
  {S.}~\bibnamefont {Prestel}}, \bibinfo {author} {\bibfnamefont {C.~O.}\
  \bibnamefont {Rasmussen}}, \ and\ \bibinfo {author} {\bibfnamefont {P.~Z.}\
  \bibnamefont {Skands}},\ }\href {\doibase 10.1016/j.cpc.2015.01.024}
  {\bibfield  {journal} {\bibinfo  {journal} {Comput. Phys. Commun.}\ }\textbf
  {\bibinfo {volume} {191}},\ \bibinfo {pages} {159} (\bibinfo {year}
  {2015})},\ \Eprint {http://arxiv.org/abs/1410.3012} {arXiv:1410.3012
  [hep-ph]} \BibitemShut {NoStop}%
\bibitem [{\citenamefont {Alioli}\ \emph {et~al.}(2010)\citenamefont {Alioli},
  \citenamefont {Nason}, \citenamefont {Oleari},\ and\ \citenamefont
  {Re}}]{Alioli:2010xd}%
  \BibitemOpen
  \bibfield  {author} {\bibinfo {author} {\bibfnamefont {S.}~\bibnamefont
  {Alioli}}, \bibinfo {author} {\bibfnamefont {P.}~\bibnamefont {Nason}},
  \bibinfo {author} {\bibfnamefont {C.}~\bibnamefont {Oleari}}, \ and\ \bibinfo
  {author} {\bibfnamefont {E.}~\bibnamefont {Re}},\ }\href {\doibase
  10.1007/JHEP06(2010)043} {\bibfield  {journal} {\bibinfo  {journal} {JHEP}\
  }\textbf {\bibinfo {volume} {06}},\ \bibinfo {pages} {043} (\bibinfo {year}
  {2010})},\ \Eprint {http://arxiv.org/abs/1002.2581} {arXiv:1002.2581
  [hep-ph]} \BibitemShut {NoStop}%
\bibitem [{\citenamefont {Aad}\ \emph {et~al.}(2015{\natexlab{b}})\citenamefont
  {Aad} \emph {et~al.}}]{Aad:2015eua}%
  \BibitemOpen
  \bibfield  {author} {\bibinfo {author} {\bibfnamefont {G.}~\bibnamefont
  {Aad}} \emph {et~al.} (\bibinfo {collaboration} {ATLAS}),\ }\href {\doibase
  10.1007/JHEP11(2015)172} {\bibfield  {journal} {\bibinfo  {journal} {JHEP}\
  }\textbf {\bibinfo {volume} {11}},\ \bibinfo {pages} {172} (\bibinfo {year}
  {2015}{\natexlab{b}})},\ \Eprint {http://arxiv.org/abs/1509.05276}
  {arXiv:1509.05276 [hep-ex]} \BibitemShut {NoStop}%
\bibitem [{\citenamefont {Campbell}\ \emph {et~al.}(2015)\citenamefont
  {Campbell}, \citenamefont {Ellis}, \citenamefont {Nason},\ and\ \citenamefont
  {Re}}]{Campbell:2014kua}%
  \BibitemOpen
  \bibfield  {author} {\bibinfo {author} {\bibfnamefont {J.~M.}\ \bibnamefont
  {Campbell}}, \bibinfo {author} {\bibfnamefont {R.~K.}\ \bibnamefont {Ellis}},
  \bibinfo {author} {\bibfnamefont {P.}~\bibnamefont {Nason}}, \ and\ \bibinfo
  {author} {\bibfnamefont {E.}~\bibnamefont {Re}},\ }\href {\doibase
  10.1007/JHEP04(2015)114} {\bibfield  {journal} {\bibinfo  {journal} {JHEP}\
  }\textbf {\bibinfo {volume} {04}},\ \bibinfo {pages} {114} (\bibinfo {year}
  {2015})},\ \Eprint {http://arxiv.org/abs/1412.1828} {arXiv:1412.1828
  [hep-ph]} \BibitemShut {NoStop}%
\bibitem [{\citenamefont {Re}(2011)}]{Re:2010bp}%
  \BibitemOpen
  \bibfield  {author} {\bibinfo {author} {\bibfnamefont {E.}~\bibnamefont
  {Re}},\ }\href {\doibase 10.1140/epjc/s10052-011-1547-z} {\bibfield
  {journal} {\bibinfo  {journal} {Eur. Phys. J.}\ }\textbf {\bibinfo {volume}
  {C71}},\ \bibinfo {pages} {1547} (\bibinfo {year} {2011})},\ \Eprint
  {http://arxiv.org/abs/1009.2450} {arXiv:1009.2450 [hep-ph]} \BibitemShut
  {NoStop}%
\bibitem [{\citenamefont {Melia}\ \emph {et~al.}(2011)\citenamefont {Melia},
  \citenamefont {Nason}, \citenamefont {R{\"o}ntsch},\ and\ \citenamefont
  {Zanderighi}}]{Melia:2011tj}%
  \BibitemOpen
  \bibfield  {author} {\bibinfo {author} {\bibfnamefont {T.}~\bibnamefont
  {Melia}}, \bibinfo {author} {\bibfnamefont {P.}~\bibnamefont {Nason}},
  \bibinfo {author} {\bibfnamefont {R.}~\bibnamefont {R{\"o}ntsch}}, \ and\
  \bibinfo {author} {\bibfnamefont {G.}~\bibnamefont {Zanderighi}},\ }\href
  {\doibase 10.1007/JHEP11(2011)078} {\bibfield  {journal} {\bibinfo  {journal}
  {JHEP}\ }\textbf {\bibinfo {volume} {11}},\ \bibinfo {pages} {078} (\bibinfo
  {year} {2011})},\ \Eprint {http://arxiv.org/abs/1107.5051} {arXiv:1107.5051
  [hep-ph]} \BibitemShut {NoStop}%
\bibitem [{\citenamefont {Nason}\ and\ \citenamefont
  {Zanderighi}(2014)}]{Nason:2013ydw}%
  \BibitemOpen
  \bibfield  {author} {\bibinfo {author} {\bibfnamefont {P.}~\bibnamefont
  {Nason}}\ and\ \bibinfo {author} {\bibfnamefont {G.}~\bibnamefont
  {Zanderighi}},\ }\href {\doibase 10.1140/epjc/s10052-013-2702-5} {\bibfield
  {journal} {\bibinfo  {journal} {Eur. Phys. J.}\ }\textbf {\bibinfo {volume}
  {C74}},\ \bibinfo {pages} {2702} (\bibinfo {year} {2014})},\ \Eprint
  {http://arxiv.org/abs/1311.1365} {arXiv:1311.1365 [hep-ph]} \BibitemShut
  {NoStop}%
\bibitem [{\citenamefont {Cacciari}\ \emph {et~al.}(2008)\citenamefont
  {Cacciari}, \citenamefont {Salam},\ and\ \citenamefont
  {Soyez}}]{Cacciari:2008gp}%
  \BibitemOpen
  \bibfield  {author} {\bibinfo {author} {\bibfnamefont {M.}~\bibnamefont
  {Cacciari}}, \bibinfo {author} {\bibfnamefont {G.~P.}\ \bibnamefont {Salam}},
  \ and\ \bibinfo {author} {\bibfnamefont {G.}~\bibnamefont {Soyez}},\ }\href
  {\doibase 10.1088/1126-6708/2008/04/063} {\bibfield  {journal} {\bibinfo
  {journal} {JHEP}\ }\textbf {\bibinfo {volume} {04}},\ \bibinfo {pages} {063}
  (\bibinfo {year} {2008})},\ \Eprint {http://arxiv.org/abs/0802.1189}
  {arXiv:0802.1189 [hep-ph]} \BibitemShut {NoStop}%
\bibitem [{\citenamefont {Cacciari}\ \emph {et~al.}(2012)\citenamefont
  {Cacciari}, \citenamefont {Salam},\ and\ \citenamefont
  {Soyez}}]{Cacciari:2011ma}%
  \BibitemOpen
  \bibfield  {author} {\bibinfo {author} {\bibfnamefont {M.}~\bibnamefont
  {Cacciari}}, \bibinfo {author} {\bibfnamefont {G.~P.}\ \bibnamefont {Salam}},
  \ and\ \bibinfo {author} {\bibfnamefont {G.}~\bibnamefont {Soyez}},\ }\href
  {\doibase 10.1140/epjc/s10052-012-1896-2} {\bibfield  {journal} {\bibinfo
  {journal} {Eur. Phys. J.}\ }\textbf {\bibinfo {volume} {C72}},\ \bibinfo
  {pages} {1896} (\bibinfo {year} {2012})},\ \Eprint
  {http://arxiv.org/abs/1111.6097} {arXiv:1111.6097 [hep-ph]} \BibitemShut
  {NoStop}%
\bibitem [{\citenamefont {Aad}\ \emph {et~al.}(2008)\citenamefont {Aad} \emph
  {et~al.}}]{Aad:2008zzm}%
  \BibitemOpen
  \bibfield  {author} {\bibinfo {author} {\bibfnamefont {G.}~\bibnamefont
  {Aad}} \emph {et~al.} (\bibinfo {collaboration} {ATLAS}),\ }\href {\doibase
  10.1088/1748-0221/3/08/S08003} {\bibfield  {journal} {\bibinfo  {journal}
  {JINST}\ }\textbf {\bibinfo {volume} {3}},\ \bibinfo {pages} {S08003}
  (\bibinfo {year} {2008})}\BibitemShut {NoStop}%
\bibitem [{\citenamefont {Aad}\ \emph {et~al.}(2009)\citenamefont {Aad} \emph
  {et~al.}}]{Aad:2009wy}%
  \BibitemOpen
  \bibfield  {author} {\bibinfo {author} {\bibfnamefont {G.}~\bibnamefont
  {Aad}} \emph {et~al.} (\bibinfo {collaboration} {ATLAS}),\ }\href@noop {} {\
  (\bibinfo {year} {2009})},\ \Eprint {http://arxiv.org/abs/0901.0512}
  {arXiv:0901.0512 [hep-ex]} \BibitemShut {NoStop}%
\bibitem [{\citenamefont {Haisch}\ \emph {et~al.}(2017)\citenamefont {Haisch},
  \citenamefont {Pani},\ and\ \citenamefont {Polesello}}]{Haisch:2016gry}%
  \BibitemOpen
  \bibfield  {author} {\bibinfo {author} {\bibfnamefont {U.}~\bibnamefont
  {Haisch}}, \bibinfo {author} {\bibfnamefont {P.}~\bibnamefont {Pani}}, \ and\
  \bibinfo {author} {\bibfnamefont {G.}~\bibnamefont {Polesello}},\ }\href
  {\doibase 10.1007/JHEP02(2017)131} {\bibfield  {journal} {\bibinfo  {journal}
  {JHEP}\ }\textbf {\bibinfo {volume} {02}},\ \bibinfo {pages} {131} (\bibinfo
  {year} {2017})},\ \Eprint {http://arxiv.org/abs/1611.09841} {arXiv:1611.09841
  [hep-ph]} \BibitemShut {NoStop}%
\bibitem [{\citenamefont {Pani}\ and\ \citenamefont
  {Polesello}(2018)}]{Pani:2017qyd}%
  \BibitemOpen
  \bibfield  {author} {\bibinfo {author} {\bibfnamefont {P.}~\bibnamefont
  {Pani}}\ and\ \bibinfo {author} {\bibfnamefont {G.}~\bibnamefont
  {Polesello}},\ }\href {\doibase 10.1016/j.dark.2018.04.006} {\bibfield
  {journal} {\bibinfo  {journal} {Phys. Dark Univ.}\ }\textbf {\bibinfo
  {volume} {21}},\ \bibinfo {pages} {8} (\bibinfo {year} {2018})},\ \Eprint
  {http://arxiv.org/abs/1712.03874} {arXiv:1712.03874 [hep-ph]} \BibitemShut
  {NoStop}%
\bibitem [{\citenamefont {Hinchliffe}\ \emph {et~al.}(1997)\citenamefont
  {Hinchliffe}, \citenamefont {Paige}, \citenamefont {Shapiro}, \citenamefont
  {Soderqvist},\ and\ \citenamefont {Yao}}]{Hinchliffe:1996iu}%
  \BibitemOpen
  \bibfield  {author} {\bibinfo {author} {\bibfnamefont {I.}~\bibnamefont
  {Hinchliffe}}, \bibinfo {author} {\bibfnamefont {F.~E.}\ \bibnamefont
  {Paige}}, \bibinfo {author} {\bibfnamefont {M.~D.}\ \bibnamefont {Shapiro}},
  \bibinfo {author} {\bibfnamefont {J.}~\bibnamefont {Soderqvist}}, \ and\
  \bibinfo {author} {\bibfnamefont {W.}~\bibnamefont {Yao}},\ }\href {\doibase
  10.1103/PhysRevD.55.5520} {\bibfield  {journal} {\bibinfo  {journal} {Phys.
  Rev.}\ }\textbf {\bibinfo {volume} {D55}},\ \bibinfo {pages} {5520} (\bibinfo
  {year} {1997})},\ \Eprint {http://arxiv.org/abs/hep-ph/9610544}
  {arXiv:hep-ph/9610544 [hep-ph]} \BibitemShut {NoStop}%
\bibitem [{\citenamefont {Gripaios}(2011)}]{Gripaios:2011na}%
  \BibitemOpen
  \bibfield  {author} {\bibinfo {author} {\bibfnamefont {B.}~\bibnamefont
  {Gripaios}},\ }\href {\doibase 10.1142/S0217751X11054826} {\bibfield
  {journal} {\bibinfo  {journal} {Int. J. Mod. Phys.}\ }\textbf {\bibinfo
  {volume} {A26}},\ \bibinfo {pages} {4881} (\bibinfo {year} {2011})},\ \Eprint
  {http://arxiv.org/abs/1110.4502} {arXiv:1110.4502 [hep-ph]} \BibitemShut
  {NoStop}%
\bibitem [{\citenamefont {Paige}(1996)}]{Paige:1996nx}%
  \BibitemOpen
  \bibfield  {author} {\bibinfo {author} {\bibfnamefont {F.~E.}\ \bibnamefont
  {Paige}},\ }\bibfield  {booktitle} {\emph {\bibinfo {booktitle} {{1996 DPF /
  DPB Summer Study On New Directions For High-Energy Physics: Proceedings,
  Snowmass 1996}}},\ }\href@noop {} {\bibfield  {journal} {\bibinfo  {journal}
  {eConf}\ }\textbf {\bibinfo {volume} {C960625}},\ \bibinfo {pages} {SUP114}
  (\bibinfo {year} {1996})},\ \bibinfo {note} {[,710(1996)]},\ \Eprint
  {http://arxiv.org/abs/hep-ph/9609373} {arXiv:hep-ph/9609373 [hep-ph]}
  \BibitemShut {NoStop}%
\bibitem [{\citenamefont {Read}(2002)}]{Read:2002hq}%
  \BibitemOpen
  \bibfield  {author} {\bibinfo {author} {\bibfnamefont {A.~L.}\ \bibnamefont
  {Read}},\ }\bibfield  {booktitle} {\emph {\bibinfo {booktitle} {{Advanced
  Statistical Techniques in Particle Physics. Proceedings, Conference, Durham,
  UK, March 18-22, 2002}}},\ }\href {\doibase 10.1088/0954-3899/28/10/313}
  {\bibfield  {journal} {\bibinfo  {journal} {J. Phys.}\ }\textbf {\bibinfo
  {volume} {G28}},\ \bibinfo {pages} {2693} (\bibinfo {year} {2002})},\
  \bibinfo {note} {[,11(2002)]}\BibitemShut {NoStop}%
\bibitem [{\citenamefont {Moneta}\ \emph {et~al.}(2010)\citenamefont {Moneta},
  \citenamefont {Belasco}, \citenamefont {Cranmer}, \citenamefont {Kreiss},
  \citenamefont {Lazzaro}, \citenamefont {Piparo}, \citenamefont {Schott},
  \citenamefont {Verkerke},\ and\ \citenamefont {Wolf}}]{Moneta:2010pm}%
  \BibitemOpen
  \bibfield  {author} {\bibinfo {author} {\bibfnamefont {L.}~\bibnamefont
  {Moneta}}, \bibinfo {author} {\bibfnamefont {K.}~\bibnamefont {Belasco}},
  \bibinfo {author} {\bibfnamefont {K.~S.}\ \bibnamefont {Cranmer}}, \bibinfo
  {author} {\bibfnamefont {S.}~\bibnamefont {Kreiss}}, \bibinfo {author}
  {\bibfnamefont {A.}~\bibnamefont {Lazzaro}}, \bibinfo {author} {\bibfnamefont
  {D.}~\bibnamefont {Piparo}}, \bibinfo {author} {\bibfnamefont
  {G.}~\bibnamefont {Schott}}, \bibinfo {author} {\bibfnamefont
  {W.}~\bibnamefont {Verkerke}}, \ and\ \bibinfo {author} {\bibfnamefont
  {M.}~\bibnamefont {Wolf}},\ }\bibfield  {booktitle} {\emph {\bibinfo
  {booktitle} {{Proceedings, 13th International Workshop on Advanced computing
  and analysis techniques in physics research (ACAT2010): Jaipur, India,
  February 22-27, 2010}}},\ }\href@noop {} {\bibfield  {journal} {\bibinfo
  {journal} {PoS}\ }\textbf {\bibinfo {volume} {ACAT2010}},\ \bibinfo {pages}
  {057} (\bibinfo {year} {2010})},\ \Eprint {http://arxiv.org/abs/1009.1003}
  {arXiv:1009.1003 [physics.data-an]} \BibitemShut {NoStop}%
\bibitem [{\citenamefont {Gunion}\ and\ \citenamefont
  {Haber}(2003)}]{Gunion:2002zf}%
  \BibitemOpen
  \bibfield  {author} {\bibinfo {author} {\bibfnamefont {J.~F.}\ \bibnamefont
  {Gunion}}\ and\ \bibinfo {author} {\bibfnamefont {H.~E.}\ \bibnamefont
  {Haber}},\ }\href {\doibase 10.1103/PhysRevD.67.075019} {\bibfield  {journal}
  {\bibinfo  {journal} {Phys. Rev.}\ }\textbf {\bibinfo {volume} {D67}},\
  \bibinfo {pages} {075019} (\bibinfo {year} {2003})},\ \Eprint
  {http://arxiv.org/abs/hep-ph/0207010} {arXiv:hep-ph/0207010 [hep-ph]}
  \BibitemShut {NoStop}%
\bibitem [{\citenamefont {Bauer}\ \emph {et~al.}(2018)\citenamefont {Bauer},
  \citenamefont {Klassen},\ and\ \citenamefont {Tenorth}}]{Bauer:2017fsw}%
  \BibitemOpen
  \bibfield  {author} {\bibinfo {author} {\bibfnamefont {M.}~\bibnamefont
  {Bauer}}, \bibinfo {author} {\bibfnamefont {M.}~\bibnamefont {Klassen}}, \
  and\ \bibinfo {author} {\bibfnamefont {V.}~\bibnamefont {Tenorth}},\ }\href
  {\doibase 10.1007/JHEP07(2018)107} {\bibfield  {journal} {\bibinfo  {journal}
  {JHEP}\ }\textbf {\bibinfo {volume} {07}},\ \bibinfo {pages} {107} (\bibinfo
  {year} {2018})},\ \Eprint {http://arxiv.org/abs/1712.06597} {arXiv:1712.06597
  [hep-ph]} \BibitemShut {NoStop}%
\bibitem [{\citenamefont {Seymour}(1995)}]{Seymour:1995np}%
  \BibitemOpen
  \bibfield  {author} {\bibinfo {author} {\bibfnamefont {M.~H.}\ \bibnamefont
  {Seymour}},\ }\href {\doibase 10.1016/0370-2693(95)00699-L} {\bibfield
  {journal} {\bibinfo  {journal} {Phys. Lett.}\ }\textbf {\bibinfo {volume}
  {B354}},\ \bibinfo {pages} {409} (\bibinfo {year} {1995})},\ \Eprint
  {http://arxiv.org/abs/hep-ph/9505211} {arXiv:hep-ph/9505211 [hep-ph]}
  \BibitemShut {NoStop}%
\bibitem [{\citenamefont {Goria}\ \emph {et~al.}(2012)\citenamefont {Goria},
  \citenamefont {Passarino},\ and\ \citenamefont {Rosco}}]{Goria:2011wa}%
  \BibitemOpen
  \bibfield  {author} {\bibinfo {author} {\bibfnamefont {S.}~\bibnamefont
  {Goria}}, \bibinfo {author} {\bibfnamefont {G.}~\bibnamefont {Passarino}}, \
  and\ \bibinfo {author} {\bibfnamefont {D.}~\bibnamefont {Rosco}},\ }\href
  {\doibase 10.1016/j.nuclphysb.2012.07.006} {\bibfield  {journal} {\bibinfo
  {journal} {Nucl. Phys.}\ }\textbf {\bibinfo {volume} {B864}},\ \bibinfo
  {pages} {530} (\bibinfo {year} {2012})},\ \Eprint
  {http://arxiv.org/abs/1112.5517} {arXiv:1112.5517 [hep-ph]} \BibitemShut
  {NoStop}%
\bibitem [{\citenamefont {Passarino}\ \emph {et~al.}(2010)\citenamefont
  {Passarino}, \citenamefont {Sturm},\ and\ \citenamefont
  {Uccirati}}]{Passarino:2010qk}%
  \BibitemOpen
  \bibfield  {author} {\bibinfo {author} {\bibfnamefont {G.}~\bibnamefont
  {Passarino}}, \bibinfo {author} {\bibfnamefont {C.}~\bibnamefont {Sturm}}, \
  and\ \bibinfo {author} {\bibfnamefont {S.}~\bibnamefont {Uccirati}},\ }\href
  {\doibase 10.1016/j.nuclphysb.2010.03.013} {\bibfield  {journal} {\bibinfo
  {journal} {Nucl. Phys.}\ }\textbf {\bibinfo {volume} {B834}},\ \bibinfo
  {pages} {77} (\bibinfo {year} {2010})},\ \Eprint
  {http://arxiv.org/abs/1001.3360} {arXiv:1001.3360 [hep-ph]} \BibitemShut
  {NoStop}%
\bibitem [{\citenamefont {Anastasiou}\ \emph {et~al.}(2011)\citenamefont
  {Anastasiou}, \citenamefont {Buehler}, \citenamefont {Herzog},\ and\
  \citenamefont {Lazopoulos}}]{Anastasiou:2011pi}%
  \BibitemOpen
  \bibfield  {author} {\bibinfo {author} {\bibfnamefont {C.}~\bibnamefont
  {Anastasiou}}, \bibinfo {author} {\bibfnamefont {S.}~\bibnamefont {Buehler}},
  \bibinfo {author} {\bibfnamefont {F.}~\bibnamefont {Herzog}}, \ and\ \bibinfo
  {author} {\bibfnamefont {A.}~\bibnamefont {Lazopoulos}},\ }\href {\doibase
  10.1007/JHEP12(2011)058} {\bibfield  {journal} {\bibinfo  {journal} {JHEP}\
  }\textbf {\bibinfo {volume} {12}},\ \bibinfo {pages} {058} (\bibinfo {year}
  {2011})},\ \Eprint {http://arxiv.org/abs/1107.0683} {arXiv:1107.0683
  [hep-ph]} \BibitemShut {NoStop}%
\bibitem [{\citenamefont {Anastasiou}\ \emph {et~al.}(2012)\citenamefont
  {Anastasiou}, \citenamefont {Buehler}, \citenamefont {Herzog},\ and\
  \citenamefont {Lazopoulos}}]{Anastasiou:2012hx}%
  \BibitemOpen
  \bibfield  {author} {\bibinfo {author} {\bibfnamefont {C.}~\bibnamefont
  {Anastasiou}}, \bibinfo {author} {\bibfnamefont {S.}~\bibnamefont {Buehler}},
  \bibinfo {author} {\bibfnamefont {F.}~\bibnamefont {Herzog}}, \ and\ \bibinfo
  {author} {\bibfnamefont {A.}~\bibnamefont {Lazopoulos}},\ }\href {\doibase
  10.1007/JHEP04(2012)004} {\bibfield  {journal} {\bibinfo  {journal} {JHEP}\
  }\textbf {\bibinfo {volume} {04}},\ \bibinfo {pages} {004} (\bibinfo {year}
  {2012})},\ \Eprint {http://arxiv.org/abs/1202.3638} {arXiv:1202.3638
  [hep-ph]} \BibitemShut {NoStop}%
\end{thebibliography}

%

\end{document}